\documentclass[twocolumn]{autart}

\usepackage{mathrsfs}
\usepackage{graphicx}          
\usepackage{amsmath}

\newcommand*{\QEDA}{\hfill\ensuremath{\square}}%

\usepackage{amssymb}
\usepackage{balance}
\usepackage{color}
\usepackage{caption}
\usepackage{subcaption}
\usepackage{enumerate}
\usepackage[shortlabels]{enumitem}

\makeatletter
\renewcommand*{\@opargbegintheorem}[3]{\trivlist
  \item[\hskip \labelsep{\bfseries #1\ #2}] \textbf{(#3)}\ \itshape}
\makeatother
 
\usepackage{empheq}
\usepackage{mathrsfs}

\usepackage{mathtools}

\usepackage{tikz}

\usepackage{pgfplots}
\usetikzlibrary{shapes.geometric, arrows}
\usetikzlibrary{arrows,automata}
\usetikzlibrary{arrows,positioning}
\usetikzlibrary{shapes.multipart}

\usepackage{circuitikz}
\usetikzlibrary{matrix}
\usetikzlibrary{positioning,arrows}
\usetikzlibrary{shapes.multipart}
\renewcommand*{\thefootnote }{\textit{\alph{footnote}}}

\usepackage{graphicx,amssymb,amstext,amsmath}
\usepackage[bookmarks=true]{hyperref}
\usepackage{algorithm,algorithmicx,algpseudocode}
\algnewcommand{\algorithmicgoto}{\textbf{go to}}%
\algnewcommand{\Goto}[1]{\algorithmicgoto~\ref{#1}}%
\algnewcommand{\LineComment}[1]{\Statex \(\triangleright\) #1}
\algnewcommand{\LineCommentN}[1]{\Statex \hspace{1cm}\(\triangleright\) #1}
\newcommand{\RNum}[1]{\uppercase\expandafter{\romannumeral #1\relax}}

\newtheorem{innercustomthm}{Class}
\newenvironment{class}[1]
  {\renewcommand\theinnercustomthm{#1}\innercustomthm}
  {\endinnercustomthm}

\newtheorem{problem}{Problem}
\newtheorem{exm}{Example}
\newenvironment{proof}{{\bf \emph{Proof.} }}{\hfill $\square$ \\}
\definecolor{officegreen}{rgb}{0.0, 0.5, 0.0}
\newcommand{\moh}[1]{{\color{black} #1}}
\newcommand{\moha}[1]{{\color{black} #1}}
\newcommand{\yong}[1]{{\color{black} #1}}

\newcommand{\moham}[1]{{\color{black} #1}}
\newcommand{\syong}[1]{{\color{black} #1}}
\newcommand{\mohk}[1]{{\color{black} #1}}
\begin{document}

\begin{frontmatter}

\title{Simultaneous State and Unknown Input 
Set-Valued Observers for \moham{\syong{Some} 
Classes of} Nonlinear Dynamical Systems}


\author{Mohammad Khajenejad}\ead{mkhajene@asu.edu},    
\author{Sze Zheng Yong}\ead{szyong@asu.edu}               

\address{School for Engineering of Matter, Transport and Energy, Arizona State University, Tempe, AZ, USA}  


\begin{abstract}                          
In this paper, we propose fixed-order set-valued \moham{(in the form of $\ell_2$-norm hyperballs)} observers for \syong{some classes of} nonlinear bounded-error dynamical systems with unknown input signals that 
	 simultaneously find bounded \moham{hyperballs} of states and unknown inputs that include the true states and inputs. \moham{Necessary and} sufficient conditions in the form of Linear Matrix Inequalities (LMIs) for the stability \moham{(in the sense of quadratic stability)} of the proposed observers are derived for ($\mathcal{M},\gamma$)-Quadratically Constrained (($\mathcal{M},\gamma$)-QC) systems, which includes several 
	 classes of nonlinear systems: 
	(\RNum{1}) Lipschitz continuous, (\RNum{2}) ($
	 \mathcal{A},\gamma$)-QC* and (\RNum{3}) Linear Parameter-Varying (LPV) systems. This 
	 new quadratic constraint property 
	 is at least as general as the incremental quadratic constraint property for nonlinear systems and is proven in the paper to embody 
	 a broad range of nonlinearities. 
	 In addition, we design the optimal $\mathcal{H}_{\infty}$ observer among those that satisfy the \moham{quadratic} stability conditions \syong{and show that the design results in Uniformly Bounded-Input Bounded-State (UBIBS) estimate radii/error dynamics and uniformly bounded sequences of the estimate radii.} Furthermore, we provide \syong{closed-form upper bound sequences} for the estimate radii and sufficient condition for their convergence to steady state. 
		 Finally, the effectiveness of the proposed set-valued observers is demonstrated through illustrative examples, where we compare the performance of our observers with some existing observers.  
\end{abstract}
\end{frontmatter}
\section{Introduction}
\subsection{Motivation}
Cyber-physical systems (CPS), e.g., power grids, autonomous vehicles, medical devices, etc., are systems in which computational and communication components are deeply intertwined 
and interacting with each other in several ways to control physical entities. Such \emph{safety-critical} systems, if jeopardized or malfunctioning, can cause serious detriment to their operators, controlled physical components and the people utilizing them. A need for CPS security and for new designs of resilient estimation and control has been accentuated by recent incidents of attacks on CPS, e.g., the Iranian nuclear plant, the Ukrainian power grid and the Maroochy water service \cite{Cardenas.2008b,Farwell.2011,Richards.2008,slay2007lessons,ukraine.2016}. 
Specifically, false data injection attack is one of the most serious types of attacks on CPS, where malicious and/or strategic attackers inject counterfeit data signals into the sensor measurements and actuator signals to cause damage, steal energy etc. Given the strategic nature of these false data injection signals, they are not well-modeled by a zero-mean, Gaussian white noise nor by signals with known bounds. 
Nevertheless, reliable estimates of states and unknown inputs are indispensable and useful for the sake of attack identification, resilient control, etc. Similar state and input estimation problems can be found across a wide range of disciplines, from input estimation in physiological systems \cite{DeNicolao.1997}, to fault detection and diagnosis \cite{Patton.1989}, to the estimation of mean areal precipitation \cite{Kitanidis.1987}.
\subsection{Literature Review}
Much of the research focus has been on simultaneous input and state estimation for stochastic systems with unknown inputs, assuming that the noise signals are Gaussian and white, via minimum variance unbiased (MVU) estimation approaches (e.g.,  
 \cite{Gillijns.2007,Gillijns.2007b,Yong.Zhu.ea.CDC15_General,Yong.Zhu.ea.Automatica16}),
     modified double-model adaptive estimation methods (e.g, \cite{lu2016framework}), or robust regularized least square approaches as in \cite{abolhasani2018robust}. 
   However, 
       {in order to address} ``set-membership" estimation problems in bounded-error settings, as is considered in this paper, where \emph{set-valued} uncertainties are considered and \emph{sets} of states and unknown inputs that are compatible with measurements are desired (cf. \cite{yong2018simultaneous} for a comprehensive discussion), {the development of set-theoretic approaches are needed.} 
      
    In the context of attack-resilient estimation, numerous approaches were proposed for deterministic systems (e.g., \cite{chong2015observability,Fawzi.2014,Pasqualetti.2013,shoukry2015imhotep}), stochastic systems (e.g., 
\cite{kim2016attack,yong2015resilient,yong2016tcps}) and bounded-error systems 
\cite{nakahira2015dynamic,pajic2015attack,yong2016robust}, against false data injection attacks. 
     However, these approaches mainly yield point estimates, i.e, the most likely or best single estimate, as opposed to set-valued estimates. On the other hand, the work in \cite{pajic2015attack} only computes error bounds for the initial state and \cite{nakahira2015dynamic} assumes zero initial states and does not consider any optimality criteria. 
     
In addition, unknown input observer designs for different classes of discrete-time nonlinear systems are relatively scarce. The method proposed in \cite{veluvolu2008discrete} leverages discrete-time sliding mode observers for calculating state and unknown input point estimates, assuming that the unknown inputs have \emph{known} bounds  and evolve as \emph{known} functions of states, which may not be directly applicable when considering adversaries in the system. The authors in \cite{korbicz2007lmi} proposed an LMI-based state estimation approach for globally Lipschitz nonlinear discrete-time systems, but did not consider unknown input reconstruction. An LMI-based approach was also used in \cite{ha2004state} for simultaneous estimation of state and unknown input for a class of continuous-time dynamic systems with Lipschitz nonlinearities, but the authors did not address optimality nor stability properties for their observer, as well as only considered point estimates. 

The work in \cite{accikmecse2011observers} designed an asymptotic observer to calculate point estimates for a class of continuous-time systems whose nonlinear terms satisfy an \emph{incremental quadratic inequality} property. Similar work was done for the same class of discrete-time nonlinear systems in \cite{chakrabarty2016state}{, while the set-valued state estimation approach in \cite{rego2020guaranteed} 
{uses mean value and first-order Taylor extensions to efficiently propagate} constrained zonotopes through nonlinear mappings}. However, none of them addressed unknown input estimation. Moreover, the restrictive assumption of bounded unknown inputs is needed in order to obtain convergent estimates. 

Considering bounded unknown inputs, but with unknown bounds, the work in \cite{chen2018nonlinear} applied second-order series expansions to construct observer for state estimation in nonlinear discrete-time systems. The authors also provided sufficient conditions for stability and optimality of the designed estimator. However, their method does not compute unknown input estimates. On the other hand, in a recent and interesting work in \cite{chakrabarty2019estimating}, the authors designed an observer for reconstruction of unknown exogenous inputs in nonlinear continuous-time systems with unknown and potentially unbounded inputs, providing sufficient LMI conditions for $\mathcal{L}_\infty$-stability of the observer. However, their observer does not simultaneously estimate the state, the unknown input estimates are point estimates and the optimality of their approach was not analyzed.  
  
 The author in \cite{yong2018simultaneous} and references therein discussed the advantages of set-valued observers (when compared to point estimators) in terms of providing hard accuracy bounds, which are important to guarantee safety \cite{blanchini2012convex}. 
In addition, the use of \emph{fixed-order} set-valued methods can help decrease the complexity of optimal observers \cite{milanese1991optimal}, 
 which grows with time. 
Hence, 
a fixed-order set-valued observer for linear time-invariant discrete time systems with bounded errors, was presented in \cite{yong2018simultaneous}, that simultaneously finds bounded \moham{hyperballs} of compatible states and unknown inputs that are optimal in the minimum $\mathcal{H}_\infty$-norm sense, i.e., with minimum average power amplification. In our preliminary work \cite{khajenejad2019simultaneous}, we extended the approach in \cite{yong2018simultaneous} to \emph{linear parameter-varying} systems, while in \cite{khajenejadasimultaneous}, we generalized the method to \emph{switched linear} systems with unknown modes and sparse unknown inputs (attacks), and in \cite{khajenejad2020simultaneousinterval1,khajenejad2020simultaneousinterval2} we considered simultaneous input and state \emph{interval-valued} observers. In this paper, 
 we aim to further design novel 
 {set}-valued observers 
 for broader classes of nonlinear systems.
\subsection{Contribution}
The goal of this paper is to bridge the gap between set-valued state estimation without unknown inputs and point-valued state and unknown input estimation for a broad range of \moham{time-varying} nonlinear dynamical systems, \moham{with nonlinear observation functions}. In particular, we propose fixed-order set-valued \moham{(in the form of $\ell_2$-norm hyperball\syong{s})} observers for nonlinear discrete-time bounded-error systems that simultaneously find {uniformly} bounded sets of states and unknown inputs that contain the true state and unknown input, are compatible/consistent with measurement outputs \moham{as well as nonlinear observation function}, and are optimal in the minimum $\mathcal{H}_{\infty}$-norm sense, i.e., with minimum average power amplification.

First, we introduce a novel class of \moham{time-varying} nonlinear vector fields 
{that we call $(\mathcal{M},\gamma)$-Quadratically Constrained ($(\mathcal{M},\gamma)$-QC) functions} 
and show that they include a broad range of nonlinearities. We also derive some results on the relationship between {$(\mathcal{M},\gamma)$-QC} functions with 
other classes of nonlinearities, such as {the} incremental{ly} quadratic{ally} constrain{ed}, Lipschitz continuous and linear parameter-varying (LPV) functions. 

Then, we present our three-step recursive set-valued observer for nonlinear discrete-time systems. In particular, {in the absence of noise,}
we derive 
sufficient \moham{and necessary} conditions in the form of LMIs for the \syong{existence and} stability of the observer \moham{in the sense of quadratic stability} 
for {this class of $(\mathcal{M},\gamma)$-QC functions, as well as three} 
classes of nonlinearities: 
(I) Lipschitz continuous, (II) {($\mathcal{A},\gamma$)}-QC* and (III) LPV systems. 

Furthermore, we design $\mathcal{H}_{\infty}$ observers among those that satisfy the {quadratic} stability conditions, using semi-definite programs with additional LMIs constraints for each of the aforementioned {system classes.} 
\moham{Then, we show that our $\mathcal{H}_{\infty}$ observer design \syong{leads to estimate radii \syong{dynamics} that are Uniformly Bounded-Input Bounded-State (UBIBS),} 
which equivalently results in 
uniformly bounded sequences of the estimate radii \syong{in the presence of noise}. \syong{Moreover, we derive closed-form expressions for upper bound sequences for the estimate radii  as well as}
		 sufficient conditions for the convergence \moham{of these radii upper \syong{bound} sequences to steady state.}} 
 
Note that we consider \emph{completely unknown} inputs (different from noise signals) without imposing any assumptions on them (such as being norm bounded, with limited energy or being included in some known set). 
Considering resilient estimation in cyber-physical systems, our set-valued observers are applicable for achieving attack-resiliency against false data injection attacks on both actuator and sensor signals. It is worth mentioning that in our preliminary work \cite{khajenejad2019simultaneous}, we designed \moham{hyperball}-valued $\mathcal{H}_{\infty}$ observers for the special case of LPV systems. 
\section{Preliminary Material}
\subsection{Notation}
$\mathbb{R}^n$ denotes the $n$-dimensional Euclidean space, {$\mathbb{Z}$ nonnegative integers} and 
$\mathbb{N}$ {positive} integers, while $\mathbb{R}_+$ and $\mathbb{R}_{++}$ denote the sets of non-negative and positive real numbers {and $0_{n \times m}$ denotes a zero matrix in $\mathbb{R}^{n \times m}$}. For a vector $v \triangleq [v_1,\dots, v_n] \in \mathbb{R}^n$ and a matrix $M \in \mathbb{R}^{p \times q}$, $\|v\|\triangleq \sqrt{v^\top v}$ and $\|M\|$ denote their (induced) 2-norm, {while $\|v\|_{\infty} \triangleq \max\limits_{i \in \{1,\dots,n\}} |v_i|$ denotes the $\infty$-norm of $v$.} 
Moreover, the transpose, inverse, 
Moore-Penrose pseudoinverse 
and rank of $M$ are given by $M^\top$, $M^{-1}$, $M^\dagger$ 
and ${\rm rk}(M)$. For symmetric matrices $S$ and $S'$, $S \succ 0$, $S \succeq 0$, $S \prec 0$, and $S \preceq 0$ 
mean that $S$ is positive semi-definite, positive definite, positive semi-negative and positive negative, respectively. 
Moreover, $S \succeq S'$ and $S \preceq S'$ mean $S-S'$ is positive semi-definite and positive semi-negative, respectively. 
\subsection{Structural Properties} \label{sec:properties}
\begin{figure}[!h]
\begin{center}
\begin{tikzpicture}[every text node part/.style={align=center}]
  \node (a) {LPV};
  \node[below=1cm of a] (c) {Lipschitz};
  \node[right=2.7cm of c] (f) {($\mathcal{M},\gamma$)-QC};
  \node[below=1cm of f] (g) {$\delta$-QC};
   \node[below=1cm of c] (d) {Incrementally \\Sector \\ Bounded};
    \node[below=1cm of g] (e) {Matrix Parametrized};
  \node[right=3cm of a] (b) {($\mathcal{A},\gamma$)-QC*};
  \draw[->]
    (c) edge[double] node[fill=white] {\tiny{Prop. \ref{prop:LiptodeltaQC}--\ref{prop:weakerness}}} (f)
    (c) edge[] node[fill=white] { \tiny{+Bounded Domain }} (b)
    (a) edge[double] node[fill=white] {\tiny{Prop. \ref{prop:LPVtoLip}}} (c)
   (b)  edge [bend right]  node [fill=white]  {\tiny{$\gamma \geq 0$+Prop. \ref{prop:qcstar_Lip}}}(c)
    (f)  edge [bend right] node [right] {\tiny{+Prop. \ref{prop:AA}}}(b)
    (e) edge [double] node [fill=white] {\tiny{\cite[Section 5.2]{accikmecse2011observers}}}(g)
    (d) edge [double] node [fill=white] {\tiny{\cite[Section 5.1]{accikmecse2011observers}}}(g)
    (g) edge [double] node [right] {\tiny{Prop. \ref{prop:weakerness}}} (f)
    (c) edge [double] node[fill=white] {\tiny{Prop. \ref{prop:LiptodeltaQC}}} (g)
    (a) edge [bend left] node [fill=white] {\tiny{+Bounded Domain}}  (b);
  \draw[->,bend right, double] (b) to (f);
\end{tikzpicture}
 \caption{Relationships between different classes of nonlinearities. $\Rightarrow$ denotes direct implication, while $\rightarrow$ denotes implication with addition assumptions. {Lipschitz, LPV, $\delta$-QC, {($\mathcal{M},\gamma$)}-QC and {($\mathcal{A},\gamma$)}-QC* nonlinearities are defined in Definitions \ref{def:Lipschitz}--\ref{defn:A}. Incrementally Sector Bounded nonlinearities can be characterized by four fixed matrices $K_{11}$, $K_{12}$, $K_{21}$, and $K_{22}$, and a set  of matrices, {$\mathcal{X}$}. In particular, they satisfy $(K_{11} \Delta x + K_{12} \Delta f_k)^\top X (K_{21} \Delta x + K_{22} \Delta f_k) \geq 0, \ \forall X \in \mathcal{X}$, $\moham{\forall k \in \mathbb{Z}}$. Matrix Parametrized nonlinearities can be characterized by some known set of matrices, {${\Im}$}. Specifically, \moham{at each time step $k \in \mathbb{Z}$,} for any $\Delta x$ and corresponding $\Delta f_k$, there exists a $\Theta_k \in \Im$ such that $\Delta f_k = \Theta_k \Delta x$. The reader is referred to \cite[Sections 5.1 and 5.2]{accikmecse2011observers} for detailed discussions about these two classes of nonlinearities, which are omitted here {for the sake of brevity.}} 
 \label{fig:relations} }  
 \end{center}
 \end{figure}
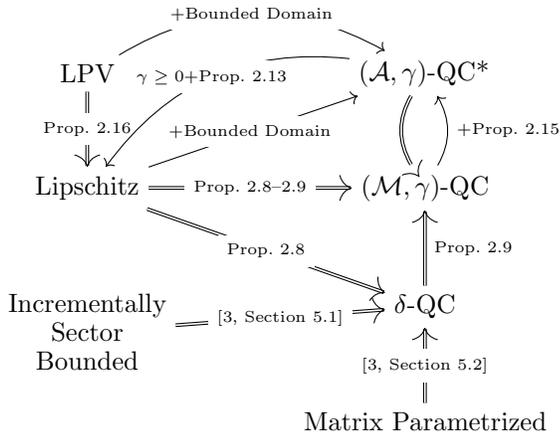

Here, we briefly introduce the structural properties that we will consider for our different classes of systems, so that we will be able to refer to them later when needed.
\begin{defn}[Strong Detectability \cite{yong2018simultaneous}] \label{def:strongdetectable}
 The following bounded-error Linear Time Invariant (LTI) system:
 \begin{align} \label{eq:LTI}
\begin{array}{ll}
x_{k+1}&=A x_k+B u_k+G d_k + w_k,\\
y_k&=C x_k +D u_k + H d_k + v_k, \end{array}
\end{align}
 i.e., the tuple $(A,G,C,H)$, is strongly detectable if 
$y_k=0 \ \forall \, k \geq 0 $ implies $x_k \to 0$ as $k \to \infty$, 
for all initial states and input sequences $\{d_i\} _{i\in \mathbb{N}}$, where $A,B,G,C,D,H$ are known constant matrices with appropriate dimensions, and $x_k$, $u_k$, $y_k$, $d_k$, $w_k$ and $v_k$ are system state, known input, output, unknown input, bounded norm process noise and measurement noise signals, respectively.
\end{defn}
\begin{rem} \label{rem:strongdetectable}
 Several necessary and sufficient rank conditions are provided in \cite[Theorem 1]{yong2018simultaneous} to check the strong detectability of
 system \eqref{eq:LTI}, i.e., $(A,G,C,H)$, including  ${\rm rk}\,\mathcal{R}_S(z)\hspace{-0.05cm}\triangleq\hspace{-0.05cm}{\rm rk}\begin{bmatrix} zI-A & -G \\ C & H \end{bmatrix}\hspace{-0.05cm}=\hspace{-0.05cm}n+p$ for all $z \in \mathbb{C}, |z| \geq 1$. 
It is worth mentioning that all the aforementioned conditions are equivalent to 
the system being minimum-phase (i.e., the invariant zeros of $\mathcal{R}_S(z)$ are stable).
Moreover, strong detectability implies that the pair $(A,C)$ is detectable, and if $l=p$, then strong detectability implies that the pair $(A,G)$ is stabilizable (cf. \cite[Theorem 1]{yong2018simultaneous} for more details).
\end{rem}
\begin{defn}[\moham{Time-Varying Lipschitzness}] \label{def:Lipschitz}
A \moham{time-varying} vector field $f_k(\cdot):\mathcal{D}_{f_k} \rightarrow \mathbb{R}^m$ is globally $L_f$-Lipschitz continuous on $\mathcal{D}_{f_k} \subseteq \mathbb{R}^n$, if there exists $L_f \in \mathbb{R}_{++}$, such that \moham{for any time step $k \in \mathbb{Z}$,} $\|f_k(x_1)-f_k(x_2)\| \leq L_f \|x_1-x_2\|$, $ \forall x_1,x_2 \in D_{f_k}$.
\end{defn}
\begin{defn} [\moham{Time-Varying} LPV Functions]\label{defn:convexcomb}
A \moham{time-varying} vector field $f_k(\cdot):\mathbb{R}^p \rightarrow \mathbb{R}^q $ is Linear Parameter-Varying (LPV), if at each time step $k \in \mathbb{Z}$, $f_k(x_k)$ can be decomposed into a convex combination of linear functions with \emph{known} coefficients, i.e., $\forall k \geq 0, \exists N \in \mathbb{N}$ such that $\forall i \in \{1,\dots ,N\}$, there exist known  $ \lambda_{i,k} \in [0,1]$ and $A^i \in \mathbb{R}^{p \times q}$ such that $\textstyle{\sum}_{i=1}^N \lambda_{i,k}=1$ and $f_k(x_k)=\textstyle{\sum}_{i=1}^N \lambda_{i,k}A^ix_k$. Each linear function $A^ix_k$ is called a \emph{constituent} function of the original nonlinear \moham{time-varying  vector field $f_k(x_k)$ at $x_k \in \mathcal{D}_{f_k}$}. 
\end{defn}
\moham{Next, through the following definition, we slightly generalize the class of $\delta$-QC functions introduced in \cite{accikmecse2011observers} to time-varying $\delta$-QC vector fields.}
\begin{defn}[\moham{Time-Varying} $\delta$-QC Mappings] \label{def:deltaQC}
A symmetric matrix $M \in \mathbb{R}^{(n_q +n_f ) \times (n_q +n_f )}$ is an incremental multiplier matrix ($\delta MM$) for \moham{time-varying vector field} $f_k(\cdot) \in \mathcal{D}_{f_k} \subseteq \mathbb{R}^{n_q}$, if the following incremental quadratic constraint ($\delta$-QC) is satisfied for all $q_1, q_2 \in \mathcal{D}_{f_k}$ \moham{and \syong{for} all time step\syong{s} $k \in \mathbb{Z}$}: $\begin{bmatrix} \Delta f_k^\top & \Delta q^\top \end{bmatrix} {M} \begin{bmatrix} \Delta f_k^\top & \Delta q^\top \end{bmatrix}^\top \geq 0$, where $\Delta q \triangleq q_2-q_1$ and $\Delta f_k \triangleq f_k(q_2)-f_k(q_1)$.
\end{defn}
Now, we introduce a new class of systems we call {($\mathcal{M},\gamma$)-
quadratic{ally} constrained 
(($\mathcal{M},\gamma$)-QC}) {systems} that is at least as general as $\delta$-QC {systems} and includes a broad range of nonlinearities. 
\begin{defn}[\moham{Time-Varying {$(\mathcal{M},\gamma$)}-QC} Functions] \label{def:DQM}
A \moham{time-varying} vector field $f_k(\cdot):\mathcal{D}_{f_k} \subseteq \mathbb{R}^p \rightarrow \mathbb{R}^q $ is $(\mathcal{M},\gamma)$-
Quadratic{ally} Constrained, i.e., $(\mathcal{M},\gamma)$-QC, if there exist symmetric matrix $\mathcal{M} \in \mathbb{R}^{(p+q) \times (p+q)} $ and $\gamma \in {\mathbb{R}}$ such that 
\begin{align} \label{eq:dqc-main}
\begin{bmatrix} \Delta f_k^\top & \Delta x^\top \end{bmatrix} \mathcal{M} \begin{bmatrix} \Delta f_k^\top & \Delta x^\top \end{bmatrix}^\top \geq \gamma,
\end{align} 
for all $x_1,x_2 \in \mathcal{D}_{f_k}$ \moham{and \syong{for} all time step\syong{s} $k \in \mathbb{Z}$}, where $\Delta x \triangleq x_2-x_1$ and $\Delta f_k \triangleq f_k(x_2)-f_k(x_1)$. We call $\mathcal{M}$ a 
multiplier matrix for function $f_k(\cdot)$.
\end{defn}
First of all, we show that a vector field may satisfy ($\mathcal{M},\gamma$)-QC property with different pairs of $(\mathcal{M},\gamma)$'s. 
For clarity, all proofs are provided in the Appendix.
\begin{prop} \label{prop:DQCextend}
Suppose $f_k(\cdot)$ is $(\mathcal{M},\gamma)$-QC. Then it is also $(\kappa \mathcal{M},\kappa \gamma)$-QC, $(\nu \mathcal{M}, \gamma)$-QC, $( \mathcal{M}, \rho)$-DQC and $(\mathcal{M}',\gamma)$-QC for every $\kappa \geq 0$, $ \nu \geq 1$, $\rho \leq \gamma$ and $\mathcal{M}' \succeq \mathcal{M}$. 
\end{prop}
Moreover, we next show that the ($\mathcal{M},\gamma$)-QC property includes Lipschitz continuity and is at least as general as the incremental quadratic constrain{t} ($\delta$-QC) 
property (cf. Definition \ref{def:deltaQC}), which recently has received considerable attention in nonlinear system state and input estimation (e.g., in \cite{accikmecse2011observers,chakrabarty2019estimating,chakrabarty2016state}). 
Consequently, the class of ($\mathcal{M},\gamma$)-QC functions is a generalization of several types of nonlinearities (cf. Corollary \ref{cor:generalization} \moham{and Figure \ref{fig:relations}}). 
 \begin{prop} \label{prop:LiptodeltaQC}
 Every globally $L_f$-Lipschitz continuous function is $\delta$-QC with multiplier matrix $M=\begin{bmatrix} -I & 0 \\ 0 & L^2_f I\end{bmatrix}$.
 \end{prop}
 \begin{prop} \label{prop:weakerness}
  Every nonlinearity 
  {that} is $\delta$-QC with multiplier matrix $M$ is ($M,\gamma$)-QC for any $\gamma \leq 0$. 
\end{prop}
\begin{cor} \label{cor:generalization}
Lipschitz nonlinearities, incrementally sector bounded nonlinearities   and nonlinearities with matrix parameterizations, 
etc., which are $\delta$-QC {(cf. Figure \ref{fig:relations} and \cite[Sections 5.1--5.2]{accikmecse2011observers})}, are also ($\mathcal{M},\gamma$)-QC ({the reader is referred to}  \cite{accikmecse2011observers,chakrabarty2019estimating,chakrabarty2016state} for definitions, demonstrations and more detailed examples). 
\end{cor}
Next, we provide some instances of nonlinear ($\mathcal{M},\gamma$)-QC vector fields, that to our best knowledge, have not been shown to be $\delta$-QC. 
\begin{exm}
Consider any monotonically increasing vector-filed $f_k(\cdot):\mathbb{R}^n \rightarrow \mathbb{R}^n$, which is not necessarily globally Lipschitz. By monotonically increasing, we mean that $\Delta f_k^\top \Delta x \geq 0$, for all $x_1,x_2 \in \mathcal{D}_{f_k}$, where $\Delta f_k$ and $\Delta x$ are defined in Definition \ref{def:DQM}. As simple examples, the reader can consider $g(x)=x^5$ with $\mathcal{D}_g=\mathbb{R}$ or $h(x)=\tan(x)$ with $\mathcal{D}_h=(-\frac{\pi}{2},\frac{\pi}{2})$. It can be easily validated that such functions are $(\mathcal{M},\gamma)$-QC with $\mathcal{M}=\begin{bmatrix} 0_{n \times n} & I_{n \times n} \\ I_{n \times n} & 0_{n \times n} \end{bmatrix}$ and any $\gamma \leq 0$. Similarly, any monotonically decreasing vector field is $(\mathcal{M},\gamma)$-QC. 
\end{exm}
\begin{exm} \label{exm:DQC2}
Now, consider $f(x)=x^2$ with $\mathcal{D}_f=[ -\overline{x}, \overline{x}] \in \mathbb{R}$, $\overline{x} \geq 0.5$, which is not a monotone function. Let $\mathcal{M}_0=\begin{bmatrix} -1 & 1 \\ 1 & -1 \end{bmatrix}$. It can be verified that $\begin{bmatrix} (\Delta f)^\top & (\Delta x)^\top \end{bmatrix} \mathcal{M}_0 \begin{bmatrix} (\Delta f)^\top & (\Delta x)^\top \end{bmatrix}^\top=\| \Delta f - \Delta x \|^2 = \| (\Delta x)^2-\Delta x \|^2 \geq -[2\overline{x}(2\overline{x}+1)]^2=-9$, for $x_1,x_2 \in \mathcal{D}_f$. Hence, $f(x)=x^2$ for all $x \in [ -\overline{x}, \overline{x}] \in \mathbb{R}$ with $\overline{x} \geq 0.5$ is $(\mathcal{M}_0,-9)$-QC.
\end{exm}
Furthermore, considering a specific structure for the 
multiplier matrix $\mathcal{M}$, we introduce a new class of functions that is a subset of the ($\mathcal{M},\gamma$)-QC class.   
\begin{defn}[\moham{Time-Varying} {($\mathcal{A},\gamma$)}-QC* Functions] \label{defn:A}
A \moham{time-varying} vector field $f_k(\cdot)$ is a{n} ($\mathcal{A},\gamma$)-QC* function, if it is ($\mathcal{M},\gamma$)-QC for some ${\gamma \in \mathbb{R}}$ and 
there exists a known $\mathcal{A}\in \mathbb{R}^{n \times n}$, such that $ \mathcal{M} = \begin{bmatrix} -I_{n \times n} & \mathcal{A} \\ \mathcal{A}^\top & -\mathcal{A}^\top \mathcal{A} \end{bmatrix} $. {We call $\mathcal{A}$ a{n auxiliary} multiplier matrix for function $f_k(\cdot)$.}  
\end{defn}
Now we present some results that establish the relationships between the aforementioned classes of nonlinearities.  
\begin{prop} \label{prop:assumptions12}
Suppose $f_k(\cdot)$ is globally $L_f$-Lipschitz continuous and the state space, $\mathcal{X}$, is bounded, i.e., there exists $r \in \mathbb{R}_{+}$ such that for all $x \in \mathcal{X}$, $\|x\| \leq r$. Then, $f_k(\cdot)$ is a ($\mathcal{A},\gamma$)-QC* function with $\mathcal{A}=0_{n \times n}$, 
and $\gamma = -4r^2 L^2_f$.  
\end{prop}
\moham{
\begin{prop}\label{prop:qcstar_Lip}
Suppose $f_k(\cdot)$ is ($\mathcal{A},\gamma$)-QC* with some $\gamma \geq 0$ and $\mathcal{A} \neq 0_{n \times n}$. Then, $f_k(\cdot)$ is globally $L_f$-Lipschitz continuous with $L_f=\sqrt{\lambda_{\max}(\mathcal{A}^\top \mathcal{A})}$.   
\end{prop}
}
\begin{lem} \label{lem:decomposition}
Suppose vector field $f_k(\cdot)$ can be decomposed as the sum of an affine and a bounded nonlinear function $g_k(\cdot)$ {at each time step $k \in \mathbb{Z}$}, i.e., $f_k(x)=Ax+h+g_k(x), \forall k \in \mathbb{Z}$, where $A \in \mathbb{R}^{n \times n}$, $h \in \mathbb{R}^n$ and $\|g_k(x)\| \leq r \in \mathbb{R}_{+}$ for all $x \in \mathcal{D}_{g_k}$ {and all $k \in  \mathbb{Z}$}. Then, $f_k(\cdot)$ is a{n} ($\mathcal{A},\gamma$)-QC* function with $\mathcal{A}=A$ 
and any $\gamma \leq -(2r)^2$.
\end{lem}
Note that some ($\mathcal{M},\gamma$)-QC systems are also ($\mathcal{A},\gamma$)-QC*. The following Proposition \ref{prop:AA} helps with finding such an $\mathcal{A}$ for some specific structures of $\mathcal{M}$.
\begin{prop} \label{prop:AA}
Suppose $f_k(\cdot):\mathbb{R}^{2n} \rightarrow \mathbb{R}^{2n}$ is a ($\mathcal{M},\gamma$)-QC vector field, with $\mathcal{M}=\begin{bmatrix} \mathcal{M}_{11} & \mathcal{M}_{12} \\ \mathcal{M}^\top_{12} & \mathcal{M}_{22} \end{bmatrix}$, where $\mathcal{M}_{11}, \mathcal{M}_{12}, \mathcal{M}_{22} \in \mathbb{R}^{n \times n}$, $\mathcal{M}_{11}+I_{n \times n} \preceq 0$ and $\mathcal{M}_{22}+\mathcal{M}^\top_{12}\mathcal{M}_{12} \preceq 0$. Then, $f_k(\cdot)$ is a{n} ($\mathcal{A},\gamma$)-QC* function with $\mathcal{A}=\mathcal{M}_{12}$.
\end{prop}
The reader can verify that such sufficient conditions in Proposition \ref{prop:AA} hold for the function in Example \ref{exm:DQC2}.
\begin{prop} \label{prop:LPVtoLip}
Every LPV function $f_k(\cdot)$ with constituent matrices $A^i, \forall i \in {1\dots N}$, is $\|A^m\|$-globally Lipschitz continuous, where $\|A^m\|= \max_{i \in {1\dots N}} \|A^i\|$. 
\end{prop}
\begin{cor}
As a direct corollary of Propositions \ref{prop:assumptions12} and \ref{prop:LPVtoLip}, any bounded domain LPV function is a{n $(\mathcal{A},\gamma$)}-QC* function.
\end{cor}
Figure \ref{fig:relations} summarizes all the above results on the relationships between several classes of nonlinearities.
 
 \moham{We conclude this section by stating a prior result on affine abstraction, which will be used in the next section.
\begin{prop}\cite[Affine Abstraction]{singh2018mesh}\label{prop:affine abstractions}
Consider the vector field $q(\cdot):\mathcal{B} \subset \mathbb{R}^{n_q} \to \mathbb{R}^{m_q}$, where $\mathcal{B}$ is an interval 
with $\overline{z},\underline{z},\mathcal{V}_{\mathcal{B}}$ being its maximal, minimal and set of vertices, respectively. Suppose $\overline{A}_{\mathcal{B}},\underline{A}_{\mathcal{B}}, \overline{e}_{\mathcal{B}}, \underline{e}_{\mathcal{B}},\theta_{\mathcal{B}}$ \moham{is a solution of} the following {linear program (LP):}   
\begin{align} \label{eq:abstraction}
&\min\limits_{\theta,\overline{A},\underline{A},\overline{e},\underline{e}} {\theta} \\
\nonumber & \quad \quad s.t \ \underline{A} {z}_{s}+\underline{e}+\sigma \leq q({z}_{s}) \leq \overline{A} {z}_{s}+\overline{e}-\sigma, \\
\nonumber &\quad \quad \quad \ (\overline{A}-\underline{A}) {z}_{s}+\overline{e}-\underline{e}-2\sigma \leq \theta \mathbf{1}_{m_q} , \ \forall z_s \in \mathcal{V}_{\mathcal{B}},
\end{align}  
where $\mathbf{1}_{m_q} \in \mathbb{R}^{m_q}$ is a vector of ones and $\sigma$ can be computed via {\cite[Proposition 1]{singh2018mesh}} 
{for different function classes}. Then, $\underline{A} {z}+\underline{e} \leq q(z) \leq  \overline{A} {z}+\overline{e}, \forall z \in \mathcal{B}$. We call $\overline{A},\underline{A}$ upper and lower affine abstraction 
\syong{gradients} of function $q(\cdot)$ on 
$\mathcal{B}$. 
\end{prop}
}
\section{Problem Statement}\label{sec:problemstatement}
In this section, we describe the system, vector field and unknown input signal assumptions as well as formally state the observer design problem.

\noindent\textbf{\emph{System Assumptions.}} 
Consider the following nonlinear time-varying discrete-time bounded-error system 
\begin{align} \label{eq:system}
\hspace{-0.1cm}\moham{\begin{array}{rl}
x_{k+1}&=f_k(x_k)+ \hat{B}_k\hat{u}_k+\hat{G}\hat{g}_k(x_k,\hat{u}_k,d^s_k)+ Ww_k,\\
y_k&=\mu(x_k,\hat{u}_k)+\hat{D}_k\hat{u}_k+\hat{H}\hat{h}_k(x_k,\hat{u}_k,d^o_k)+ \hat{v}_k, \end{array}}\hspace{-0.1cm}
\end{align}
where $x_k \in \mathcal{X} \subseteq \mathbb{R}^n$ is the state vector at time $k \in \mathbb{N}$, 
$\hat{u}_k \in \mathcal{U} \subseteq \mathbb{R}^m$ is a known input vector and $y_k \in \mathbb{R}^l$ is the measurement vector. The process noise $w_k \in \mathbb{R}^n$ and the measurement noise $\hat{v}_k \in \mathbb{R}^l$ are assumed to be bounded, with $\|w_k\| \leq \eta_w$ and $\|\hat{v}_k\|\leq \eta_{\hat{v}}$ (thus, they are $\ell_\infty$ sequences) and $W$ is known and of appropriate dimension. We also assume an estimate $\hat{x}_0$ of the initial state $x_0$ is available, where $\|\hat{x}_0-x_0\|\leq \delta_0^x$. 

The mapping $ f_k(.):\mathbb{R}^n \to \mathbb{R}^n$ is a known \emph{time-varying} nonlinear function, \moham{\syong{while} $d^s_k \in \mathbb{R}^p_s$ and $d^o_k \in \mathbb{R}^p_o$ can be interpreted as arbitrary {(and \emph{different})}
unknown inputs \syong{that affect} 
the state and observation equations through the known \emph{time-varying nonlinear} vector fields $\hat{g}_k(.,.):\mathbb{R}^n \times \mathbb{R}^m \times \mathbb{R}^{p_s} \to \mathbb{R}^{n_{\hat{G}}}$ and $\hat{h}_k(.,.):\mathbb{R}^n \times \mathbb{R}^m \times \mathbb{R}^{p_o} \to \mathbb{R}^{n_{\hat{H}}}$, respectively.} {Moreover,} 
$\hat{G} \in \mathbb{R}^{n \times n_{\hat{G}}}$ and $\hat{H} \in \mathbb{R}^{l \times n_{\hat{H}}}$ are known {time-invariant} matrices, {whereas} 
$\hat{B}_k \in \mathbb{R}^{n \times m}$ and $\hat{D}_k \in \mathbb{R}^{l \times m}$ are known \emph{time-varying}  matrices at each time step $k$.

\syong{On the other hand,} 
\moham{$\mu(\cdot,\cdot):\mathbb{R}^n \times \mathbb{R}^m \to \mathbb{R}^l$ is 
 \syong{a} known observation mapping 
for which we 
consider two cases:
\begin{case}
$\mu(x_k,\hat{u}_k)=Cx_k+\tilde{D}\hat{u}_k$, i.e., $\mu(\cdot,\cdot)$ is linear in $x_k$ and $\hat{u}_k$.
\end{case}
\begin{case}
$\mu(\cdot,\cdot)$ is nonlinear \syong{with bounded interval domains, i.e., there exist known intervals $\overline{\mathcal{X}}$ and $\overline{\mathcal{U}}$ such that $\mathcal{X} \subseteq \overline{\mathcal{X}} \subset \mathbb{R}^n$ and $\mathcal{U} \subseteq \overline{\mathcal{U}} \subset \mathbb{R}^m$.} 
\end{case}
\syong{In the second case,} we can apply our previously developed \emph{affine abstraction} tools in \cite{singh2018mesh} (cf. Proposition \ref{prop:affine abstractions})  to derive \emph{affine upper and lower abstractions} for $\mu(\cdot,\cdot)$  using Proposition \ref{prop:affine abstractions} and the linear program therein to \syong{obtain} 
$\overline{C},\underline{C},\overline{D},\underline{D},\overline{e}$ and $\underline{e}$  with appropriate dimensions, such that for all $x_k \in \mathcal{X}$ and $\hat{u}_k \in \mathcal{U}$: 
\begin{align} \label{eq:abstraction1}
\underline{C}x_k +\underline{D}\hat{u}_k+\underline{e} \leq \mu(x_k,\hat{u}_k) \leq \overline{C}x_k +\overline{D}\hat{u}_k+\overline{e}.
\end{align}
Next, by taking the average of \syong{the} upper and lower affine abstractions in \eqref{eq:abstraction1} and adding \syong{an additional bounded} 
disturbance/perturbation term $v^a_k$ (with its $\infty$-norm being less than half of the maximum distance), it is straightforward to express the inequalities in \eqref{eq:abstraction1} as the following equality:
 \begin{align} \label{eq:ffine_observation}
 \mu(x_k,\hat{u}_k)= {C}x_k +\tilde{D}\hat{u}_k+{e}+v^a_k,
\end{align} 
\syong{with} $C \triangleq \frac{1}{2}(\overline{C}+\underline{C})$, $\tilde{D} \triangleq \frac{1}{2}(\overline{D}+\underline{D})$, $e \triangleq \frac{1}{2}(\overline{e}+\underline{e})$, \syong{$\|v^a_k\|_{\infty} \leq \eta_{v^a} \triangleq \frac{1}{2} \theta^*$, where} 
$\theta^*$ is the solution to the LP \syong{in} \eqref{eq:abstraction}. \syong{In a nutshell, the above procedure ``approximates'' $ \mu(x_k,\hat{u}_k)$ with an appropriate linear term and accounts for the ``approximation error" using an additional disturbance/noise term.} 

Then, using \eqref{eq:ffine_observation}, the system in \eqref{eq:system} can be rewritten as: 
\begin{align} \label{eq:system2}
\hspace{-0.1cm}\begin{array}{rl}
x_{k+1}&=f_k(x_k)+ {B}_k{u}_k+\hat{G}{g}_k(x_k,{u}_k,d^s_k)+ Ww_k,\\
y_k&=Cx_k+{D}_k{u}_k+\hat{H}{h}_k(x_k,{u}_k,d^o_k)+ {v}_k, \end{array}\hspace{-0.1cm}
\end{align}
where $B_k \triangleq [\hat{B}_k \ 0]$, $u_k \triangleq [\hat{u}^\top_k \ e^\top]^\top$, ${g}_k(x_k,{u}_k,d^s_k) \triangleq \hat{g}_k(x_k,\hat{u}_k,d^s_k)$, \syong{$D_k \triangleq [\hat{D}_k\hspace{-0.15cm}+\hspace{-0.15cm}\tilde{D} \ I]$}, ${h}_k(x_k,{u}_k,d^o_k) \triangleq \hat{h}_k(x_k,\hat{u}_k,d^o_k)$ and $v_k \triangleq \hat{v}_k+v^a_k$ with $\|v_k\| \leq \eta_v \triangleq \eta_{\hat{v}}+\eta_{v^a}$. \syong{Similarly, it is straightforward to notice that Case 1 can also be represented by \eqref{eq:system2} with $B_k \triangleq \hat{B}_k$, $u_k\triangleq\hat{u}_k$, $D_k \triangleq \hat{D}_k\hspace{-0.05cm}+\hspace{-0.05cm}\tilde{D}$ and $v_k \triangleq \hat{v}_k$ with $\|v_k\| \leq \eta_v \triangleq \eta_{\hat{v}}$.}

{Now, courtesy of the fact that the unknown input signals $d^s_k$ and $d^o_k$ in \eqref{eq:system2} can be completely arbitrary, we can lump} 
the nonlinear functions with the unknown inputs in \eqref{eq:system2} into a {newly} defined unknown input signal $d_k \in \mathbb{R}^p$ {to} obtain an equivalent representation of the system \eqref{eq:system2} as follows:
\begin{align} \label{eq:system3}
\begin{array}{ll}
x_{k+1}&=f_k(x_k)+ {B}_k{u}_k+{G}d_k+ Ww_k,\\
y_k&=Cx_k+{D}_k{u}_k+{H}d_k+ {v}_k, \end{array}
\end{align}
where 
${G}\triangleq \begin{bmatrix} \hat{G} & 0_{n \times n_{\hat{H}}}\end{bmatrix}$, ${H}\triangleq \begin{bmatrix}  0_{l \times n_{\hat{G}}} & \hat{H} \end{bmatrix}$ and ${d}_k \triangleq \begin{bmatrix} g_k(x_k,u_k,d^s_k) \\ h_k(x_k,u_k,d^o_k)\end{bmatrix}$, correspondingly. 
Note that without loss of generality, we assume that ${\rm rk}[G^\top\; H^\top ]=p$, $n \geq l \geq 1$, $l \geq p \geq 0$ and $m \geq 0$. 

\begin{rem} \label{rem1}
\syong{From the discussion above, we can conclude that} \syong{set}-valued state and input observer \syong{designs} for system \eqref{eq:system3} 
\syong{are also applicable} to 
system \eqref{eq:system}, with the slight difference in input estimates that the former returns set-valued estimates for ${d}_k \triangleq \begin{bmatrix} g_k(x_k,u_k,d^s_k) \\ h_k(x_k,u_k,d^o_k)\end{bmatrix}$, where we can apply any 
\emph{pre-image} set computation techniques in the literature such as \cite{nien1998algorithm,sheng2003efficient,chandrasekar2006implicit} 
to find set estimates for $d^s_k$ and $d^o_k$ using the set-valued estimate for $x_k$ and the known $u_k$. Given this, 
\syong{throughout} the rest of the paper, we \syong{will} consider \syong{the design of} 
\syong{set}-valued state and unknown input \syong{observers} for system \eqref{eq:system3} \syong{(with sets in the form of hyperballs)}. 
\end{rem}
}

\noindent\textbf{\emph{Vector Field Assumptions.}} 
Here, we formally state the classes of nonlinear systems, related to the assumptions about the nonlinear, \moham{time-varying} vector field $f_k(\cdot): \mathbb{R}^n \rightarrow \mathbb{R}^n=\begin{bmatrix} f^\top_{k,1}(\cdot) & \dots & f^\top_{k,j}(\cdot) & \dots &  f^\top_{k,n}(\cdot) \end{bmatrix}^\top$, $\forall j \in \{1,\dots , n\}$, {$\forall k \in \mathbb{Z}$}, that we consider in this paper. 
{
\begin{class}{0}\label{class:general}
($\mathcal{M},\gamma$)-QC systems, with some known $\mathcal{M} \in \mathbb{R}^{2n \times 2n}$ and ${\gamma \in \mathbb{R}_+}$. 
\end{class}
}
\begin{class}{I} \label{class:Lip}
Globally $L_f$-Lipschitz continuous systems. 
\end{class}
\begin{class}{II} \label{class:A}
($\mathcal{A},\gamma$)-QC* systems, with some known $\mathcal{A}\in \mathbb{R}^{n \times n}$ and ${\gamma \in \mathbb{R}_+}$. 
\end{class}
\begin{class}{III} \label{class:convexcomb}
LPV systems with constituent matrices $A^i \in \mathbb{R}^{n \times n}, \forall i \in \{1,\dots ,N\}$.
\end{class}
For Class \ref{class:convexcomb} of systems, the system dynamics is governed by an LPV system with known parameters at run-time. We call each tuple $(A^i,C,G,H), \forall i \in \{1\dots N\}$, an LTI constituent of system {\eqref{eq:system3}}.

\noindent \textbf{\emph{Unknown Input (or Attack) Signal Assumptions.}} 
The unknown inputs {$d^s_k$, $d^o_k$ and consequently, $d_k$} are not constrained to be a signal of any type (random or strategic) nor to follow any model, thus no prior `useful' knowledge of the dynamics of $d_k$ is available (independent of $\{d_\ell\}$ $\forall k\neq \ell$, $\{w_\ell\}$ and $\{v_\ell\}$ $ \forall  \ell$). We also do not assume that $d_k$ is bounded or has known bounds and thus, $d_k$ is suitable for representing adversarial 
attack signals.

The simultaneous input and state set-valued observer design problem is twofold and 
 can be stated as follows:
\begin{problem}
Given the nonlinear discrete-time bounded-error system with unknown inputs \eqref{eq:system3} (cf. Remark \ref{rem1}),
\begin{enumerate}[1)]
\item \label{item:first-goal} Design stable observers that simultaneously find bounded sets 
of compatible states and unknown inputs for the four classes of nonlinear systems. 
\item Among the observers that satisfy \ref{item:first-goal}, find the optimal observer in the minimum $\mathcal{H}_\infty$-norm sense, i.e., with minimum average power amplification.
\end{enumerate}
\end{problem}
\section{Fixed-Order Simultaneous Input and State Set-Valued Observer Framework} \label{sec:structure}
{In this paper, we propose recursive 
{set-valued} observers 
 that consist of three steps: (i) an \emph{unknown input estimation} step that returns the set 
  of compatible unknown inputs using the current measurement and the set of compatible states, 
(ii) a \emph{time update} step in which the compatible set of states is propagated based on the system dynamics, and (iii) a \emph{measurement update} step where the set of compatible states is updated according to the current measurement. 
Since the complexity of optimal 
observers increases with time, we 
{will only focus on} \emph{fixed-order} recursive filters, similar to \cite{blanchini2012convex,chen2005observer,yong2018simultaneous}, and {in particular, we consider}  
set-valued estimates in the form \syong{of \emph{hyperballs},} as follows:
\begin{align*}
\begin{array}{rl}
\hat{D}_{k-1}&=\{d \in \mathbb{R}^p: \|d_{k-1}-\hat{d}_{k-1}\|\leq \delta^d_{k-1}\},\\
\hat{X}^\star_k&=\{x \in \mathbb{R}^n: \|x_k-\hat{x}^\star_{k|k}\|\leq \delta^{x,\star}_{k}\},\\
\hat{X}_k&=\{x \in \mathbb{R}^n: \|x_k-\hat{x}_{k|k}\|\leq \delta^x_k\},
\end{array}
\end{align*}
where $\hat{D}_{k-1}$, $\hat{X}^\star_{k}$ and $\hat{X}_k$ are the \moham{hyperballs} of compatible unknown inputs at time $k-1$, propagated, and updated states at time $k$, correspondingly. 
In other words, we restrict the estimation errors to \moham{hyperballs} of norm $\delta$. In this setting, the observer design problem is equivalent to finding the centroids $\hat{d}_{k-1}$, $\hat{x}^\star_{k|k}$ and $\hat{x}_{k|k}$ as well as the radii $\delta^d_{k-1}$, $\delta_k^{x,\star}$ and $\delta_k^x$ of the sets $\hat{D}_{k-1}$, $\hat{X}^\star_{k}$ and $\hat{X}_k$, respectively. 
In addition, we limit our attention to observers for the centroids $\hat{d}_{k-1}$, $\hat{x}^\star_{k|k}$ and $\hat{x}_{k|k}$ that belong to the class of three-step recursive filters given in \cite{Gillijns.2007b} and \cite{Yong.Zhu.ea.Automatica16}, with $\hat{x}_{0|0}=\hat{x}_0$.
\subsection{System Transformation} \label{sec:transformation}
To aid the observer design, we first carry out a transformation 
to decompose the unknown input signal $d_k$ {
{of} system \eqref{eq:system3}} into two components $d_{1,k} $ and $d_{2,k}$, as well as to decouple the output equation {in \eqref{eq:system3}} into two components, $z_{1,k}$ and $z_{2,k}$, 
one with a full rank direct feedthrough matrix and the other without direct feedthrough, {as follows:}
\begin{align}
\hspace{-0.1cm}\begin{array}{rl} x_{k+1}
&= f_k(x_k) +B_ku_k+ G_{1} d_{1,k} +\hspace{-0.05cm} G_{2} d_{2,k}+Ww_k, 
\\ z_{1,k} &= {C}_{1} x_k + \Sigma d_{1,k} + {D}_{k,1} u_k + {v}_{1,k},\\
z_{2,k} &=  {C}_{2}  x_k + {D}_{k,2} u_k + {v}_{2,k}. \end{array} \hspace{-0.1cm}  \label{eq:sys2}
\end{align}
{For the sake of increasing readability and completeness, 
the reader is referred to Appendix \ref{app:transformation} for details of this similarity transformation{, where the transformed system 
matrices $G_1,G_2,C_1,C_2,D_{k,1},D_{k,2}$ and noise signals $v_{1,k},v_{2,k}$ are {defined}}.}}

{\begin{rem}
 It is important to note that $d_{2,k}$ cannot be estimated from $y_k$ since it does not affect $z_{1,k}$ and $z_{2,k}$.
Thus, 
in light of \eqref{eq:sys2}, we can only obtain 
a (one-step) delayed estimate of $\hat{D}_{k-1}$. The reader may refer to \cite{Yong.Zhu.ea.CDC15_General} for a 
{more detailed} discussion on when a delay is absent or when we can expect further delays.
\end{rem}}

\subsection{Observer Structure} \label{sec:obsv}
Using the above transformation, we propose the following three-step recursive observer structure to compute the state and input estimate sets:

\noindent \emph{Unknown Input Estimation} (UIE): \vspace{-0.1cm}
\begin{align}
&\hat{d}_{1,k} =M_{1} (z_{1,k}-{C}_{1}\hat{x}_{k|k} -{D}_{k,1} u_k), \label{eq:variant1}\\
&\hat{d}_{2,k-1}=M_{2} (z_{2,k}-{C}_{2}\hat{x}_{k|k-1} -{D}_{k,2} u_k), \label{eq:d2}\\
&\hat{d}_{k-1}= V_{1} \hat{d}_{1,k-1} + V_{2} \hat{d}_{2,k-1} \label{eq:d}. 
\end{align}
\emph{Time Update} (TU): \vspace{-0.1cm}
\begin{align}
    \hat{x}_{k|k-1}&= f_{k-1}( \hat{x}_{k-1 | k-1}) +{B}_{k-1} u_{k-1} + G_{1} \hat{d}_{1,k-1},\label{eq:time} \\
\hat{x}^\star_{k|k}&=\hat{x}_{k|k-1}+G_{2} \hat{d}_{2,k-1}. \label{eq:xstar}
\end{align}
\emph{Measurement Update} (MU): \vspace{-0.1cm}
\begin{align}
\begin{array}{rl}
 \hat{x}_{k|k}&=\hat{x}^\star_{k|k} + L(y_{k} - {C} \hat{x}^\star_{k|k}- {D}_ku_k) \\
&= \hat{x}^\star_{k|k} +\tilde{L}(z_{2,k}-{C}_{2} \hat{x}^\star_{k|k}- {D}_{k,2} u_k),   \hspace{-0.2cm} \label{eq:stateEst}
\end{array}
\end{align}
where $L \in \mathbb{R}^{n \times l}$, $\tilde{L} \triangleq L U_{2}\in \mathbb{R}^{n \times (l-p_{H})}$, {$M_{1} \in \mathbb{R}^{p_{H} \times p_{H}}$ and $M_{2} \in \mathbb{R}^{(p-p_{H}) \times (l-p_{H})}$} are observer gain matrices that are designed according to {Lemma \ref{lem:error-dynamics}  and} Theorem \ref{thm:noise-attenuation-general}, 
to minimize the ``volume" of the set of compatible states and unknown inputs, quantified by the radii $\delta^d_{k-1}$, $\delta_k^{x,\star}$ and $\delta_k^x$.  
Note also that we applied $L=L U_{2} U_{2}^\top=\tilde{L} U_2^\top$ from Lemma \ref{lem:error-dynamics} 
into \eqref{eq:stateEst}{, where $U_2$ is 
{defined} in Appendix \ref{app:transformation}.} The resulting fixed-order set-valued observer {(that will be further described in the following section)} is summarized in Algorithm \ref{algorithm1}. 
\section{Observer Design and Analysis} \label{seq:design}
In this section, we derive LMI conditions for designing observers that are \syong{quadratically} stable \syong{in the absence of noise} 
(Section \ref{sec:stability}) and optimal in the $\mathcal{H}_{\infty}$ sense \syong{in the presence of noise} (Section \ref{sec:Hinf})  \syong{with uniformly bounded estimate radii (Section \ref{sec:convergence})}. 

To {design and analyze the observer}, we first 
derive our observer error dynamics via the following Lemma \ref{lem:error-dynamics}. For conciseness, all proofs are provided in the Appendix.

\begin{lem}\label{lem:error-dynamics} 
Consider system \eqref{eq:system3} \syong{(cf. Remark \ref{rem1})} 
and the observer \eqref{eq:variant1}-\eqref{eq:stateEst}. Suppose ${\rm rk}(C_{2} G_{2})=p-p_{H}${, where $C_2$ and $G_2$ are given in Appendix \ref{app:transformation}}. Then, designing observer matrix gains as $M_1=\Sigma^{-1}$, $M_2=(C_2 G_2)^\dagger$, $ L U_{1}=0$ and $L=L U_{2} U_{2}^\top=\tilde{L} U_2^\top${, with $U_1$ and $U_2$ given in Appendix \ref{app:transformation},} yields $M_{1} \Sigma=I $ and $ M_{2}C_{2} G_{2} = I $, and leads to the following difference equation for the state estimation error  dynamics {(i.e., {the dynamics of} $\tilde{x}_{k|k} \triangleq x_k-\hat{x}_{k|k}$)}:
\begin{align}\label{eq:errors-dynamics}
\tilde{x}_{k+1|k+1}=(I-\tilde{L}C_2)\Phi (\Delta f_{k}-\Psi \tilde{x}_{k|k})+\mathcal{W}(\tilde{L})\overline{w}_k,
\end{align}
where 
\begin{align*}
\Delta f_k &\triangleq f_k(x_k)-f_k(\hat{x}_k), \quad \Phi \triangleq I-G_2 M_2 C_2,\\
 \overline{w}_k &\triangleq \begin{bmatrix} (\frac{1}{\sqrt{2}})v^\top_k & w^\top_k & (\frac{1}{\sqrt{2}})v^\top_{k+1}  \end{bmatrix}^\top,\\
  R &\triangleq \begin{bmatrix} -\sqrt{2}\Phi G_1M_1T_1 & -\Phi W & -\sqrt{2}G_2M_2T_2  \end{bmatrix},\\
   Q &\triangleq \begin{bmatrix} 0_{(l-p_H) \times l} & 0_{(l-p_H) \times n} & -\sqrt{2}T_2 \end{bmatrix}, \\
    \Psi &\triangleq G_1M_1C_1, 
     \quad \mathcal{W}(\tilde{L}) \triangleq (I-\tilde{L}C_2)R+\tilde{L}Q.
    \end{align*}
\end{lem}  
Note that $\overline{w}_k$ is chosen such that $\lim \limits_{k \rightarrow \infty} \frac{1}{k+1}\textstyle{\sum}_{i=0}^k \overline{w}^\top_i \overline{w}_i=\lim \limits_{k \rightarrow \infty} \frac{1}{k+1}\textstyle{\sum}_{i=0}^k {w}^\top_i {w}_i+{v}^\top_i {v}_i$. The result in \eqref{eq:errors-dynamics} shows that we successfully decoupled/canceled out $d_k$ from the error dynamics, otherwise there would be a potentially unbounded and unknown term in the error dynamics.    
  \begin{algorithm}[t] \small
	\caption{Simultaneous Input and State Observer 
	}\label{algorithm1}
		\begin{algorithmic}[1]
		\State Initialize:  
		\Statex 
		Compute $\mohk{P}, M_1,M_2,\tilde{{L}},\mohk{\rho}$ via Theorem \ref{thm:noise-attenuation-general} and \mohk{$\theta_1$, $\theta_2$}, $\overline{\eta}$, $\beta$, $\overline{\alpha}$ 
		 via Theorem \ref{thm:error_bound}; $\Phi=I-G_2 M_2 C_2$, \mohk{$\Psi= G_1M_1C_1$;}
		\Statex  $\hat{x}_{0|0}=\hat{x}_0=\text{centroid}(\hat{X}_0)$;
		\Statex $\hat{d}_{1,0}=M_1 (z_{1,0}-C_{1} \hat{x}_{0|0}-D_{1} u_0)$; 
		\Statex  $\overline{\delta}^x_0= \mohk{\delta^x_0}=\min\limits_{\delta} \{\|x-\hat{x}_{0|0}\| \leq \delta, \forall x \in \hat{X}_0\}$; 
		\For {$k =1$ to $K$}
	\LineComment{Estimation of $d_{2,k-1}$ and $d_{k-1}$}
		\State $ \hat{x}_{k|k-1}= f_k( \hat{x}_{k-1 | k-1}) + {B}_k u_{k-1} + G_{1} \hat{d}_{1,k-1};$
		\State  $\hat{d}_{2,k-1}=M_{2} (z_{2,k}-{C}_{2}\hat{x}_{k|k-1} -{D}_{k,2} {u}_k);$
		\State $\hat{d}_{k-1} =V_{1} \hat{d}_{1,k-1} + V_{2} \hat{d}_{2,k-1}$; 
		\State \mohk{$\overline{\delta}^d_{k-1}=\beta \overline{\delta}^x_{k-1}+ \overline{\alpha}$};
		\State $\hat{D}_{k-1}=\{d \in \mathbb{R}^l : \|d-\hat{d}_{k-1}\| \leq \mohk{\overline{\delta}^d_{k-1}}\}$;
		\LineComment{Time update}
		\State
		$\hat{x}^\star_{k|k}=\hat{x}_{k|k-1}+G_{2} \hat{d}_{2,k-1}$;
		\LineComment{Measurement update}
		\State 
		$\hat{x}_{k|k}=\hat{x}^\star_{k|k}+\tilde{{L}}(z_{2,k}-C_{2} \hat{x}^\star_{k|k}- {D}_{k,2} {u}_k)$;
	\mohk{	\State $\overline{\delta}^x_{k,1}=\sqrt{(\delta^x_0)^2 \theta_1^k +  \frac{\rho^2}{\lambda_{\min}(P)}(\eta_w^2+\eta_v^2) \textstyle\sum_{i=1}^k \theta_1^{i-1}}$;
		\State $\overline{\delta}^x_{k,2}=\delta^x_0 \theta_2^k +  \overline{\eta} \textstyle\sum_{i=1}^k \theta_2^{i-1}$;
		\State $\overline{\delta}^x_k=\min(\overline{\delta}^x_{k,1},\overline{\delta}^x_{k,2})$;
		}
		\State  $\hat{X}_{k}=\{x \in \mathbb{R}^n : \|x-\hat{x}_{k|k}\| \leq \overline{\delta}^x_{k}\}$;
		\LineComment{Estimation of $d_{1,k}$}
		\State $\hat{d}_{1,k}=M_{1} (z_{1,k}-C_{1} \hat{x}_{k|k}- {D}_{k,1} {u}_k)$;
		\EndFor

	\end{algorithmic}
\end{algorithm}   

\subsection{Stable Observer Design} \label{sec:stability}
\moham{
In this \syong{section,} 
we first 
\syong{investigate} the existence of a \emph{stable} observer in the form of \eqref{eq:variant1}--\eqref{eq:stateEst} by providing \emph{necessary and sufficient} conditions for quadratic stability of the observer 
for 
\syong{the system classes described in Section \ref{sec:problemstatement} by supposing for the moment that there is no exogenous bounded noise $w_k$ and $v_k$.} 
Inspired by the definition of quadratic stability for nonlinear continuous-time systems in \cite{acikmese2003stability}, we formally define our considered notion of quadratic stability for nonlinear discrete-time systems.
\begin{defn}[Quadratic Stability]\label{def:quad-stab}
The nonlinear discrete-time dynamical system $x_{k+1}=f_k(x_k)$, with the vector field $f_k(\cdot):\mathbb{R}^n \to \mathbb{R}^n$, is quadratically stable, if it admits a quadratic positive definite Lyapunov function $V_k=x^\top_k P x_k >0$, with $P$ being a positive definite matrix in $\mathbb{R}^{n \times n}$, such that the Lyapunov function increment $\Delta V_k \triangleq V_{k+1}-V_k$ satisfies the following inequality for some $\alpha \in \syong{[0,1]}$, for all $k \in \mathbb{Z}$: 
\begin{align}\label{eq:quadratic}
\Delta V_k \leq-\alpha x^\top_k P x_k.
\end{align}
\end{defn}
\begin{rem}
\syong{It can be shown} that \eqref{eq:quadratic} implies that $\| x_k \| \leq \frac{\lambda_{\max}(P)}{\lambda_{\min}(P)} (1-\alpha)^k \|x_0 \|$, \syong{which is exponentially decreasing with $\alpha\in (0,1)$, non-increasing with $\alpha=0$ or deadbeat with $\alpha=1$. Note that there is also a} 
 slightly different notion of quadratic stability in the literature, e.g., in \cite{xie2004quadratic,garcia1994robust}, with $\Delta V_k \leq-\alpha x^\top_k  x_k$, which implies $\|x_k\| \leq (\frac{\lambda_{\max}(P)-\alpha}{\lambda_{\min}(P)})^k \| x_0 \|$, \syong{where the required condition on $\alpha$ for stability is dependent on $P$, making it slightly more complicated to perform a line search over $\alpha$. Hence, in this paper, we} 
 selected the notion of quadratic stability in \eqref{eq:quadratic}, similar to \cite{acikmese2003stability}.  
\end{rem}
Now we are ready to state our first set of main results on \emph{necessary and sufficient} conditions for the existence of quadratic\syong{ally} stable observers \syong{for noiseless systems} through the following theorem.
\begin{thm}[\syong{Necessary and Sufficient Conditions for Quadratically} 
Stable Observers]\label{thm:quadratic_stability}
Consider system \eqref{eq:system3} \syong{(cf. Remark \ref{rem1})}. 
Suppose there is no bounded noise $w_k$ and $v_k$ and all the conditions in Lemma \ref{lem:error-dynamics} hold.	
	Then, there exists a quadratically stable observer in the form of \eqref{eq:variant1}--\eqref{eq:stateEst}, \emph{if and only if} there exist $\alpha \in \syong{[0,1]}$, \mohk{$\kappa >0$} and matrices $ {P},\tilde{\Gamma},\breve{Q},\breve{Z} \in \mathbb{R}^{n \times n}$ and $Y \in \mathbb{R}^{n \times (l-p_H)}$  
	such that the following feasibility conditions hold:
\begin{align}\label{eq:quadratic-stable-LMI}
\begin{array}{rl}	
& P \succ 0,\tilde{\Gamma} \succeq 0,\breve{Q} \succeq 0,\\
&\begin{bmatrix} P & \tilde{Y}_1 \\ \tilde{Y}^\top_1 & \tilde{\mathbf{M}}_{1}  \end{bmatrix} \succeq 0,\begin{bmatrix} P & \tilde{Y}_2 \\ \tilde{Y}^\top_2 & \tilde{\mathbf{M}}_{2} \end{bmatrix} \succeq 0, \begin{bmatrix} P & \tilde{Y}_1 \\ \tilde{Y}^\top_1 & \tilde{\mathbf{M}}_{3} \end{bmatrix} \succeq 0,\\
&\begin{bmatrix} P & \tilde{Y}_2 \\ \tilde{Y}^\top_2 & \ \breve{Z}+\tilde{\mathbf{M}}_{4} \Psi \end{bmatrix} \succeq 0, \  \ \begin{bmatrix} \tilde{\Gamma} & \breve{Z} \\ \breve{Z}^\top  & \Psi^\top \breve{Q} \Psi \end{bmatrix} \succeq 0,
\end{array}
\end{align}
where $\tilde{Y}_1 \triangleq (P-YC_2)\Phi$, $\tilde{Y}_2 \triangleq -(P-YC_2)\Phi\Psi$, $\Phi,\Psi$ are defined in Lemma \ref{lem:error-dynamics} and
$\syong{\tilde{\mathbf{M}}_{1},\tilde{\mathbf{M}}_{2}},\tilde{\mathbf{M}}_{3},\tilde{\mathbf{M}}_{4}$ \syong{for Class 0 systems are given} 
as follows:
\begin{enumerate}[start=0] 
	 \item \label{item:general-stability}
	 If $f_k(\cdot)$ is a Class \ref{class:general} function with 
	 multiplier matrix $\mathcal{M}\triangleq \begin{bmatrix} M_{11} & M_{12} \\ M^\top_{12} & M_{22} \end{bmatrix}$ and some $\gamma \geq 0$, then
	 \begin{align} \label{eq:general}
	 \begin{array}{rl}
\tilde{\mathbf{M}}_{1} &\triangleq -\mohk{\kappa}M_{11}-\breve{Q}, \\
\tilde{\mathbf{M}}_{2} &\triangleq -\mohk{\kappa}M_{22}+(1-\alpha)P-\tilde{\Gamma}, \\
 \tilde{\mathbf{M}}_{3} &\triangleq -\mohk{\kappa}M_{11}, \ \tilde{\mathbf{M}}_{4} \triangleq \mohk{\kappa}M^\top_{12}. 
\end{array}
\end{align}
\end{enumerate}
\syong{Moreover, for system classes I--III, the $M_{11}$, $M_{12}$
 and $M_{22}$ matrices in \eqref{eq:general} are given as follows: } 
 \begin{enumerate}[label=(\Roman*)]
	  \item \label{item:lip-stability}
	  If $f_k(\cdot)$ is a Class \ref{class:Lip} function with Lipschitz constant $L_f$, then
	  \begin{align}\label{eq:quad_st_lip}
	 \begin{array}{rl}
	\syong{ M_{11}=-I, M_{12}=0, M_{22}=L^2_f I.}
\end{array}
\end{align}
 \item \label{item:dqc-stability}
	  If $f_k(\cdot)$ is a Class \ref{class:A} function with multiplier matrix $\mathcal{A}$, then
	  \begin{align}\label{eq:quad_st_dqc}
	 \begin{array}{rl}
	\syong{ M_{11}=-I, M_{12}=\mathcal{A}, M_{22}=-\mathcal{A}^\top \mathcal{A}.}
\end{array}
\end{align}
 \item \label{item:lpv-stability}
	  If $f_k(\cdot)$ is a Class \ref{class:convexcomb} function with constituent matrices $A^i \in \mathbb{R}^{n \times n}, \forall i \in \{1,\dots ,N\}$ and $\tilde{\sigma}_m \triangleq \max_{i \in {1\dots N}} \|A^i\|$, then
	  \begin{align}\label{eq:quad_st_lpv}
	 \begin{array}{rl}
	 \syong{ M_{11}=-I, M_{12}=0, M_{22}=\tilde{\sigma}_m^2I.}
\end{array}
\end{align}
	  \end{enumerate}
	  \syong{Furthermore,} no quadratically stable estimator can be designed if $\gamma<0$. 
	\end{thm} 
	\begin{rem} 
	The feasibility conditions in Theorem \ref{thm:quadratic_stability} can be easily verified by applying line search/bisection over $\alpha$ and solving the corresponding LMIs for $ \mohk{\kappa},{P},\tilde{\Gamma},\breve{Q},\breve{Z}$ and $Y$, given $\alpha$. 
	\end{rem}
}
\moham{
	Theorem \ref{thm:quadratic_stability} provides powerful tools in terms of necessary and sufficient conditions 
	\syong{for designing quadratically} stable observers. 
	\syong{When} the LMIs in \eqref{eq:quadratic-stable-LMI} do not hold, 
	it equivalently implies that there does not exist any \syong{quadratically} stable observer for that particular system. However, in such cases, one may still be able to design a Lyapunov stable observer, given the fact that quadratic and Lyapunov stability are not equivalent for general nonlinear systems (since Lyapunov stability, in its most general sense, hinges \syong{upon} admitting any form of Lyapunov function\syong{s} and not necessarily a quadratic form). This motivates us to derive \emph{necessary conditions} in terms of LMI ``infeasibility" conditions for Lyapunov stability of the observer. \syong{If these necessary conditions are feasible,} 
	then we 
	know \syong{for certain} that no stable observer, in the most general sense of stability, can be designed. 
	}
\begin{prop}[\syong{Necessary Condition for} Observer \moham{Lyapunov} Stability]
	\label{thm:stability-general}
	Consider system \eqref{eq:system3} \syong{(cf. Remark \ref{rem1})} and the observer \eqref{eq:variant1}--\eqref{eq:stateEst}. Suppose there is no bounded noise $w_k$ and $v_k$ and all the conditions in Lemma \ref{lem:error-dynamics} hold.	
	Then, the observer error dynamics is Lyapunov stable, \emph{\moham{only if}} 
	  \moham{
	  the following LMIs are always infeasible} 
	\moham{for all $0 \prec \tilde{P} \in \mathbb{R}^{n \times n}$, $\tilde{Y} \in \mathbb{R}^{n \times (l-p_H)}$, $0 \prec \tilde{\Gamma} \in \mathbb{R}^{(l-p_H) \times (l-p_H)}$ and $0 < \tilde{\eta} < 1$.
\begin{align} \label{eq:sability-necessity-1}
	 \hat{\Pi} \triangleq  \begin{bmatrix} I-\tilde{\Gamma} & 0 & 0 \\  0 &\tilde{\Gamma} & \tilde{Y}^\top   \\ 0 & \tilde{Y} & \tilde{P}\end{bmatrix} \succeq 0, \tilde{\Pi} \triangleq \begin{bmatrix} \tilde{\Pi}_{11} &  \tilde{\Pi}_{12}\\   \tilde{\Pi}^\top_{12} & \tilde{\Pi}_{22} \end{bmatrix} \succ 0, 
	\end{align}
where $\tilde{S} \triangleq \tilde{P}-C^\top_2\tilde{Y}^\top -\tilde{Y}C_2$ and $\tilde{\Pi}_{11}$, $\tilde{\Pi}_{12}$ and $\tilde{\Pi}_{22}$ \syong{for Class 0 systems are defined}  
as follows: 
\begin{enumerate}[start=0] 
	 \item \label{item:general-stability-nec}  
	 If $f_k(\cdot)$ is a Class \ref{class:general} function with multiplier matrix $\mathcal{M}\triangleq \begin{bmatrix} M_{11} & M_{12} \\ M^\top_{12} & M_{22} \end{bmatrix}$ and some $\gamma \geq 0$, then 
	 \begin{align}\label{eq:gen-nec-lmi}
	 \begin{array}{rl}
	 &\tilde{\Pi}_{11}\triangleq \Phi^\top ( \tilde{S} -(1-\tilde{\eta}) C^\top_2  C_2 )\Phi-M_{11}, \\ 
	 &\tilde{\Pi}_{12} \triangleq -\Phi^\top \tilde{S} \Phi \Psi-M_{12}, \\
	  &\tilde{\Pi}_{22} \triangleq \Psi^\top \Phi^\top (\tilde{S}-(1-\tilde{\eta})C^\top_2  C_2)\Phi \Psi-\tilde{P}-M_{22}.
	  \end{array}
	  \end{align}
	  \end{enumerate}
	  \syong{Moreover, for system classes I--III, the $M_{11}$, $M_{12}$
 and $M_{22}$ matrices in \eqref{eq:gen-nec-lmi} are given by \eqref{eq:quad_st_lip}, \eqref{eq:quad_st_dqc} and \eqref{eq:quad_st_lpv}, respectively.}
	}
	
	\end{prop} 
	It is worth mentioning that if $f_k(\cdot)$ is a Class \ref{class:convexcomb} function, then we can provide 
	sufficient conditions for the \emph{existence} of {Lyapunov} stable observers {as well as necessary conditions that} 
	are conveniently testable. 
	{The latter} are 
	beneficial in the sense that if they are \emph{not} satisfied, the designer knows \emph{a priori} that there does not exist any $\mathcal{H}_\infty$-observer for 
{those LPV} systems with unknown inputs/attacks.  The conditions are formally derived in the following Lemma \ref{lem:stability-existence}.  
	\begin{lem}\label{lem:stability-existence}
	Suppose $f_k(\cdot)$ is a Class \ref{class:convexcomb} function and all the conditions in Lemma \ref{lem:error-dynamics} hold. Then, there exists a stable observer for the system \eqref{eq:system}, with any sequence $\{\lambda_{i,k}\}_{k=0}^\infty$ for all $i \in \{1,2,\dots,N\}$ that satisfies $ 0 \leq \lambda_{i,k} \leq1 , \sum_{i=1}^N \lambda_{i,k} =1, \forall k $, 
	 if $ (\overline{A}_k , C_{2}  ) $ be uniformly detectable\footnote{The readers are referred to \cite[Section 2]{Anderson.Moore.1981} for the concise definition of uniform detectability. A spectral test can be found in \cite{Peters.Iglesias.1999}. } for each $k$, and only if all constituent LTI systems $(A^i,G,C,H), \forall i \in \{1\dots N \},$ are strongly detectable (cf. Definition \ref{def:strongdetectable}), where $\overline{A}_k \triangleq \Phi ( \sum_{i=1}^N \lambda_{i,k} {A}^{i} - \Psi )$, with $\Phi$ and $\Psi$ defined in Lemma \ref{lem:error-dynamics}.   
	  \end{lem}
	  \begin{cor}\label{cor:LTI_stable}
	   There exists a stable simultaneous state and input set-valued observer for the LTI system \eqref{eq:LTI}, through \eqref{eq:variant1}--\eqref{eq:stateEst}, if and only if the tuple $(A,G,C,H)$ is strongly detectable and only if ${\rm rk}(C_{2} G_{2})=p-p_{H}$. Moreover, the observer gain matrices can be designed as $M_1=\Sigma^{-1}$, $M_2=(C_2 G_2)^\dagger$ and $L=\tilde{L}U^\top_2$ and $\tilde{L}=P^{-1}Y$, where $P \succ 0$ and $Y$ solve the following feasibility program with LMI constraints:
	   \begin{align*}
	   \begin{array}{rl}
\nonumber & {\rm Find} \hspace{.2 cm} (P \succ 0,Y)
\\ & s.t. \quad \begin{bmatrix} P & \Lambda \\ \Lambda^\top & P \end{bmatrix} \succeq 0,
\end{array}
	   \end{align*} 
	   with $\Lambda \triangleq ({A}-\Psi)^{\top} \Phi^\top ( P -  C^{\top}_2 Y^\top )$ and $\Phi$ and $\Psi$ defined in Lemma \ref{lem:error-dynamics}.
	  \end{cor}	
\subsection{ $\mathcal{H}_\infty$ Observer Design}\label{sec:Hinf}
The goal of this section is to provide additional sufficient conditions to guarantee optimality of the observers in the $\mathcal{H}_{\infty}$ sense \syong{in the presence of exogenous noise}. We first define our considered notion of optimality via the following Definition \ref{def:attenuation}.  
\begin{defn}[$\mathcal{H}_\infty$-Observer]\label{def:attenuation}
Let $T_{\tilde{x},w,v}$ denote the transfer function matrix that maps the noise signals $\vec{w}_k \triangleq\begin{bmatrix} w_k^\top & v_k^\top\end{bmatrix}^\top$ to the updated state estimation error $\tilde{x}_{k|k}\triangleq x_k-\hat{x}_{k|k}$.
For a given {\emph{noise attenuation level}} $\rho \in \mathbb{R}_+$, the observer performance satisfies $\mathcal{H}_{\infty}$ norm bounded by $\rho$, if $\|T_{\tilde{x},w,v}\|_\infty \leq \rho$, i.e., the maximum average signal power amplification is upper-bounded by $\rho^2$:
\begin{align}
  \frac{\lim_{k \to \infty}\frac{1}{k+1} \sum_{i=0}^k \tilde{x}_{i|i}^\top \tilde{x}_{i|i} }{\lim_{k \to \infty} \frac{1}{k+1} \sum_{i=0}^k \vec{w}_{i}^\top \vec{w}_{i} } \triangleq \|T_{\tilde{x},w,v}\|_\infty^2 \leq \rho^2.
\end{align}
\end{defn}
Now we present our second set of main results, on designing stable and optimal observers in the minimum $\mathcal{H}_{\infty}$ sense. 
\moham{
\begin{thm}[$\mathcal{H}_\infty$-Observer Design]\label{thm:noise-attenuation-general}
Consider system \eqref{eq:system3} \syong{(cf. Remark \ref{rem1})}, 
the observer \eqref{eq:variant1}--\eqref{eq:stateEst} and \syong{a} given $\rho >0$. Suppose all the conditions in Theorem \ref{thm:quadratic_stability} hold and 
\syong{let} $\Phi$, $\Psi$, $Q$ and $R$ \syong{be defined as} in Lemma \ref{lem:error-dynamics} and 
$\Omega \triangleq C_2R-Q$. 
 Then, with the gain $\tilde{L}=P^{-1}Y$, we obtain a \moham{quadratically} stable observer with $\mathcal{H}_\infty$ norm bounded by ${\rho} $, if \syong{the LMIs in \eqref{eq:quadratic-stable-LMI} hold with some $P \succ 0$ and $Y=P \tilde{L}$, and} there exist $0 \preceq \Gamma \in \mathbb{R}^{n \times n}$ and $\varepsilon_1,\varepsilon_2 >0$ such that:
\begin{align} \label{eq:noise-attenuation-general}
\Pi \triangleq \begin{bmatrix} I-\Gamma & 0 & 0 \\ 0 & P & Y \\ 0 & Y^\top & I \end{bmatrix} \succeq 0, \
\mathcal{N} \triangleq \begin{bmatrix} \mathcal{N}_{11} & * & *  \\ \mathcal{N}_{21} & \mathcal{N}_{22} & *  \\ \mathcal{N}_{31} & \mathcal{N}_{32} & \mathcal{N}_{33}  \end{bmatrix} \succeq 0,
\end{align}
where
\begin{align} \label{eq:elements-gen}
\nonumber\mathcal{N}_{11} &\hspace{-0.1cm}\triangleq \hspace{-0.1cm}\rho^2 I \hspace{-0.1cm}+\hspace{-0.1cm} 2R^\top Y\Omega\hspace{-0.1cm}-\hspace{-0.1cm}R^\top P R\hspace{-0.1cm}-\hspace{-0.1cm}\Omega^\top (\Gamma\hspace{-0.1cm}+\hspace{-0.1cm}(\varepsilon^{-1}_1\hspace{-0.1cm}+\hspace{-0.1cm}\varepsilon^{-1}_2)I) \Omega \\
\mathcal{N}_{21} &\triangleq \Psi^\top \Phi^\top (PR-Y\Omega-C^\top_2 Y^\top R),\\
\nonumber \mathcal{N}_{31} &\triangleq \Phi^\top(Y \Omega+C^\top_2 Y^\top R -PR), 
\end{align}
and $\mathcal{N}_{22}$, $\mathcal{N}_{32}$ and $\mathcal{N}_{33}$ 
are defined for \syong{Class 0 systems} 
as follows:
 \begin{enumerate}[start=0] 
	\item \label{item:general-optimality}
	If $f(\cdot)$ is a Class \ref{class:general} function with multiplier matrix $\mathcal{M}\triangleq \begin{bmatrix} M_{11} & M_{12} \\ M^\top_{12} & M_{22} \end{bmatrix}$ and some $\gamma \geq 0$, then
	\begin{align}\label{eq:general_class_opt}
	\nonumber \mathcal{N}_{22} &\triangleq  -I+\alpha P-\varepsilon_1\Psi^\top \Phi^\top C^\top_2 C_2 \Phi \Psi-M_{22},\\
	\mathcal{N}_{32} &\triangleq -M_{12}, \\
	\nonumber \mathcal{N}_{33} &\triangleq -\varepsilon_2\Phi^\top C^\top_2 C_2 \Phi-M_{11}.
	\end{align}
\end{enumerate}	
 \syong{Moreover, for system classes I--III, the $M_{11}$, $M_{12}$
 and $M_{22}$ matrices in \eqref{eq:general_class_opt} are given by \eqref{eq:quad_st_lip}, \eqref{eq:quad_st_dqc} and \eqref{eq:quad_st_lpv}, respectively.} 
  Finally, the minimum $\mathcal{H}_{\infty}$ bound can be found by solving the following semi-definite program \syong{(SDP)} (with line searches over $\varepsilon_1>0$ and $\varepsilon_2>0$): 
\begin{align*}
(\rho^{\star})^2=&\min_{P \succ 0,\Gamma \succ 0,Y,\rho^2>0,0 \leq \alpha \leq1,\mohk{\kappa>0}} \rho^2 \\ 
& \ \ \ \     s.t. \quad \eqref{eq:quadratic-stable-LMI}, \eqref{eq:noise-attenuation-general} \ \mathrm{ hold},
\end{align*}
where $\rho^2$ is a decision variable. \syong{If this SDP} 
\syong{is feasible, then the infinity norm of the transfer function matrix $T_{\tilde{x},w,v}$ satisfies} 
$\|T_{\tilde{x},w,v}\|_\infty \leq {\rho^{\star}}$. This bound is obtained by applying the observer gain $\tilde{L}^\star = {P^{\star}}^{ -1}Y^\star$, where $(P^\star,Y^\star,\Gamma^\star)$ solves the above SDP.
 \end{thm}
}
 \subsection{Radii of Estimates and Convergence of Errors} \label{sec:convergence}
In this section, we are interested in \syong{(i)} \moham{proving the existence of uniformly bounded estimate radii,} \syong{(ii)} computing closed-form expressions for \moham{\syong{upper bounds/}over-approximations of} the estimate radii and \syong{(iii)} finding sufficient conditions for the convergence \syong{of the upper bound sequences}, as well as their steady-state values (if they exist). 

It is worth mentioning that 
 for linear time-invariant systems,
   strong detectability of the system (cf. Definition \ref{def:strongdetectable}) is a sufficient condition for the convergence of the radii $\delta^x_{k}$ and $\delta^d_{k-1}$ to steady state \cite{yong2018simultaneous}, but it is less clear for general nonlinear systems. Notice that if $f_k(\cdot)$ is a Class \ref{class:convexcomb} function, i.e., in the LPV case, even strong detectability of all constituent LTI systems does not guarantee that the radii converge. The reason is that the convergence hinges on the stability of the product of \emph{time-varying} matrices (cf. proof of Theorem \ref{thm:error_bound}), which is not guaranteed even if all the multiplicands are stable.

\moham{To address the existence of uniformly bounded radii \syong{for the proposed observer designs for the nonlinear systems we consider}, 
\syong{we first} define the notion of uniformly \syong{bounded-input bounded-state} systems.} 
\moham{
\begin{defn}[UBIBS Systems]\cite[Section 3.2]{jiang2001input}\label{defn:UBIBS}
A dynamic system is uniformly \syong{bounded-input bounded-state}  (UBIBS), if bounded initial states $x_0$ and 
\syong{bounded (disturbance/noise)}
inputs $u$ produce uniformly bounded trajectories, i.e., there exist two $\mathcal{K}$-functions\footnote{A function $\sigma:\mathbb{R}_+ \to \mathbb{R}_+$ is a $\mathcal{K}$-function if it is continuous, strictly increasing and $\sigma(0)=0.$} $\sigma_1$ and $\sigma_2$ such that
\begin{align*}
\sup\limits_k \|x(k,x_0,u)\| \leq \max \{\sigma_1(\|x_0\|),\sigma_2(\|u\|)\}.
\end{align*} 
\end{defn}
Now, we are ready to state our results on the uniform boundedness of the \syong{estimate} radii.
\begin{thm}[Uniformly Bounded \syong{Estimate} Radii]\label{prop:radii-existence}
Consider system \eqref{eq:system3} \syong{(cf. Remark \ref{rem1})} 
and the observer \eqref{eq:variant1}--\eqref{eq:stateEst}. Suppose all the conditions in Theorem \ref{thm:noise-attenuation-general} hold. Then, the state estimation \syong{radii/}error dynamics \eqref{eq:errors-dynamics} is a UBIBS system 
\syong{with} noise as 
\syong{exogenous} inputs. In other words, bounded initial state errors and noise  produce uniformly bounded trajectories of errors, i.e.,  there exist $\mathcal{K}$-functions $\sigma_1$ and $\sigma_2$ such that 
\begin{align*}
\sup\limits_k \|\tilde{x}_{k|k}\| \leq \max \{\sigma_1(\|\tilde{x}_{0|0}\|),\sigma_2(\sqrt{\eta^2_v+\eta^2_w})\}.
\end{align*} 
Moreover, \eqref{eq:errors-dynamics} admits a $\mathcal{K}$-asymptotic gain, i.e., there exist $\mathcal{K}$-function $\gamma_a$ such that
\begin{align*}
\limsup\limits_{k \to \infty} \|\tilde{x}_{k|k}\| \leq \gamma_a (\limsup_{k \to \infty}\|\overline{w}_k\|),
\end{align*}
where $\limsup\limits_{n \to \infty}x_n \triangleq \lim\limits_{n \to \infty}(\sup\limits_{m \geq n}x_m)$ denotes the \emph{limit superior} of the sequence $\{x_n\}_{n=1}^{\infty}$. 
\end{thm}
}
\moham{\syong{The above} Theorem \ref{prop:radii-existence} guarantees uniform boundedness of the \syong{estimate} radii, if an $\mathcal{H}_{\infty}$ observer in the form of \eqref{eq:variant1}--\eqref{eq:stateEst} exists and can be designed through Theorem \ref{thm:noise-attenuation-general}. 
Next, we are interested in deriving closed-form expressions for the \syong{upper bound/over-approximation of the uniformly bounded estimate} radii,} 
%
\moham{ i.e., \emph{upper \syong{bound} sequences} for the resulting sequences of radii $\{\delta^x_{k}\}_{k=1}^{\infty}$ and $\{\delta^d_{k-1}\}_{k=1}^{\infty}$, when using our proposed observer for \syong{the} different classes of systems. We also discuss some sufficient conditions for the convergence of the over-approximations of the \syong{estimate} radii to steady state. 
} 
 \begin{thm}[\syong{Upper Bounds of the} Radii of Estimates]\label{thm:error_bound}
	Consider system \eqref{eq:system3} \syong{(cf. Remark \ref{rem1})} 
	along with the observer \eqref{eq:variant1}--\eqref{eq:stateEst}. Suppose the conditions of Theorem \ref{thm:noise-attenuation-general} hold. Let $\Re \triangleq -(\Psi \Phi G_1 M_1 T_1 +\Psi G_2 M_2 T_2 + \tilde{L} T_2)$, $\overline{\alpha} \triangleq \| V_2 M_2 C_2 \| \eta_w
 +\big{[}\|(V_2 M_2 C_2G_1-V_1)M_1T_1\|+\|V_2M_2T_2\| \big{]} \eta_v$ and $\overline{\eta} \triangleq \| \Re \| \eta_v + \| \Psi \Phi W\| \eta_w$, with $\Phi$ and $\Psi$ defined in Lemma \ref{lem:error-dynamics} {and $T_1,T_2$ given in Appendix \ref{app:transformation}}. Then, \syong{the upper bound sequences for} 
  the \syong{estimate} radii, denoted $\overline{\delta}^x_{k}$ and $\overline{\delta}^d_{k-1}$, can be obtained as:
\moham{
	\begin{align}
	\label{eq:stateradius}\delta^x_{k} &\leq \overline{\delta}^x_k \quad \triangleq \min(\overline{\delta}^x_{k,1},\overline{\delta}^x_{k,2}),\\
	\label{eq:inputradius} \delta^d_{k-1} &\leq \overline{\delta}^d_{k-1} \triangleq \beta \overline{\delta}^x_{k-1}+ \overline{\alpha},
	\end{align}
	where
	\begin{align}
	 \overline{\delta}^x_{k,1} &\triangleq \sqrt{(\delta^x_0)^2 \theta_1^k + \textstyle \syong{\frac{\rho^2}{\lambda_{\min}(P)}}(\eta_w^2+\eta_v^2) \textstyle\sum_{i=1}^k \theta_1^{i-1}}, \label{eq:delta1}\\
\theta_1 &\triangleq \mohk{{\textstyle\frac{|\lambda_{\max}(P)-1|}{\lambda_{\min}(P)}}},\\
	\label{eq:stateradius} \overline{\delta}^x_{k,2} &\triangleq \delta^x_0 \theta_2^k +  \overline{\eta} \textstyle\sum_{i=1}^k \theta_2^{i-1}, 
	\end{align}
}
and $\theta_2$ and $\beta$ are defined for the different function classes as follows:
\begin{enumerate}[(I)]
\item If $f(\cdot)$ is a Class \ref{class:Lip} function, then 
\begin{align}\label{eq:errors1}
 \syong{\theta_2} &\triangleq (L_f+\| \Psi \|)\|(I-\tilde{L}C_2)\Phi \|, \\
\nonumber \beta &\triangleq \|V_1M_1C_1-V_2M_2C_2\Psi \|+L_f \|V_2M_2C_2 \|,
 \end{align}
\item If $f(\cdot)$ is a Class \ref{class:A} function, then
\begin{align} \label{eq:errors2}
 \syong{\theta_2} &\triangleq (\lambda_{\max}(\mathcal{A}^\top \mathcal{A})\hspace{-.1cm}+\hspace{-.1cm}\| \Psi \|)\|(I\hspace{-.1cm}-\hspace{-.1cm}\tilde{L}C_2)\Phi \|, \\
\nonumber \beta &\triangleq \|V_1M_1C_1\hspace{-.1cm}-\hspace{-.1cm}V_2M_2C_2\Psi \|\hspace{-.1cm}+\hspace{-.1cm}\lambda_{\max}(\mathcal{A}^\top \hspace{-.1cm}\mathcal{A}) \|V_2M_2C_2 \|
   \end{align}
\item If $f(\cdot)$ is a Class \ref{class:convexcomb} function, then 
\begin{align}\label{eq:errors3}
  \syong{\theta_2} &\triangleq \textstyle\max_{i \in \{1,2,\dots ,N\}} \| A_{e,i} \|, \\
 \nonumber  \beta &\triangleq \textstyle\max_{i\in \{1,2,\dots,N\}} \|V_1M_1C_1+ V_2 M_2 C_2 A_{e,i}\|, 
   \end{align}
 with $A_{e,i} \triangleq (I-\tilde{L}C_2) \Phi (A^i-\Psi)$, for all $i \in \{1\dots N\}$ {and $V_1,V_2$ given in Appendix \ref{app:transformation}}.
\end{enumerate}
Furthermore, \moham{the upper \syong{bound} sequences} \syong{for the estimate radii} are convergent if \mohk{$\theta \triangleq \min (\theta_1,\theta_2)<1$ (equivalently, if $\theta_1 <1$ or $\theta_2 <1$)}, 
and \syong{in this case,} 
the steady\syong{-state upper bounds of the estimate} 
radii are given by: 
\begin{align*}
\begin{array}{l}
\displaystyle\lim_{k \to \infty} \overline{\delta}^x_k  = \overline{\delta}_\infty^x \triangleq 
\begin{cases} \syong{\overline{\delta}_{\infty,1}^x}, & \text{if} \ \theta_1 <1,\theta_2 \geq 1, \\ \syong{\overline{\delta}_{\infty,2}^x}, & \text{if} \ \theta_1 \geq 1,\theta_2 < 1, \\ \min(\syong{\overline{\delta}_{\infty,1}^x,\overline{\delta}_{\infty,2}^x}
), &  \text{if} \ \theta_1 < 1,\theta_2 < 1, \end{cases} \\
\displaystyle \lim_{k \to \infty} \overline{\delta}^d_{k}= \beta \overline{\delta}^x +\overline{\alpha},
 \end{array}
\end{align*} 
\syong{where $\overline{\delta}_{\infty,1}^x\triangleq \rho\sqrt{\frac{\eta_w^2+\eta_v^2}{\syong{\lambda_{\min} (P)(1-\theta_1)}}}$ and $\overline{\delta}_{\infty,2}^x\triangleq \frac{ \overline{\eta}}{1-\theta_2}$.}
\end{thm}

\begin{cor} \label{cor:convergence-lpv}
If $f_k(\cdot)$ is a Class \ref{class:convexcomb} function and the conditions of Theorem \ref{thm:noise-attenuation-general} hold, then, the \moham{upper \syong{bound} sequences for the \syong{estimate}} radii, \syong{i.e., $\overline{\delta}^x_{k}$ and $\overline{\delta}^d_{k-1}$}, computed in \eqref{eq:stateradius} and \eqref{eq:inputradius}, are convergent if $\|A_{e,i}\|<1$ for all $i \in\left\{1,2,\dots ,N\right\}$, where $A_{e,i} \triangleq (I-\tilde{L}C_2) \Phi (A^i-\Psi)$, with $\Phi$ and $\Psi$ defined in Lemma \ref{lem:error-dynamics}.
\end{cor}

\moh{
\begin{rem}
According to \moham{Theorem \ref{prop:radii-existence},}
if the necessary and sufficient conditions in Theorem \moham{\ref{thm:noise-attenuation-general}} 
hold, \yong{i.e., when the observer is \moham{\syong{quadratically} stable and optimal \syong{in the sense of $\mathcal{H}_{\infty}$}},} 
the sequences of \syong{estimate} radii, $\{\delta^x_k,\delta^d_{k-1}\}_{k=1}^{\infty}$, \yong{are uniformly bounded, \moha{regardless of the value of \syong{$\theta_1$ or $\theta_2$}}. Consequently, the sequences of errors, $\{\tilde{x}_{k|k},\tilde{d}_{k-1|k-1}\}_{k=1}^{\infty}$, are also uniformly bounded and do not diverge. \syong{On the other hand, the closed-form (potentially conservative) upper bound sequences,  $\{\overline{\delta}^x_k,\overline{\delta}^d_{k-1}\}_{k=1}^{\infty}$, may diverge even when \mohk{$\{\tilde{x}_{k|k},\tilde{d}_{k-1|k-1}\}_{k=1}^{\infty}$} are uniformly bounded, and a sufficient condition for their convergence is that \mohk{$\theta=\min(\theta_1,\theta_2) <1$, i.e.,} $\theta_1<1$ and/or $\theta_2<0$.}
}
\end{rem}
}

We conclude this section by {stating a proposition with which} 
we can trade off between observer \emph{optimality} {(i.e., the noise attenuation level)} 
and \emph{convergence} of the \moham{upper \syong{bound} sequences for the} error radii 
by adding some additional LMIs to 
\syong{the} conditions in \eqref{eq:quadratic-stable-LMI} \syong{and \eqref{eq:noise-attenuation-general},} and solving the corresponding mixed-integer SDP. 
\moham{ 
\begin{prop}[Convergence of Upper \syong{Bound} Sequences] \label{lem:theta}
Consider system \eqref{eq:system3} \syong{(cf. Remark \ref{rem1})} 
and suppose that the assumptions in Lemma \ref{lem:error-dynamics} hold. Then, solving the following mixed-integer SDP:
\begin{align*}
&(\rho^{\star\star})^2=\hspace{-.35cm}\min_{\{P \succ 0,\Gamma \succ 0,Y,\rho^2>0,0 \leq\alpha \leq 1,\varepsilon_1>0,\varepsilon_2>0,\mohk{\kappa>0},\kappa_1 >0,\kappa_2>0\}} \rho^2 \\ 
&       s.t. \quad \eqref{eq:quadratic-stable-LMI}, \eqref{eq:noise-attenuation-general} \ \mathrm{ hold}, \quad \kappa_1I \preceq P \preceq \kappa_2 I, \\
&       \phantom{s.t.} \quad (\kappa_1 \geq 1, \kappa_2-\kappa_1 <1) \ \vee \ (\kappa_2 \leq 1, \kappa_1 >0.5),
\end{align*}
guarantees that \syong{$\theta_1 <1$ and thus,} $\mohk{\theta <1}$, where \syong{$\theta_1$ \mohk{and $\theta$ are}} 
given in Theorem \ref{thm:error_bound}, and results in a \syong{quadratically} stable  observer in the form of \eqref{eq:variant1}--\eqref{eq:stateEst}, with \emph{convergent} upper \syong{bound} sequences for the radii $\{\delta^x_k,\delta^d_{k-1}\}_{k=1}^{\infty}$  in the form of $\overline{\delta}^x_k$ and $\overline{\delta}^d_{k-1}$ in \eqref{eq:stateradius} and \eqref{eq:inputradius}, respectively, with 
noise attenuation level $\rho^{\star\star}$ 
\syong{when using} the observer gain $\tilde{L}^{\star\star}=P^{\star\star-1}Y^{\star\star}$, where $(P^{\star\star},Y^{\star\star})$ are solutions to the above mixed-integer SDP.
\end{prop}
}
\syong{Note that the above mixed-integer SDP can also be solved using two independent SDPs with each the disjunctive constraints (denoted with $\vee$) and selecting the solution corresponding to the smaller $\rho^{\star\star}$. Moreover,}
it is worth mentioning that although the designed observer may not be optimum in the minimum $\mathcal{H}_\infty$ sense when using the mixed-integer 
 SDP in 
Proposition \ref{lem:theta}, we can instead guarantee the steady-state convergence of the \syong{closed-form upper bound} sequences of the estimate radii. 
	\renewcommand{\theenumi}{\roman{enumi}}
\section{Simulation Results and Comparison with Benchmark Observers}
Two simulation examples are considered in this section to demonstrate the performance of the proposed observer. In the first example, where the dynamic system belongs to Classes \ref{class:Lip} and \ref{class:A}, we consider simultaneous input and state estimation problem and design observers for each class 
to study their performances. Our second example is a benchmark dynamical Lipschitz continuous (i.e., Class \ref{class:Lip}) system, where we compare the results of our observer with two other existing observers in the literature, \cite{chakrabarty2016state,chen2018nonlinear}. We consider two different scenarios, one with a bounded unknown input, and the other with an unbounded unknown input. The results show that in the unbounded input scenario, when applying the observers in \cite{chakrabarty2016state,chen2018nonlinear}, the estimation errors diverge, while as expected from our theoretical results, the estimation errors of our proposed observer converge to steady state values. 
\subsection{Single-Link Flexible-Joint Robotic System}
\begin{figure}[!h]
\begin{center}
\includegraphics[scale=0.146,trim=32mm 0mm 0mm 0mm,clip]{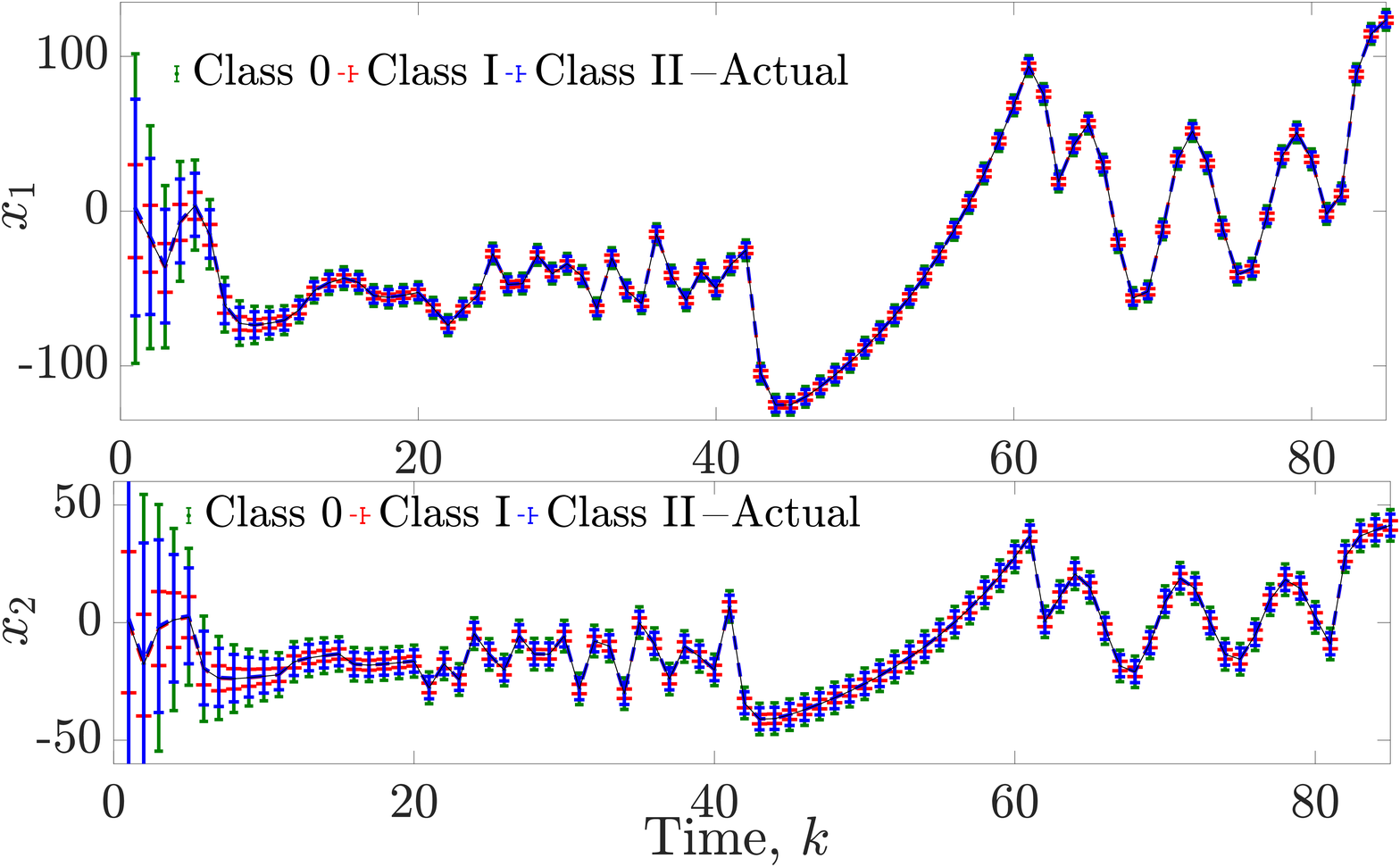} 
\\ 
\includegraphics[scale=0.146,trim=32mm 0mm 0mm 0mm,clip]{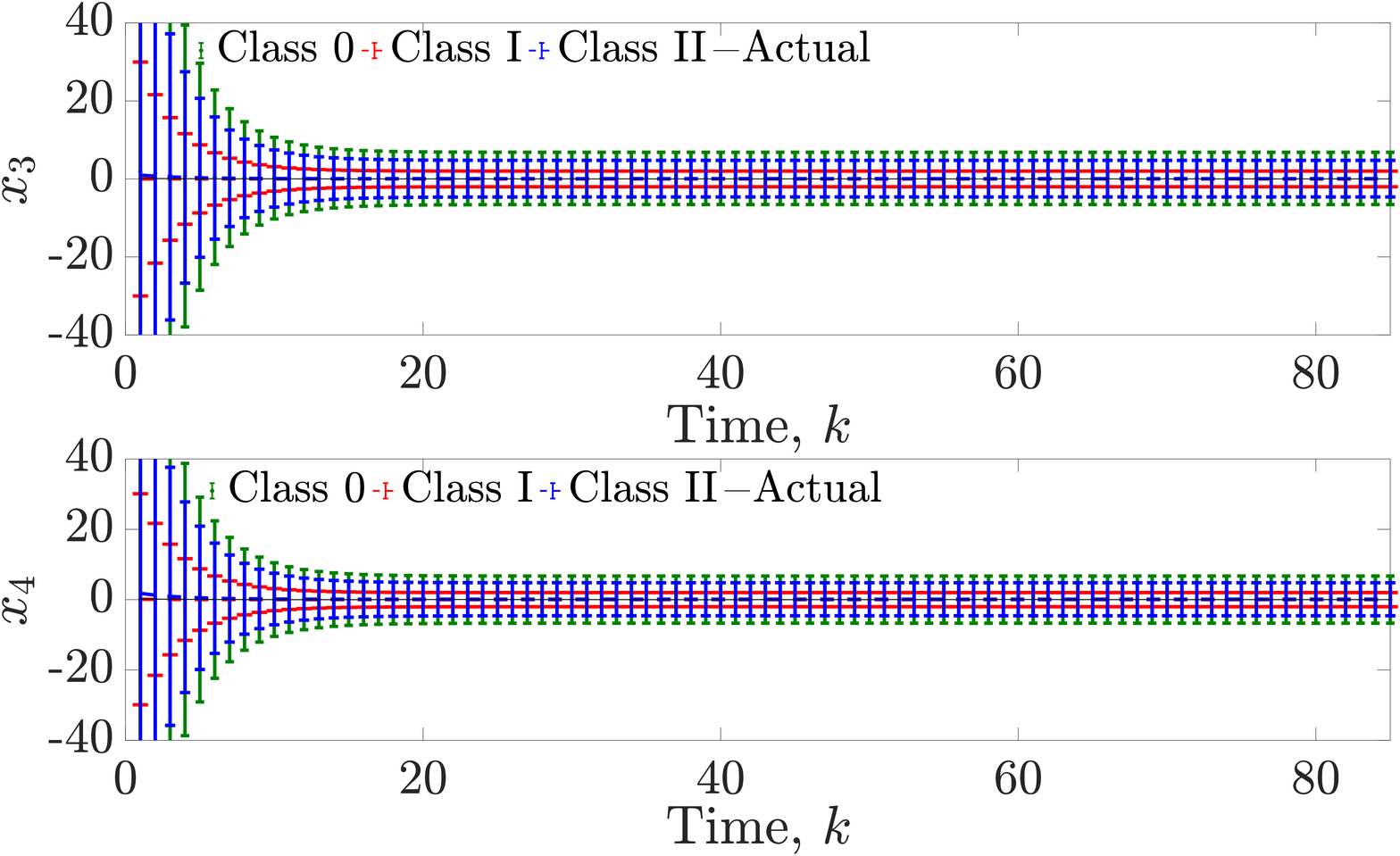} 
\includegraphics[scale=0.148,trim=30mm 0mm 0mm 0mm,clip]{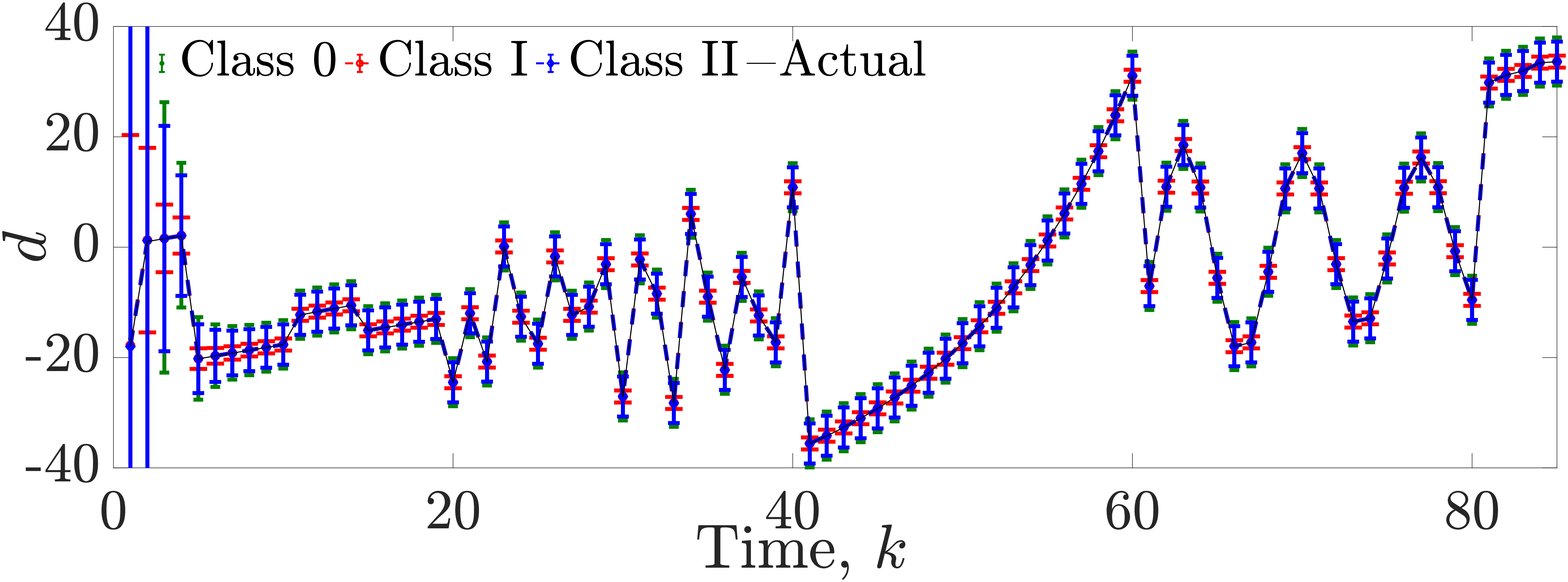} \vspace{-2.4cm}
\caption{Actual states $x_1$,$x_2$,$x_3$,$x_4$ and input $d$, as well as their {Class $0$}, 
{Class I} 
and {Class II} estimates (i.e., the obtained estimates by applying the corresponding gains for {Classes 0--II} 
in Theorem \ref{thm:noise-attenuation-general}). \label{fig:LipDQCestimates} }  
\end{center}
\end{figure}

\begin{figure}[!h]
\begin{center}
\includegraphics[scale=0.155,trim=47mm 35mm 0mm 0mm,clip]{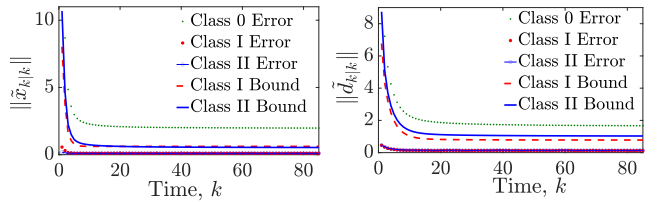}\vspace{-2.2cm}
\caption{Estimation errors and their upper bounds for {Class $0$} ({${(\mathcal{M},\gamma)}$-QC}), {Class I} (Lipschitz) and {Class II} ({${(\mathcal{A},\gamma)}$-QC*}) functions. \label{fig:LipDQCerrors} }  
\end{center}
\end{figure}

We consider a single-link manipulator with flexible joints \cite{abolhasani2018robust,raghavan1994observer}, where the system has 4 states. We slightly modify the dynamical system described in \cite{abolhasani2018robust}, by ignoring the dynamics for the unknown inputs (different from the existing bounded disturbances) to make them \emph{completely unknown} input signals. We also consider bounded-norm disturbances (instead of stochastic noise signals in \cite{abolhasani2018robust,raghavan1994observer}). So, we have the dynamical system \eqref{eq:system} with $n=4$, $f(x)=Ax+\begin{bmatrix} 0 & 2.16T_s & 0 & -3.33 T_s \sin (x_3) \end{bmatrix}^\top$, $A=\begin{bmatrix} 1 & T_s & 0 & 0 \\ -48.6T_s & 1-1.25T_s & 48.6T_s & 0 \\ 0 & 0 & 1 & T_s \\ 19.5T_s & 0 & -19.5T_s & 1 \end{bmatrix}$, $p=m=1$, $l=2$, $B=0_{4 \times 1}$, $G=T_s \begin{bmatrix} 5 & 5 & 2 & 1 \end{bmatrix}^\top$, $C=\begin{bmatrix} 1 & 0 & 0 & 0 \\ 0 & 1 & 0 & 0 \end{bmatrix}$, $W=I$, $D=0_{2 \times 1}$, $T_s=0.01$, $H=T_s\begin{bmatrix}1.1 & 2  \end{bmatrix}^\top$ and $\eta_w=\eta_v=0.1$. The unknown input signal is depicted in Figure \ref{fig:LipDQCestimates}. Vector field $f(\cdot)$ is a Class \ref{class:Lip} function with $L_f=3.33T_s \| diag \{0, 0, 0, 1\} \|=3.33T_s$ (cf. \cite{abolhasani2018robust}), as well as 
a Class \ref{class:A} function, 
with $\mathcal{A}=A$ and $\gamma=0$ (cf. Lemma \ref{lem:decomposition}). \moham{It is also a Class \ref{class:general} function with $\mathcal{M}=\begin{bmatrix} I & 0 \\ 0 & (3.333T_s)^2I \end{bmatrix} \succeq \begin{bmatrix} -I & 0 \\ 0 & (3.333T_s)^2I \end{bmatrix}$, by Propositions \ref{prop:DQCextend}--\ref{prop:weakerness}}. 
Solving the \syong{SDPs in Theorem \ref{thm:noise-attenuation-general} corresponding to system classes 0--II} return \moham{$P^\star_{0}=\begin{bmatrix} 1.4458 & -1.5232 & -0.3419 & -0.2265 \\ -1.5232 & 2.5753& -0.2546 & -0.1828 \\ -0.3419 & -0.2546 & 1.2159 &-0.1475 \\ -0.2265 & -0.1828 &-0.1475 & 1.2605 \end{bmatrix}$, $Y^{\star}_{0}=\begin{bmatrix} 0.4348 & 0.3372 & 0.1376 & 0.1243 \end{bmatrix}^\top$, $\alpha^\star_{0}=0.750$, ${\rho^\star_{0}}={1.1781}$ and $\tilde{L}^\star_{0}=\begin{bmatrix} 1.2471 & 0.8705 & 0.4783 & 0.2947 \end{bmatrix}^\top$ for \syong{Class} 
\ref{item:general-optimality}, 
$P^\star_{I}=\begin{bmatrix} 1.6684 & -1.8242 & -0.4606 & -0.2268 \\ -1.8242 & 2.8284& -0.2860 & -0.0424 \\ -0.4606 & -0.2860 & 1.2086 &-0.0628 \\ -0.2208 & -0.0424 &-0.0628 & 1.2088 \end{bmatrix}$, $Y^{\star}_{I}=\begin{bmatrix} 0.6464 & 0.4422 & 0.0420 & 0.0264 \end{bmatrix}^\top$, $\alpha^\star_{I}=0.825$, ${\rho^\star_{I}}={0.9436}$ and $\tilde{L}^\star_{I}=\begin{bmatrix} 1.2620 & 0.4288 & 0.4244 & 0.2667 \end{bmatrix}^\top$ for 
\syong{Class I}, 
$P^\star_{II}\hspace{-.1cm}=\hspace{-.15cm}\begin{bmatrix} 2.2823 & -2.7731 & -1.0065 & -0.5036 \\ -2.7731 & 4.7225& -0.1605 & -0.9806 \\ -1.0065 & -0.1605 & 4.5760 &-0.3148 \\ -0.5036 & -0.9806 &-0.3148 & 4.4577 \end{bmatrix}$, $Y^{\star}_{II}=\begin{bmatrix} -0.9692 & 1.0644 & 0.0259 & 0.0661 \end{bmatrix}^\top$, $\alpha^\star_{II}=0.793$, $\rho^{\star}_{II}={0.9783}$ and $\tilde{L}^\star_{II}\hspace{-.1cm}=\hspace{-.15cm}\begin{bmatrix} 0.4605 & 0.1837 & 0.9321 & 0.3519 \end{bmatrix}^\top$ for \syong{Class \ref{class:A}}.} 
 Further, we observe from Figure \ref{fig:LipDQCestimates} that our proposed $\mathcal{H}_\infty$ 
 observer, i.e., Algorithm \ref{algorithm1}, is able to find set-valued estimates of the states and unknown inputs, for $(\mathcal{M},\gamma)$-QC (Class \ref{class:general}), Lipschitz continuous (Class \ref{class:Lip}) and $(\mathcal{A},\gamma)$-QC* (Class \ref{class:A}) functions. The actual estimation errors are also within the predicted upper bounds (cf. Figure \ref{fig:LipDQCerrors}), which converge to steady-state values as established in Theorem \ref{thm:error_bound}. Furthermore, Figures \ref{fig:LipDQCestimates} and \ref{fig:LipDQCerrors} show that in this specific example system, estimation errors and their radii are tighter when applying the obtained observer gains for Class \ref{class:Lip} (i.e., Lipschitz) functions, when compared to applying the ones corresponding to the Class  \ref{class:general} (i.e., $(\mathcal{M},\gamma)$-QC) and Class  \ref{class:A} (i.e., $(\mathcal{A},\gamma)$-QC*) functions. 

\subsection{Comparison with Benchmark Observers}
\begin{figure}[!h]
\begin{center}
\includegraphics[scale=0.145,trim=4mm 2mm 0mm 0mm,clip]{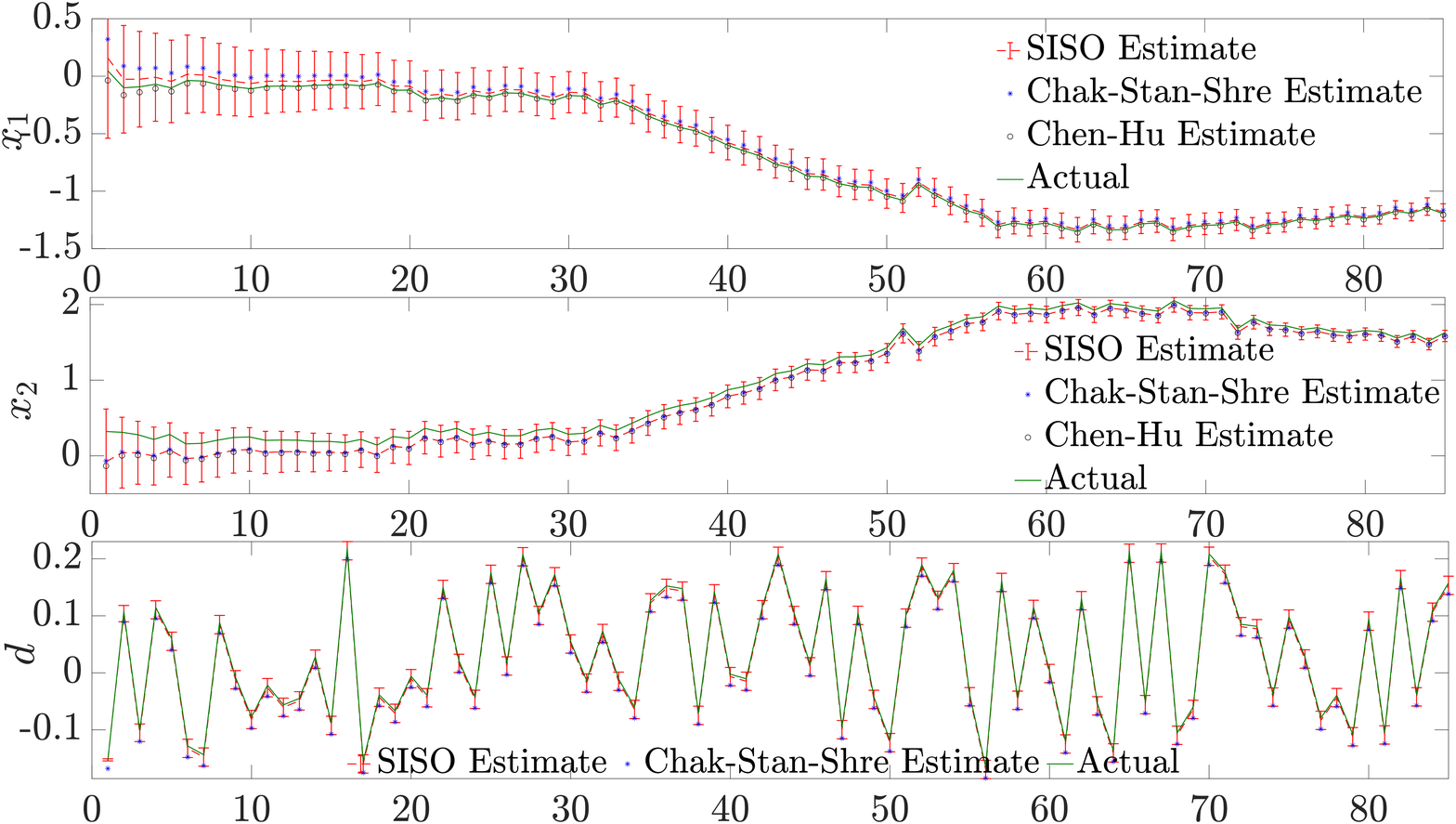}%
\caption{Actual states $x_1$, $x_2$, and their estimates, as well as unknown input $d$ and its estimates in the bounded unknown input scenario, {obtained by applying the observer {in \cite{chen2018nonlinear} (Chen-Hu Estimate), the observer in \cite{chakrabarty2016state}} (Chak-Stan-Shre Estimate) and our {proposed} 
observer (SISO Estimate)}.\label{fig:scenario1est}}
\end{center}
\end{figure}

\begin{figure}[!h]
\begin{center}
\includegraphics[scale=0.158,trim=42mm  4mm 30mm 7mm,clip]{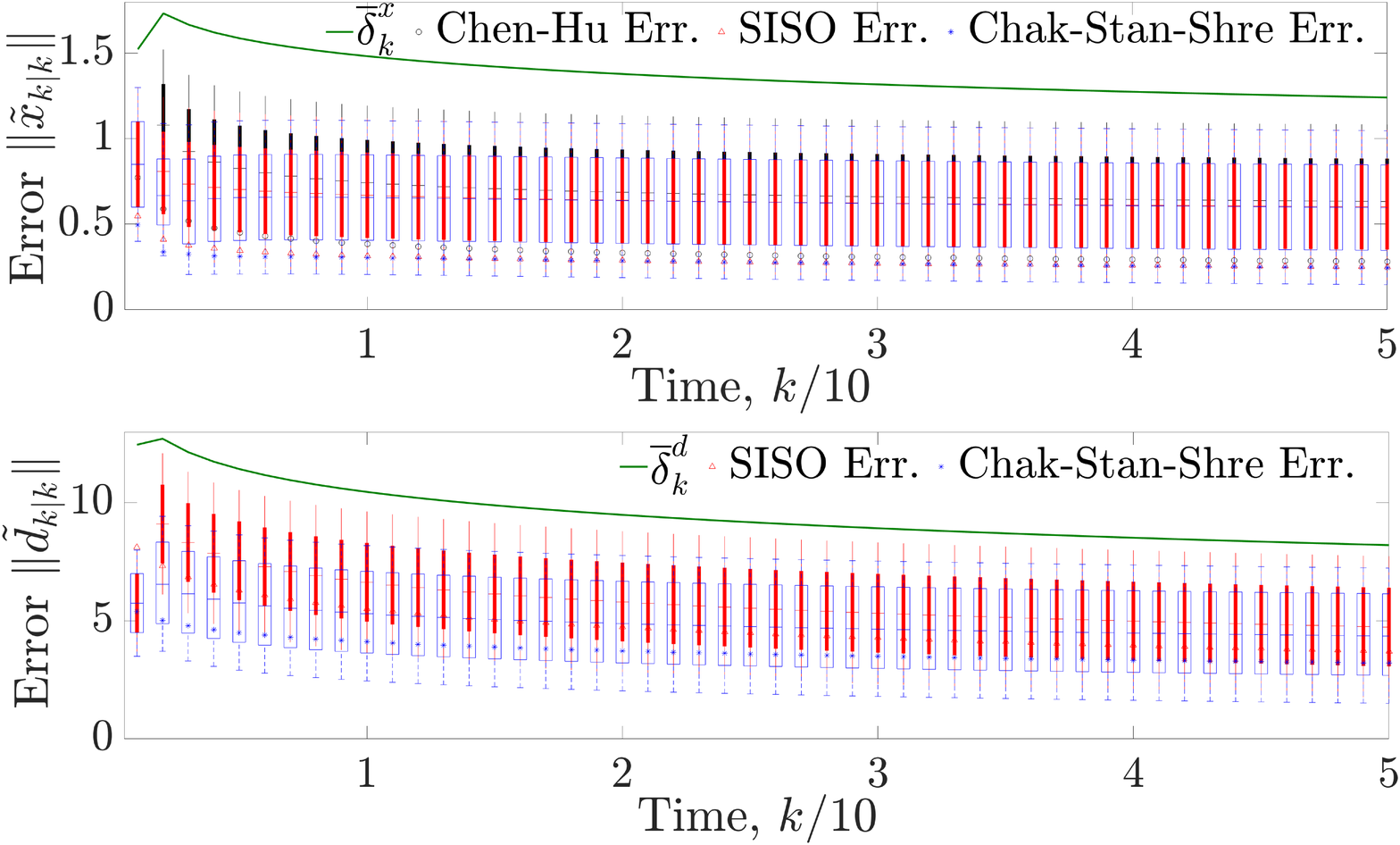} \vspace{-0.55cm}%
\caption{Estimation errors in the bounded unknown input scenario {for {50} different initial values (using box plots), obtained by applying the observer {in \cite{chen2018nonlinear} (Chen-Hu Err.), the   observer in \cite{chakrabarty2016state}}  (Chak-Stan-Shre  Err.) and our {proposed} 
observer (SISO Err.), as well as the computed upper bounds for the state and input errors ({$\overline{\delta}^x_k$ and $\overline{\delta}^d_k$)}}. \label{fig:scenario1errors} }
\end{center}
\end{figure}

In this section, we illustrate the effectiveness of our Simultaneous Input and State Set-Valued Observer (SISO), by comparing its performance with two benchmark observers in \cite{chakrabarty2016state} and \cite{chen2018nonlinear}. 
The designed estimator in \cite{chakrabarty2016state} calculates both (point) state and unknown input estimates, while the observer in \cite{chen2018nonlinear}, only obtains (point) state estimates. For comparison, we apply all the three observers on a benchmark dynamical system in \cite{chakrabarty2016state}, which is in the form of \eqref{eq:system} with $n=2$, $m=l=p=1$, $f(x)=\begin{bmatrix} -0.42x_1+x_2 & \ & -0.6x_1-1.25\tanh (x_1) \end{bmatrix}^\top$, $G=\begin{bmatrix} 1 & -0.65 \end{bmatrix}^\top$,$B=D=H=0_{1 \times 1}$, $C=\begin{bmatrix} 0 & 1 \end{bmatrix}$, $W=I$, $\eta_w=0.2$ and $\eta_v=0.1$. The vector field $f(\cdot)$ is Lipschitz continuous (i.e., Class \ref{class:Lip}) with $L_f=1.1171$. We consider two scenarios for the unknown input. In the first, we consider a random signal with bounded norm, i.e., $\|d_k\| \leq0.2$ for the unknown input $d_k$, while $d_k$ in the second scenario is a {time-varying signal} that becomes unbounded 
{as} time increases. As is demonstrated in Figures \ref{fig:scenario1est} and \ref{fig:scenario1errors}, in the first scenario, i.e., {with} bounded unknown inputs, the set estimates of our approach (i.e., SISO estimates) converge to steady{-state} values and the point estimates of the two benchmark approaches \cite{chakrabarty2016state,chen2018nonlinear} are within the predicted upper bounds and exhibit a convergent behavior {for {all 50} {randomly chosen} 
initial  values (cf. Figure \ref{fig:scenario1errors})}. In this scenario, the two benchmark approaches result in slightly better performance than SISO, since they 
benefit from the additional assumption of \emph{bounded} input. 

\begin{figure}[!h]
\begin{center}
\includegraphics[scale=0.165,trim=15mm 2mm 3mm 0mm,clip]{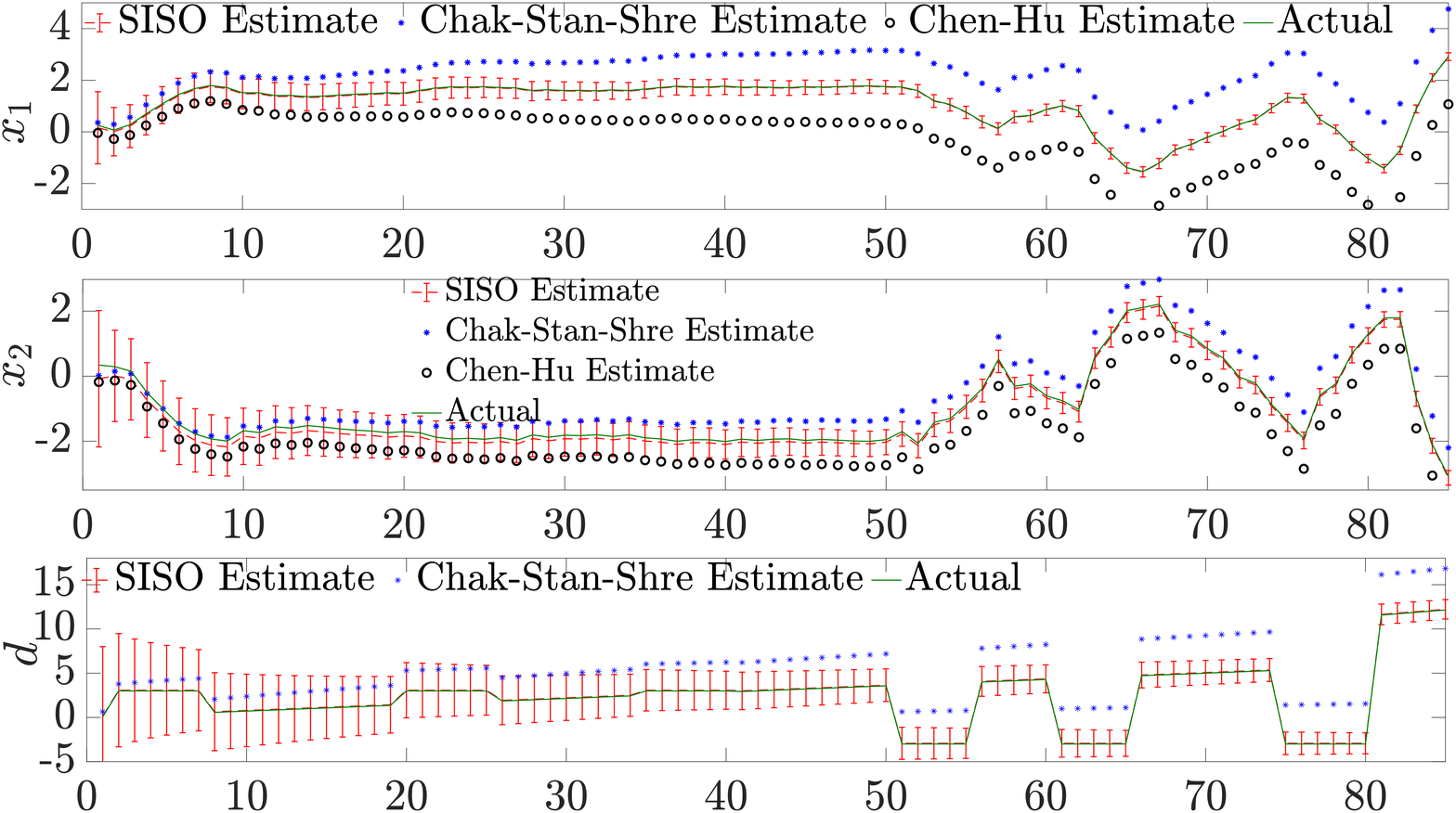}%
\vspace{-0.15cm}
\caption{Actual states $x_1$, $x_2$, and their estimates, as well as unknown input $d$ and its estimates in the unbounded unknown input scenario, {obtained by applying the observer {in \cite{chen2018nonlinear} (Chen-Hu Estimate), the   observer in \cite{chakrabarty2016state}}  (Chak-Stan-Shre Estimate) and our {proposed} 
observer (SISO Estimate)}. \label{fig:scenario2est}}
\end{center}
\end{figure}

\begin{figure}[!h]
\begin{center}
\includegraphics[scale=0.158,trim=48mm  4mm 30mm 7mm,clip]{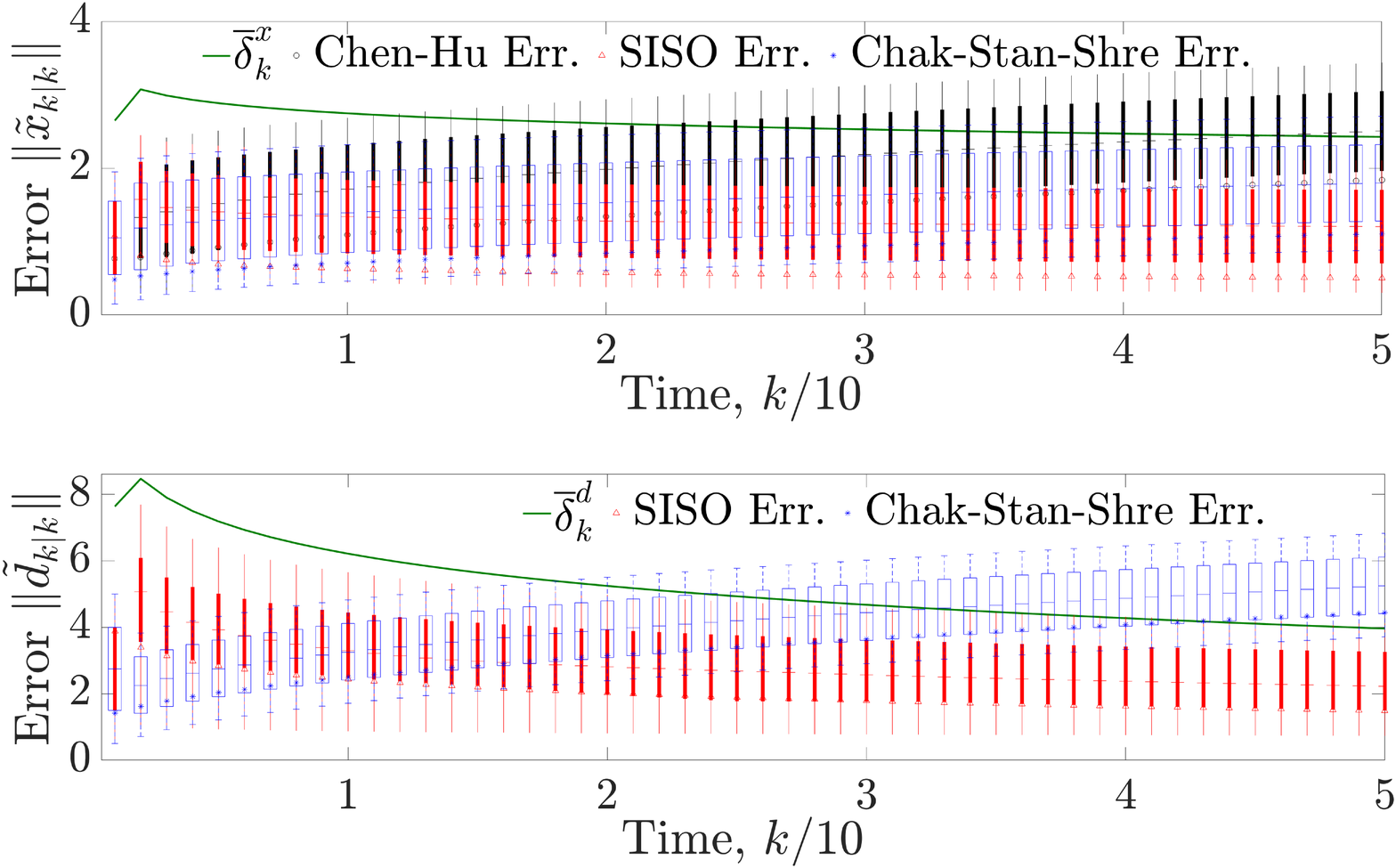} \vspace{-0.55cm}%
\caption{Estimation errors in the unbounded unknown input scenario {for {50} different initial values (using box plots), obtained by applying the observer {in \cite{chen2018nonlinear} (Chen-Hu Err.), the   observer in \cite{chakrabarty2016state}} (Chak-Stan-Shre  Err.) and our {proposed} 
observer (SISO Err.), as well as the computed upper bounds for the state and input errors (\moham{$\overline{\delta}^x_k$ and $\overline{\delta}^d_k$})}.\label{fig:scenario2errors} }
\end{center}
\end{figure}

More interestingly, considering the second scenario, i.e., with unbounded unknown inputs, Figures \ref{fig:scenario2est} and \ref{fig:scenario2errors} 
demonstrate that our set-valued estimates still converge, i.e., our observer remains stable {for all {50} 
randomly chosen initial values}, with \moham{$P^\star=\begin{bmatrix} 1.8086 & 1.4022 \\ 1.4022 & 5.2068 \end{bmatrix}$, $Y^\star=\begin{bmatrix} -0.2282 & 0.6664 \end{bmatrix}^\top$, $\tilde{L}^\star=\begin{bmatrix} -0.2604 & 0.2064 \end{bmatrix}$, $\alpha^\star =0.8875$ and ${\rho^\star}={1.7336}$}, 
 while the estimates of the two benchmark approaches exceed  the boundaries of the compatible sets of states and inputs after some time steps of our approach and display a divergent behavior {for all the initial values} (cf. Figure \ref{fig:scenario2errors}). 
\section{Conclusion and Future Work}
 We presented fixed-order set-valued $\mathcal{H}_\infty$-observers for nonlinear bounded-error discrete-time dynamic systems with unknown inputs. \moham{Necessary and} sufficient Linear Matrix Inequalities for \moham{quadratic} stability of the designed observer were derived for different classes of nonlinear systems, including {${(\mathcal{M},\gamma)}$-QC} systems, Lipschitz continuous systems, {${(\mathcal{A},\gamma)}$-QC*} systems and Linear Parameter-Varying systems. 
  Moreover, we derived additional LMI conditions and corresponding tractable semi-definite programs for obtaining the minimum $\mathcal{H}_{\infty}$-norm for the transfer function that maps the noise signal to the state error of the stable observers. 
 
  In addition, we \moham{showed that 
  the sequences of estimate radii \syong{of our $\mathcal{H}_\infty$-observer are} 
  uniformly bounded and derived closed-\syong{form} expressions for their upper \syong{bound} 
  sequences.} Further, we obtained sufficient conditions for the convergence of the radii \moham{upper \syong{bound} sequences} and derived their steady-state values. 
  Finally,  using two illustrative examples, we 
demonstrated the effectiveness of our proposed design, as well as its advantages over two existing benchmark observers. 
For future work, we plan to generalize this framework to hybrid and switched nonlinear systems and consider other forms of CPS attacks. 


\bibliographystyle{plain}        
\bibliography{autosam}           

\begin{thebibliography}{10}

\bibitem{abolhasani2018robust}
M.~Abolhasani and M.~Rahmani.
\newblock Robust deterministic least-squares filtering for uncertain
  time-varying nonlinear systems with unknown inputs.
\newblock {\em Systems \& Control Letters}, 122:1--11, 2018.

\bibitem{acikmese2003stability}
A.~Acikmese and M.~Corless.
\newblock Stability analysis with quadratic lyapunov functions: A necessary and
  sufficient multiplier condition.
\newblock In {\em Proceedings of the Annual Allerton Conference on
  Communication, Control and Computing}, volume~41, pages 1546--1555. Citeseer,
  2003.

\bibitem{accikmecse2011observers}
B.~A{\c{c}}{\i}kme{\c{s}}e and M.~Corless.
\newblock Observers for systems with nonlinearities satisfying incremental
  quadratic constraints.
\newblock {\em Automatica}, 47(7):1339--1348, 2011.

\bibitem{Anderson.Moore.1981}
B.D.O. Anderson and J.B. Moore.
\newblock Detectability and stabilizability of time-varying discrete-time
  linear systems.
\newblock {\em SIAM Journal on Control and Optimization}, 19(1):20--32, 1981.

\bibitem{blanchini2012convex}
F.~Blanchini and M.~Sznaier.
\newblock A convex optimization approach to synthesizing bounded complexity
  $\ell^{\infty}$ filters.
\newblock {\em IEEE Transactions on Automatic Control}, 57(1):216--221, 2012.

\bibitem{Cardenas.2008b}
A.A. C\'{a}rdenas, S.~Amin, and S.~Sastry.
\newblock Research challenges for the security of control systems.
\newblock In {\em Conference on Hot Topics in Security}, pages 6:1--6:6, 2008.

\bibitem{chakrabarty2019estimating}
A.~Chakrabarty and M.~Corless.
\newblock Estimating unbounded unknown inputs in nonlinear systems.
\newblock {\em Automatica}, 104:57--66, 2019.

\bibitem{chakrabarty2016state}
A.~Chakrabarty, S.H. {\.Z}ak, and S.~Sundaram.
\newblock State and unknown input observers for discrete-time nonlinear
  systems.
\newblock In {\em 2016 IEEE 55th Conference on Decision and Control (CDC)},
  pages 7111--7116. IEEE, 2016.

\bibitem{chandrasekar2006implicit}
K.~Chandrasekar and M.S Hsiao.
\newblock Implicit search-space aware cofactor expansion: A novel preimage
  computation technique.
\newblock In {\em 2006 International Conference on Computer Design}, pages
  280--285. IEEE, 2006.

\bibitem{chen2018nonlinear}
B.~Chen and G.~Hu.
\newblock Nonlinear state estimation under bounded noises.
\newblock {\em Automatica}, 98:159--168, 2018.

\bibitem{chen2005observer}
J.~Chen and C.M. Lagoa.
\newblock Observer design for a class of switched systems.
\newblock In {\em IEEE Conference on Decision and Control European Control
  Conference}, pages 2945--2950, 2005.

\bibitem{chong2015observability}
M.S. Chong, M.~Wakaiki, and J.P. Hespanha.
\newblock Observability of linear systems under adversarial attacks.
\newblock In {\em IEEE American Control Conference (ACC)}, pages 2439--2444,
  2015.

\bibitem{DeNicolao.1997}
G.~De~Nicolao, G.~Sparacino, and C.~Cobelli.
\newblock Nonparametric input estimation in physiological systems: Problems,
  methods, and case studies.
\newblock {\em Automatica}, 33(5):851--870, 1997.

\bibitem{Farwell.2011}
J.P. Farwell and R.~Rohozinski.
\newblock Stuxnet and the future of cyber war.
\newblock {\em Survival}, 53(1):23--40, 2011.

\bibitem{Fawzi.2014}
H.~Fawzi, P.~Tabuada, and S.~Diggavi.
\newblock Secure estimation and control for cyber-physical systems under
  adversarial attacks.
\newblock {\em IEEE Transactions on Automatic Control}, 59(6):1454--1467, June
  2014.

\bibitem{garcia1994robust}
G.~Garcia, J.~Bernussou, and D.~Arzelier.
\newblock Robust stabilization of discrete-time linear systems with
  norm-bounded time-varying uncertainty.
\newblock {\em Systems \& Control Letters}, 22(5):327--339, 1994.

\bibitem{Gillijns.2007}
S.~Gillijns and B.~De~Moor.
\newblock Unbiased minimum-variance input and state estimation for linear
  discrete-time systems.
\newblock {\em Automatica}, 43(1):111--116, January 2007.

\bibitem{Gillijns.2007b}
S.~Gillijns and B.~De~Moor.
\newblock Unbiased minimum-variance input and state estimation for linear
  discrete-time systems with direct feedthrough.
\newblock {\em Automatica}, 43(5):934--937, 2007.

\bibitem{ha2004state}
Q.P. Ha and H.~Trinh.
\newblock State and input simultaneous estimation for a class of nonlinear
  systems.
\newblock {\em Automatica}, 40(10):1779--1785, 2004.

\bibitem{horn2012matrix}
R.A. Horn and C.R. Johnson.
\newblock {\em Matrix analysis}.
\newblock Cambridge University Press, 2012.

\bibitem{jiang2001input}
Z.P. Jiang and Y.~Wang.
\newblock Input-to-state stability for discrete-time nonlinear systems.
\newblock {\em Automatica}, 37(6):857--869, 2001.

\bibitem{khajenejad2019simultaneous}
M.~Khajenejad and S.Z. Yong.
\newblock Simultaneous input and state set-valued
  $\mathcal{H}_{\infty}$-observers for linear parameter-varying systems.
\newblock In {\em American Control Conference (ACC)}, pages 4521--4526, 2019.

\bibitem{khajenejadasimultaneous}
M.~Khajenejad and S.Z Yong.
\newblock Simultaneous mode, input and state set-valued observers with
  applications to resilient estimation against sparse attacks.
\newblock In {\em IEEE Conference on Decision and Control (CDC), Accepted},
  2019.

\bibitem{khajenejad2020simultaneousinterval1}
Mohammad Khajenejad and Sze~Zheng Yong.
\newblock Simultaneous input and state interval observers for nonlinear systems
  with full-rank direct feedthrough.
\newblock {\em arXiv preprint, https://arxiv.org/abs/2002.04761}, Accepted in
  CDC 2020.

\bibitem{khajenejad2020simultaneousinterval2}
Mohammad Khajenejad and Sze~Zheng Yong.
\newblock Simultaneous input and state interval observers for nonlinear systems
  with rank-deficient direct feedthrough.
\newblock {\em arXiv preprint, https://arxiv.org/abs/2004.01861}, Submitted to
  ECC 2021, under review.

\bibitem{khalil2002nonlinear}
H.K. Khalil and J.W. Grizzle.
\newblock {\em Nonlinear systems}, volume~3.
\newblock Prentice hall Upper Saddle River, NJ, 2002.

\bibitem{kim2016attack}
H.~Kim, P.~Guo, M.~Zhu, and P.~Liu.
\newblock Attack-resilient estimation of switched nonlinear cyber-physical
  systems.
\newblock In {\em American Control Conference (ACC)}, pages 4328--4333. IEEE,
  2017.

\bibitem{Kitanidis.1987}
P.K. Kitanidis.
\newblock Unbiased minimum-variance linear state estimation.
\newblock {\em Automatica}, 23(6):775--778, November 1987.

\bibitem{korbicz2007lmi}
J.~Korbicz, M.~Witczak, and V.~Puig.
\newblock {LMI}-based strategies for designing observers and unknown input
  observers for non-linear discrete-time systems.
\newblock {\em Bulletin of the Polish Academy of Sciences: Technical Sciences},
  2007.

\bibitem{lu2016framework}
P.~Lu, E-J. Van~Kampen, C.C. De~Visser, and Q.~Chu.
\newblock Framework for state and unknown input estimation of linear
  time-varying systems.
\newblock {\em Automatica}, 73:145--154, 2016.

\bibitem{milanese1991optimal}
M.~Milanese and A.~Vicino.
\newblock Optimal estimation theory for dynamic systems with set membership
  uncertainty: An overview.
\newblock {\em Automatica}, 27(6):997--1009, 1991.

\bibitem{nakahira2015dynamic}
Y.~Nakahira and Y.~Mo.
\newblock Dynamic state estimation in the presence of compromised sensory data.
\newblock In {\em IEEE Conference on Decision and Control (CDC)}, pages
  5808--5813, 2015.

\bibitem{nien1998algorithm}
C.H. Nien and F.J Wicklin.
\newblock An algorithm for the computation of preimages in noninvertible
  mappings.
\newblock {\em International Journal of Bifurcation and Chaos}, 8(02):415--422,
  1998.

\bibitem{pajic2015attack}
M.~Pajic, P.~Tabuada, I.~Lee, and G.J. Pappas.
\newblock Attack-resilient state estimation in the presence of noise.
\newblock In {\em IEEE Conference on Decision and Control (CDC)}, pages
  5827--5832, 2015.

\bibitem{Pasqualetti.2013}
F.~Pasqualetti, F.~D{\"o}rfler, and F.~Bullo.
\newblock Attack detection and identification in cyber-physical systems.
\newblock {\em IEEE Transactions on Automatic Control}, 58(11):2715--2729,
  November 2013.

\bibitem{Patton.1989}
R.~Patton, R.~Clark, and P.M. Frank.
\newblock {\em Fault diagnosis in dynamic systems: theory and applications}.
\newblock Prentice Hall, 1989.

\bibitem{Peters.Iglesias.1999}
M.A. Peters and P.A. Iglesias.
\newblock A spectral test for observability and reachability of time-varying
  systems.
\newblock {\em SIAM Journal on Control Optimization}, 37(5):1330--1345, August
  1999.

\bibitem{raghavan1994observer}
S.~Raghavan and J.K. Hedrick.
\newblock Observer design for a class of nonlinear systems.
\newblock {\em International Journal of Control}, 59(2):515--528, 1994.

\bibitem{rego2020guaranteed}
B.S. Rego, G.V. Raffo, J.K. Scott, and D.M Raimondo.
\newblock Guaranteed methods based on constrained zonotopes for set-valued
  state estimation of nonlinear discrete-time systems.
\newblock {\em Automatica}, 111:108614, 2020.

\bibitem{Richards.2008}
G.~Richards.
\newblock Hackers vs slackers.
\newblock {\em Engineering Technology}, 3(19):40--43, November 2008.

\bibitem{sheng2003efficient}
S.~Sheng and M.~Hsiao.
\newblock Efficient preimage computation using a novel success-driven atpg.
\newblock In {\em 2003 Design, Automation and Test in Europe Conference and
  Exhibition}, pages 822--827. IEEE, 2003.

\bibitem{shoukry2015imhotep}
Y.~Shoukry, P.~Nuzzo, A.~Puggelli, A.L. Sangiovanni-Vincentelli, S.A. Seshia,
  M.~Srivastava, and P.~Tabuada.
\newblock Imhotep-{SMT}: {A} satisfiability modulo theory solver for secure
  state estimation.
\newblock In {\em 13th International Workshop on Satisfiability Modulo Theories
  (SMT)}, pages 3--13, 2015.

\bibitem{singh2018mesh}
K.R. Singh, Q.~Shen, and S.Z. Yong.
\newblock Mesh-based affine abstraction of nonlinear systems with tighter
  bounds.
\newblock In {\em Conference on Decision and Control (CDC)}, pages 3056--3061.
  IEEE, 2018.

\bibitem{slay2007lessons}
J.~Slay and M.~Miller.
\newblock Lessons learned from the {M}aroochy water breach.
\newblock In {\em International Conference on Critical Infrastructure
  Protection}, pages 73--82. Springer, 2007.

\bibitem{veluvolu2008discrete}
K.C. Veluvolu and Y.C. Soh.
\newblock Discrete-time sliding-mode state and unknown input estimations for
  nonlinear systems.
\newblock {\em IEEE Transactions on Industrial Electronics}, 56(9):3443--3452,
  2008.

\bibitem{wang1992robust}
Y.~Wang, L.~Xie, and C.E. De~Souza.
\newblock Robust control of a class of uncertain nonlinear systems.
\newblock {\em Systems \& Control Letters}, 19(2):139--149, 1992.

\bibitem{xie2004quadratic}
G.~Xie and L.~Wang.
\newblock Quadratic stability and stabilization of discrete-time switched
  systems with state delay.
\newblock In {\em 2004 43rd IEEE Conference on Decision and Control (CDC)(IEEE
  Cat. No. 04CH37601)}, volume~3, pages 3235--3240. IEEE, 2004.

\bibitem{yakubovich1997s}
V.~Yakubovich.
\newblock S-procedure in nonlinear control theory.
\newblock {\em Vestnick Leningrad Univ. Math.}, 4:73--93, 1997.

\bibitem{yong2018simultaneous}
S.Z. Yong.
\newblock Simultaneous input and state set-valued observers with applications
  to attack-resilient estimation.
\newblock In {\em American Control Conference (ACC)}, pages 5167--5174. IEEE,
  2018.

\bibitem{yong2016robust}
S.Z. Yong, M.Q. Foo, and E.~Frazzoli.
\newblock Robust and resilient estimation for cyber-physical systems under
  adversarial attacks.
\newblock In {\em American Control Conference (ACC)}, pages 308--315. IEEE,
  2016.

\bibitem{Yong.Zhu.ea.CDC15_General}
S.Z. Yong, M.~Zhu, and E.~Frazzoli.
\newblock On strong detectability and simultaneous input and state estimation
  with a delay.
\newblock In {\em IEEE Conference on Decision and Control (CDC)}, pages
  468--475, 2015.

\bibitem{yong2015resilient}
S.Z. Yong, M.~Zhu, and E.~Frazzoli.
\newblock Resilient state estimation against switching attacks on stochastic
  cyber-physical systems.
\newblock In {\em IEEE Conference on Decision and Control (CDC)}, pages
  5162--5169, 2015.

\bibitem{Yong.Zhu.ea.Automatica16}
S.Z. Yong, M.~Zhu, and E.~Frazzoli.
\newblock A unified filter for simultaneous input and state estimation of
  linear discrete-time stochastic systems.
\newblock {\em Automatica}, 63:321--329, 2016.

\bibitem{yong2016tcps}
S.Z. Yong, M.~Zhu, and E.~Frazzoli.
\newblock Switching and data injection attacks on stochastic cyber-physical
  systems: Modeling, resilient estimation, and attack mitigation.
\newblock {\em ACM Transactions on Cyber-Physical Systems}, 2(2):9, 2018.

\bibitem{ukraine.2016}
K.~Zetter.
\newblock Inside the cunning, unprecedented hack of {U}kraine's power grid.
\newblock Wired Magazine, 2016.

\end{thebibliography}


\appendix
 \section{Appendix} 
 \subsection{System Transformation} \label{app:transformation}
 
 
 Let $p_{H}\triangleq {\rm rk} (H)$. Using singular value decomposition, we rewrite the direct feedthrough matrix $H$  as
$H= \begin{bmatrix}U_{1}& U_{2} \end{bmatrix} \begin{bmatrix} \Sigma & 0 \\ 0 & 0 \end{bmatrix} \begin{bmatrix} V_{1}^{\, \top} \\ V_{2}^{\, \top} \end{bmatrix}$, 
where $\Sigma \in \mathbb{R}^{p_{H} \times p_{H}}$ is a diagonal matrix of full rank, $U_{1} \in \mathbb{R}^{l \times p_{H}}$, $U_{2} \in \mathbb{R}^{l \times (l-p_{H})}$, $V_{1} \in \mathbb{R}^{p \times p_{H}}$ and $V_{2} \in \mathbb{R}^{p \times (p-p_{H})}$, while $U\triangleq \begin{bmatrix} U_{1} & U_{2} \end{bmatrix}$ and $V\triangleq \begin{bmatrix} V_{1} & V_{2} \end{bmatrix}$ are unitary matrices. 
When there is no direct feedthrough, $\Sigma$, $U_{1}$ and $V_{1}$ are empty matrices\footnote{\ Based on the convention that the inverse of an empty matrix is an empty matrix and the assumption that operations with empty matrices are possible.}, 
and $U_{2}$ and $V_{2}$ are arbitrary unitary matrices, {while when $p_H=p=l$, $U_{2}$ and $V_{2}$ are empty matrices, and $U_{1}$ and $\Sigma$ are identity matrices}.

Then, 
we decouple the unknown input into two orthogonal components: 
\begin{align}\label{eq:dec}
d_{1,k}=V_{1}^\top d_k, \quad
d_{2,k}=V_{2}^\top d_k.
\end{align}
Considering that $V$ is unitary, 
\begin{align} \label{eq:augmenting-d}
d_k =V_{1} d_{1,k}+V_{2} d_{2,k},
\end{align}
 and we can represent the system \eqref{eq:system3} as:
\begin{align}
\hspace{-0.3cm}\begin{array}{rl} x_{k+1}
&= f_k(x_k) +B_ku_k+ G_{1} d_{1,k} +\hspace{-0.05cm} G_{2} d_{2,k}+Ww_k, 
\\  y_k
&={C} x_k + D_ku_k+ H_{1} d_{1,k}+v_k, \end{array}\hspace{-0.3cm}   \label{eq:y}
\end{align}
where $G_{1} \triangleq G V_{1}$, $G_{2} \triangleq G V_{2}$ and $H_{1} \triangleq H V_{1}=U_{1} \Sigma$. 
Next, the output $y_k$ is decoupled 
using a nonsingular transformation $T =\begin{bmatrix} T_{1}^\top & T_{2}^\top \end{bmatrix}^\top \triangleq U^\top =\begin{bmatrix} U_{1} & U_{2} \end{bmatrix}^\top $ 
to obtain $z_{1,k} \in \mathbb{R}^{p_{H}}$ and $z_{2,k} \in \mathbb{R}^{l-p_{H}}$ given by
\begin{gather} \label{eq:sysY} \hspace{-0.2cm}\begin{array}{lll}
z_{1,k} &\triangleq T_{1} y_k =U_{1}^\top y_k \\ &= {C}_{1} x_k + \Sigma d_{1,k} + {D}_{k,1} u_k + {v}_{1,k},\\
z_{2,k} &\triangleq T_{2} y_k =U_{2}^\top y_k \\ &=  {C}_{2}  x_k + {D}_{k,2} u_k + {v}_{2,k},
\end{array} 
\end{gather}
where ${C}_{1} \triangleq U_1^\top {C}$, ${C}_{2} \triangleq U_{2}^\top {C}$, ${D}_{k,1} \triangleq U_{1}^\top {D}_k$, ${D}_{k,2} \triangleq  U_{2}^\top {D}_k$, ${v}_{1,k} \triangleq U_{1}^\top {v}_k$ and ${v}_{2,k} \triangleq  U_{2}^\top {v}_k$. This transformation is also chosen such that $\|\begin{bmatrix} {{v}_{1,k}}^\top & {{v}_{2,k}}^\top \end{bmatrix}^\top\|=\| U^\top {v}_k\|=\|{v}_k\|$. {As a result, we obtain the transformed system  \eqref{eq:sys2}.}
 \subsection{Proofs}
Next, we provide proofs for our propositions, lemmas and theorems. First, for the sake of reader's convenience, we restate a lemma from \cite{wang1992robust} that we will frequently use in deriving some of our results. 
\begin{lem} \cite[Lemma 2.2]{wang1992robust} \label{lem:decouple}
Let $D$, $S$ and $F$ be real matrices of appropriate dimensions and $F^\top F \preceq I$. Then, for any scalar $\varepsilon >0$ and $x,y \in \mathbb{R}^n$, 
\begin{align*}
2x^\top DFSy \leq \varepsilon^{-1}x^\top DD^\top x+ \varepsilon y^\top S^\top S y.
\end{align*}
\end{lem}
\subsection{Proof of Proposition \ref{prop:DQCextend}}
The results follow from the facts that an inequality in $\mathbb{R}$ is preserved by multiplying the both sides by a non-negative number, or by multiplying the left hand side by a non-negative number that is not greater than 1, or by increasing the right hand side, as well as $A \preceq B \implies x^\top (A-B) x \preceq 0$. \hfill \QEDA
\subsection{Proof of Proposition \ref{prop:LiptodeltaQC}}
Considering $M=\begin{bmatrix} -I & 0 \\ 0 & L^2_f \end{bmatrix}$, we have
 $\begin{bmatrix} \Delta f_k^\top & \hspace{-0.1cm} \Delta q^\top \end{bmatrix} {M}$ $\begin{bmatrix}\hspace{-0.05cm}\Delta f_k^\top & \Delta q^\top\hspace{-0.1cm} \end{bmatrix}^\top=-\Delta f_k^\top \Delta f_k +L^2_f \Delta q^\top \Delta q \geq 0$, where the inequality is implied by the Lipschitz continuity of $f_k(\cdot)$. \hfill \QEDA
\subsection{Proof of Proposition \ref{prop:weakerness}}
By definition, $f_k(\cdot)$ is $\delta$-QC with multiplier matrix $M$ means that $\begin{bmatrix} \Delta f_k^\top & \hspace{-0.1cm} \Delta q^\top \end{bmatrix}$ ${M} \begin{bmatrix} \hspace{-0.05cm}\Delta f_k^\top & \Delta q^\top\hspace{-0.1cm} \end{bmatrix}^\top  \geq 0$. Then, it follows in a straightforward manner that  $\begin{bmatrix} \Delta f_k^\top & \Delta q^\top \end{bmatrix} ({-M}) \begin{bmatrix} \Delta f_k^\top & \Delta q^\top \end{bmatrix}^\top \leq \gamma$ for every $\gamma \geq 0$. \QEDA
\subsection{Proof of Proposition \ref{prop:assumptions12}}
We observe that $\begin{bmatrix} \Delta f_k^\top & \Delta x^\top \end{bmatrix} \mathcal{M} \begin{bmatrix} \Delta f_k^\top & \Delta x^\top \end{bmatrix}^\top$ $= -\Delta f_k^\top \Delta f_k$$\geq$ $-L^2_f\|\Delta x\|^2$$\geq -L^2_f (2r)^2=-4r^2L^2_f$, where the second and third inequalities hold by Lipschitz continuity of $f(\cdot)$ and boundedness of the state space, respectively. \QEDA
\moham{
\subsection{Proof of Proposition \ref{prop:qcstar_Lip}}
It follows from Definitions \ref{def:DQM} and \ref{defn:A} and $\gamma \geq 0$ that
\begin{align*}
-\Delta f_k^\top \Delta f_k +2 \Delta f_k^\top \mathcal{A}\Delta x-\Delta x^\top \mathcal{A}^\top \mathcal{A} \Delta x \geq \gamma \geq 0.
\end{align*}
This, \syong{along with} 
Lemma \ref{lem:decouple}, imply that
\begin{align*}
(1-\epsilon) &\|\Delta f_k\|^2 \leq (\epsilon^{-1}-1)\Delta x^\top \mathcal{A}^\top \mathcal{A} \Delta x,\quad \forall \epsilon >0 \\
\Rightarrow &\|\Delta f_k\|^2 \leq \frac{1}{\epsilon}\Delta x^\top \mathcal{A}^\top \mathcal{A} \Delta x, \quad \forall 0 < \epsilon <1. 
\end{align*}
By taking the limit of the both sides when $\epsilon \to 1$ and \syong{by the} continuity of \syong{the} "$\leq$" \syong{operator}, we obtain
\begin{align*}
\|\Delta f_k\|^2 \leq \Delta x^\top \mathcal{A}^\top \mathcal{A} \Delta x \leq \lambda_{\max}(\mathcal{A}^\top \mathcal{A})\|\Delta x\|^2. 
\end{align*}
\syong{Finally,} the 
\syong{result is obtained} by taking the square root \syong{of} 
both sides \syong{of the above inequality}. \QEDA
}
\subsection{Proof of Lemma \ref{lem:decomposition}} 
First, notice that $\Delta f_k=A\Delta x+\Delta g$. Given this and $\|g(x)\| \leq r$, we can conclude that  
\begin{align*}
&\begin{bmatrix} \Delta f_k^\top & \Delta x^\top \end{bmatrix} \mathcal{M} \begin{bmatrix} \Delta f_k^\top & \Delta x^\top \end{bmatrix}^\top \\
&=-\Delta f_k^\top \Delta f_k+2\Delta x^\top A^\top \Delta f_k-\Delta x^\top A^\top A \Delta x \\
&=-(\Delta f_k \hspace{-.1cm}-\hspace{-.1cm}A \Delta x)^\top (\Delta f_k \hspace{-.1cm}-\hspace{-.1cm}A \Delta x)\hspace{-.1cm}=\hspace{-.05cm}-\Delta g^\top\hspace{-.05cm} \Delta g \hspace{-.05cm}\geq \hspace{-.05cm} -(2r)^2. \ \QEDA
\end{align*}

 \subsection{Proof of Proposition \ref{prop:AA}}
By construction, we have the following condition: 
 $  \mathcal{M} - \begin{bmatrix} -I_{n \times n} & \mathcal{A} \\ \mathcal{A}^\top & -\mathcal{A}^\top \mathcal{A} \end{bmatrix} =\begin{bmatrix} \mathcal{M}_{11}+I & 0 \\ 0 & \mathcal{M}_{22}+ \mathcal{M}^\top_{12}\mathcal{M}_{12}\end{bmatrix} \preceq 0$, since both submatrices on the diagonal are negaitive semi-definite by assumption. \QEDA
\subsection{Proof of Proposition \ref{prop:LPVtoLip}}

The global Lipschitz continuity of LPV systems can be shown as follows:
\begin{align*}
\Delta f_k &\triangleq \|f_k(x_1)-f_k(x_2)\|=\|\textstyle{\sum}_{i=1}^N \lambda_{i,k}A^i \Delta x_k\| \\
&\leq \textstyle{\sum}_{i=1}^N \lambda_{i,k} \|A^i  \Delta x_k\|\leq \textstyle{\sum}_{i=1}^N \lambda_{i,k} \|A^i\| \|\Delta x_k\| \\
& \leq \|A^m\| \|\Delta x_k \|,
\end{align*}
with $\|A^m\|= \max_{i \in {1\dots N}} \|A^i \|$, where the first and second inequalities hold by sub-multiplicative inequality for norms and positivity of $\lambda_{i,k}$, the third inequality holds by the facts that $0 \leq \lambda_{i,k} \leq 1$ and $\textstyle{\sum}_{i=1}^N \lambda_{i,k}=1$. \QEDA
 \subsection{Proof of Lemma \ref{lem:error-dynamics}}
 Aiming to derive the governing equation for the evolution of the state errors, from \eqref {eq:sysY} and \eqref {eq:variant1}, 
we obtain
\begin{align}
\hat{d}_{1,k}= M_1 ( {C}_{1} \tilde{x}_{k|k} + \Sigma d_{1,k} + {v}_{1,k}). \label{eq:estimatedinput111}
\end{align} 
Moreover, from \eqref{eq:system}, \eqref{eq:sysY} and \eqref{eq:d2}--\eqref{eq:xstar}, we have
\begin{align}\label{eq:estimatedinput222} 
\nonumber \hat{d}_{2,k-1}&= M_2 [C_2(\Delta f_{k-1} + G_1 \tilde{d}_{1,k-1}
 + G_2 d_{2,k-1} 
 \\ &\quad \quad \quad + Ww_{k-1}) + v_{2,k}], 
\end{align}
and by plugging  $ M_1=\Sigma^{-1} $ into \eqref{eq:estimatedinput111}, 
we obtain
\begin{align}
\label{eq:inputerror111}  \tilde{d}_{1,k} = d_{1,k} - \hat{d}_{1,k} = - M_{1} ( {C}_{1} \tilde{x}_{k|k} + {v}_{1,k} ),
\end{align}  
where $\Delta f_k\triangleq f_k(x_k)-f_k(\hat{x}_k)$. Then, 
by setting $ M_{2}=( C_{2} G_{2} )^ \dagger $ in \eqref{eq:estimatedinput222} and using \eqref{eq:inputerror111}, 
we have
 \begin{align} \label{eq:Inputerror222}
\tilde{d}_{2,k-1}= &-M_{2}[C_{2}( \Delta f_{k-1} - G_1M_1\\ 
\nonumber  &(C_1\tilde{x}_{k-1|k-1}+v_{1,k-1})+Ww_{k-1})+{v}_{2,k}].
\end{align}
Furthermore, it follows from \eqref{eq:system},\eqref{eq:time} and \eqref{eq:xstar} that 
\begin{align}
 \hspace{-0.15cm}\textstyle \tilde{x}^\star_{k|k} \hspace{-0.1cm}=\hspace{-0.1cm}\Delta f_{k-1} \hspace{-0.05cm}+\hspace{-0.05cm} G_{1} \tilde{d}_{1,k-1} 
  \hspace{-0.05cm}+\hspace{-0.05cm}\textstyle G_{2} \tilde{d}_{2,k-1}\hspace{-0.05cm}+\hspace{-0.05cm}W{w}_{k-1}.   \label{eq:staterrorstarrr} 
\end{align}
In addition, 
by plugging $\tilde{d}_{k-1}$ and $\tilde{d}_{k-2}$ from \eqref{eq:inputerror111} and \eqref{eq:Inputerror222} into  \eqref{eq:staterrorstarrr}, by \eqref{eq:sysY} and \eqref{eq:stateEst}, we obtain
\begin{align}
\tilde{x}_{k|k} 
&= ( I - \tilde{L} C_{2} ) \tilde{x}^\star_{k|k} - \tilde{L} \tilde{v}_{k}. \label{eq:staterror} \\
\textstyle \tilde{x}^\star_{k|k} &= \Phi [ \Delta f_{k-1} - G_{1}M_1C_1 \tilde{x}_{k-1|k-1}] 
  +\textstyle  \tilde{w}_{k}, \label{eq:staterrorstarrr3}
\end{align}
where $\tilde{v}_{k} \triangleq v_{2,k}$, $\tilde{w}_{k} \triangleq -\Phi (G_1M_1v_{1,k-1}-Ww_{k-1})-G_2M_2v_{2,k}$ and $\Phi \triangleq I-G_2M_2C_2$. Finally, combining \eqref{eq:staterror} and \eqref{eq:staterrorstarrr3} returns the results. \QEDA
\moham{\subsection{Proof of Theorem \ref{thm:quadratic_stability}}
First, note that the state error dynamics \eqref{eq:errors-dynamics} without bounded noise signals $w_k$ and $v_k$ 
can be rewritten as
\begin{align} \label{eq:error-dynamics-noiseless-2}
\tilde{x}_{k+1|k+1}=\breve{A} \tilde{x}_{k|k}+\breve{B}\Delta f_{k},
\end{align} 
with $\breve{A}\triangleq -(I-\tilde{L}C_2)\Phi\Psi$ and $\breve{B}\triangleq (I-\tilde{L}C_2)\Phi$. Moreover, considering a quadratic positive definite candidate Lyapunov function 
\begin{align}\label{eq:delta_V_wn}
V^{wn}_k=\tilde{x}_{k|k}^\top P \tilde{x}_{k|k}, 
\end{align}
with $P \succ 0$, \eqref{eq:quadratic} is equivalent to:  
\begin{align*}
  (\breve{A}\tilde{x}_{k|k}\hspace{-.1cm}+\breve{B}\Delta f_k)^\top P(\breve{A}\tilde{x}_{k|k}\hspace{-.1cm}+\breve{B}\Delta f_k)\hspace{-.1cm} \leq \hspace{-.1cm} (1-\alpha) \tilde{x}_{k|k}^\top P \tilde{x}_{k|k},
\end{align*}
which can be reorganized as:
\begin{align} \label{eq:quad-matrix}
  \begin{bmatrix}\Delta f_k \\ \tilde{x}_{k|k} \end{bmatrix}^\top \begin{bmatrix} \breve{B}^\top P \breve{B} & \breve{B}^\top P \breve{A} \\ \breve{A}^\top P \breve{B} & \breve{A}^\top P \breve{A}+(\alpha-1)P \end{bmatrix} \begin{bmatrix}\Delta f_k \\ \tilde{x}_{k|k} \end{bmatrix} \leq 0.
\end{align}

\syong{Then, for Class 0 functions (i.e., $(\mathcal{M},\gamma)$-QC functions), by Definition \ref{def:DQM}, we have}
\begin{align} \label{eq:quad-DQC}
  \begin{bmatrix}\Delta f_k \\ \tilde{x}_{k|k} \end{bmatrix}^\top \begin{bmatrix} -M_{11} &  -M_{12} \\  -M^\top_{12} & -M_{22} \end{bmatrix} \begin{bmatrix}\Delta f_k \\ \tilde{x}_{k|k} \end{bmatrix} \leq -\gamma.
\end{align}
 In other words, we want \eqref{eq:quad-matrix} to hold for any pair of $(\tilde{x}_{k|k},\Delta f_k)$ that satisfy \eqref{eq:quad-DQC}. \syong{By applying} $\mathcal{S}$-procedure \cite{yakubovich1997s}, this is equivalent to \syong{the existence of} 
 $\kappa  \geq 0$, such that:
 \begin{align}\label{eq:extension}
 \kappa  \hspace{-.1cm}\begin{bmatrix} -M_{11} & - M_{12} & 0 \\  -M^\top_{12} & -M_{22} & 0 \\ 0 & 0 & \gamma \end{bmatrix} \hspace{-.15cm}- \hspace{-.15cm}\begin{bmatrix} \breve{B}^\top P \breve{B} & \breve{B}^\top P \breve{A} & 0 \\ \breve{A}^\top P \breve{B} & \breve{A}^\top P \breve{A} \hspace{-.1cm}+ \hspace{-.1cm}(\alpha \hspace{-.1cm}- \hspace{-.1cm}1)P & 0 \\ 0 & 0 & 0\end{bmatrix}  \hspace{-.15cm}\succeq  \hspace{-.1cm}0.
 \end{align}
 Since $\gamma \geq 0$ by assumption, \syong{the above} is equivalent to:
  \begin{align*}
 \exists \kappa \hspace{-.1cm}\geq 0\hspace{-.1cm}:\hspace{-.cm}\kappa  \hspace{-.1cm}\begin{bmatrix} -M_{11} &  -M_{12}  \\  -M^\top_{12} & -M_{22} \end{bmatrix} \hspace{-.15cm}- \hspace{-.15cm}\begin{bmatrix} \breve{B}^\top P \breve{B} & \breve{B}^\top P \breve{A} \\ \breve{A}^\top P \breve{B} & \breve{A}^\top P \breve{A} \hspace{-.1cm}+ \hspace{-.1cm}(\alpha \hspace{-.1cm}- \hspace{-.1cm}1)P  \end{bmatrix}  \hspace{-.15cm}\succeq  \hspace{-.1cm}0.
 \end{align*}
 Now, note that $\kappa=0$ is not a \syong{valid} choice, since if $\kappa=0$, \syong{it can be shown that} 
 there is no $P\succ 0$ that \syong{satisfies} 
  the above inequality. 
 Hence, $\kappa >0$, and 
 we equivalently obtain \mohk{$\exists P \succ 0, \kappa>0$, such that}   
 \begin{align}
\nonumber & \begin{bmatrix} \breve{B}^\top P \breve{B} & \breve{B}^\top P \breve{A} \\ \breve{A}^\top P \breve{B} & \breve{A}^\top P \breve{A}+(\alpha-1)P \end{bmatrix}  \preceq \mohk{\kappa} \begin{bmatrix} -M_{11} & - M_{12} \\  -M^\top_{12} & -M_{22} \end{bmatrix}  \\
  \nonumber  &\Leftrightarrow \begin{bmatrix} -\mohk{\kappa}M_{11}\hspace{-.1cm}-\hspace{-.1cm}\breve{B}^\top P \breve{B} & -\mohk{\kappa}M_{12}\hspace{-.1cm}-\hspace{-.1cm}\breve{B}^\top P \breve{A} \\ -\mohk{\kappa}M^\top_{12}\hspace{-.1cm}-\hspace{-.1cm}\breve{A}^\top P \breve{B} & -\mohk{\kappa}M_{22}\hspace{-.1cm}-\hspace{-.1cm}\breve{A}^\top P \breve{A}\hspace{-.1cm}+\hspace{-.1cm}(1\hspace{-.1cm}-\hspace{-.1cm}\alpha)P \end{bmatrix}\hspace{-.1cm} \succeq 0, 
  \end{align}
\syong{which is, in turn, equivalent to:}
   \begin{align*} 
\nonumber  &\forall \epsilon >0\mohk{,\exists P\succ 0,\kappa>0}:\\
  &\begin{bmatrix} \hspace{-.cm}-\mohk{\kappa}M_{11}\hspace{-.1cm}-\hspace{-.1cm}\breve{B}^\top\hspace{-.1cm} P \breve{B}+\epsilon I & -\mohk{\kappa}M_{12}\hspace{-.1cm}-\hspace{-.1cm}\breve{B}^\top\hspace{-.1cm} P \breve{A} \\ \hspace{-.1cm}-\mohk{\kappa}M^\top_{12}\hspace{-.1cm}-\hspace{-.1cm}\breve{A}^\top \hspace{-.1cm}P \breve{B} & -\mohk{\kappa}M_{22}\hspace{-.1cm}-\hspace{-.1cm}\breve{A}^\top\hspace{-.1cm} P \breve{A}\hspace{-.1cm}+\hspace{-.1cm}(1\hspace{-.1cm}-\hspace{-.1cm}\alpha)P\hspace{-.1cm}+\hspace{-.1cm} \epsilon I \hspace{-.cm}\end{bmatrix} \hspace{-.15cm}\succ \hspace{-.1cm}0.
\end{align*}
Then, \syong{by} applying Schur complement, \syong{we obtain} 
\begin{align}\label{eq:Q1}
\forall \epsilon >0: \begin{cases}  
\tilde{M}_{22} \succ \tilde{M}_{12}, \\ 
  -\mohk{\kappa}M_{11}\hspace{-.05cm}-\hspace{-.05cm}\breve{B}^\top P \breve{B}\hspace{-.cm}+\hspace{-.05cm}\epsilon I \hspace{-.05cm}\succ \hspace{-.05cm} 0,
  \end{cases}
  \end{align}
\syong{\syong{with} $\tilde{M}_{22} \triangleq -\mohk{\kappa}M_{22}-\breve{A}^\top P \breve{A}+(1-\alpha)P+\epsilon I$ and $\tilde{M}_{12} \triangleq (\mohk{\kappa}M_{12}^\top+\breve{A}^\top P \breve{B})(-\mohk{\kappa}M_{11}-\breve{B}^\top P \breve{B}+\epsilon I)^{-1}(\mohk{\kappa}M_{12}+\breve{B}^\top P \breve{A})$, where the second inequality is equivalent to:}
\begin{align*}  
   & -\mohk{\kappa}M_{11}\hspace{-.1cm}-\hspace{-.1cm}\breve{B}^\top PP^{-1}P \breve{B}\hspace{-.0cm} \succeq \hspace{-.0cm} 0
  \Leftrightarrow \begin{bmatrix} P & P\breve{B} \\ \breve{B}^\top P & -\mohk{\kappa}M_{11} \end{bmatrix} \succeq 0.
\end{align*}
On the other hand, $\forall \epsilon >0 : \tilde{M}_{22} \succ \tilde{M}_{12}$  \syong{is equivalent to:} 
\begin{align}\label{eq:tmp}
\nonumber &\forall \epsilon >0, \exists \tilde{\Gamma} \succeq 0 :\\
&\begin{cases}
\tilde{\Gamma} \succeq \tilde{M}_{12} \succeq 0, \\
 -\mohk{\kappa}M_{22}\hspace{-.1cm}-\hspace{-.1cm}\breve{A}^\top P P^{-1} P \breve{A}\hspace{-.1cm}+\hspace{-.1cm}(1\hspace{-.1cm}-\hspace{-.1cm}\alpha)P+\epsilon I \hspace{-.1cm}\succ \hspace{-.1cm} \tilde{\Gamma}.
 \end{cases}
\end{align}
\syong{Since the second inequality holds for all $\epsilon$, it is equivalent to $-\mohk{\kappa}M_{22}\hspace{-.cm}-\hspace{-.1cm}\breve{A}^\top P P^{-1} P \breve{A}\hspace{-.cm}+\hspace{-.cm}(1\hspace{-.cm}-\hspace{-.0cm}\alpha)P \hspace{-.cm}\succeq \hspace{-.cm} \tilde{\Gamma}$, which,} 
\syong{by applying}
Schur complement, 
is equivalent to:
\begin{align}\label{eq:Q2}
\begin{bmatrix} P & P\breve{A} \\ \breve{A}^\top P & -\mohk{\kappa}M_{22}+(1-\alpha)P-\tilde{\Gamma} \end{bmatrix} \succeq 0. 
\end{align}
By \syong{also applying} Schur complement {to \eqref{eq:tmp}, we obtain} 
\begin{align*}
&\forall \epsilon \syong{>0}: \tilde{\Gamma} \succeq \tilde{M}_{12} \\ 
& \Leftrightarrow \forall \epsilon >0:\begin{bmatrix} -\mohk{\kappa}M_{11}\hspace{-.1cm}-\hspace{-.1cm}\breve{B}^\top P \breve{B}\hspace{-.1cm}+\epsilon I & \mohk{\kappa}M_{12}\hspace{-.1cm}+ \hspace{-.1cm}\breve{B}^\top P \breve{A} \\ \mohk{\kappa}M_{12}^\top\hspace{-.1cm}+\hspace{-.1cm}\breve{A}^\top P \breve{B} & \tilde{\Gamma} \end{bmatrix} \hspace{-.1cm}\succeq 0\\
& \Leftrightarrow \forall \epsilon >0:\begin{bmatrix} \tilde{\Gamma} & \mohk{\kappa}M_{12}^\top\hspace{-.1cm}+\hspace{-.1cm}\breve{A}^\top P \breve{B} \\ \mohk{\kappa}M_{12}\hspace{-.1cm}+\hspace{-.1cm}\breve{B}^\top P \breve{A} &  -\mohk{\kappa}M_{11}\hspace{-.1cm}-\hspace{-.1cm}\breve{B}^\top P \breve{B}\hspace{-.1cm}+\epsilon I  \end{bmatrix} \hspace{-.1cm}\succeq 0 \\
& \Leftrightarrow \forall \epsilon,\delta >0: \\
&\quad \ \  \begin{bmatrix} \tilde{\Gamma}\hspace{-.1cm}+\hspace{-.1cm}\delta I & \mohk{\kappa}M_{12}^\top\hspace{-.1cm}+\hspace{-.1cm}\breve{A}^\top P \breve{B} \\ \mohk{\kappa}M_{12}\hspace{-.1cm}+\hspace{-.1cm}\breve{B}^\top P \breve{A} & \ -\mohk{\kappa}M_{11}\hspace{-.1cm}-\hspace{-.1cm}\breve{B}^\top P \breve{B}+(\epsilon+\delta) I  \end{bmatrix} \hspace{-.1cm}\succ 0.
\end{align*}
\syong{Applying Schur complement to the above yields}
\begin{align*}
 & \forall \epsilon,\delta >0:\\
  &\tilde{\Gamma}+\delta I \succ 0 \Leftrightarrow \tilde{\Gamma} \succeq 0 \ \mohk{\wedge -\kappa}M_{11}\hspace{-.1cm}-\hspace{-.1cm}\breve{B}^\top P \breve{B}\hspace{-.1cm}+\hspace{-.1cm}(\epsilon \hspace{-.1cm}+ \hspace{-.1cm}\delta ) I \succ \Xi, 
\end{align*}
where $\Xi \triangleq (\mohk{\kappa}M_{12}+\breve{B}^\top P \breve{A}) (\tilde{\Gamma}+\delta I)^{-1}(\mohk{\kappa}M_{12}^\top+\breve{A}^\top P \breve{B})$.
Then, from the last inequality, we equivalently obtain:
$\forall \epsilon,\delta >0:\exists \breve{Q}$ such that: 
\begin{align}\label{eq:Q3}
\nonumber &-\mohk{\kappa}M_{11}-\breve{B}^\top P \breve{B}+(\epsilon+\delta) I \succ \breve{Q}\\ &\Leftrightarrow -\mohk{\kappa}M_{11}\hspace{-.1cm}-\hspace{-.1cm}\breve{B}^\top P \breve{B} \succeq \hspace{-.1cm}\breve{Q}  
\Leftrightarrow \begin{bmatrix} P & P \breve{B} \\ \breve{B}^\top P & -\mohk{\kappa}M_{11}-\breve{Q} \end{bmatrix} \succeq 0
\end{align}
and $\breve{Q} \succeq (\mohk{\kappa}M_{12}\hspace{-.cm}+\hspace{-.cm}\breve{B}^\top P \breve{A}) (\tilde{\Gamma}+\delta I)^{-1}(\mohk{\kappa}M_{12}^\top\hspace{-.cm}+\hspace{-.cm}\breve{A}^\top P \breve{B})$, \syong{which is equivalent to} 
\begin{align} \label{eq:Q-help}
\forall \delta>0: \begin{bmatrix} \tilde{\Gamma}+\delta I & -\mohk{\kappa}M_{12}^\top\hspace{-.1cm}-\hspace{-.1cm}\breve{A}^\top P \breve{B} \\ -\mohk{\kappa}M_{12}\hspace{-.1cm}-\hspace{-.1cm}\breve{B}^\top P \breve{A} & \breve{Q}  \end{bmatrix} \succeq 0. 
\end{align}
Furthermore, pre\syong{-} and post\syong{-}multiplication of \eqref{eq:Q-help} by $\begin{bmatrix} I & 0 \\ 0 &\Psi^\top \end{bmatrix}$ and $\begin{bmatrix} I & 0 \\ 0 & \Psi  \end{bmatrix}$, as well as the fact that $-\breve{B}\Psi=\breve{A}$ by definition and applying Schur complement, \syong{lead to} 
\begin{align*} \label{eq:Q4}
\nonumber &\forall \delta>0: \begin{bmatrix} \tilde{\Gamma}+\delta I  & -\mohk{\kappa}M_{12}^\top \Psi \hspace{-.1cm}+\hspace{-.1cm}\breve{A}^\top P \breve{A} \\ -\Psi^\top \mohk{\kappa}M_{12}\hspace{-.1cm}+\hspace{-.1cm}\breve{A}^\top P \breve{A} & \Psi^\top \breve{Q} \Psi  \end{bmatrix} \succeq 0  \\
\nonumber &\Leftrightarrow \forall \delta>0:  \Psi^\top \breve{Q} \Psi \succeq \tilde{Q},
\end{align*}
where
$\tilde{Q} \triangleq \hspace{-.1cm} (\hspace{-.05cm}-\Psi^\top \mohk{\kappa}M_{12}\hspace{-.1cm}+\hspace{-.1cm}\breve{A}^\top P \breve{A})(\tilde{\Gamma}+\delta I)^{-1}(\hspace{-.05cm}-\mohk{\kappa}M_{12}^\top \Psi \hspace{-.1cm}+\hspace{-.1cm}\breve{A}^\top P \breve{A})$. 
Equivalently, \syong{we can represent this as}:
\begin{align*}
\nonumber &\forall \delta >0, \exists \breve{Z}:\\ 
&\breve{Z} \succeq -\mohk{\kappa}M_{12}^\top \Psi +\breve{A}^\top P \breve{A}, \  \Psi^\top  \breve{Q} \Psi \succeq \breve{Z}^\top (\tilde{\Gamma}+\delta I)^{-1}  \breve{Z},
\end{align*}                   
\syong{where the first inequality is equivalent to:}
\begin{align*}
\begin{bmatrix} P & P\breve{A} \\ \breve{A}^\top P & \ \breve{Z}+\mohk{\kappa}M_{12}^\top \Psi \end{bmatrix} \succeq 0,
\end{align*} 
\syong{and the second inequality can be written as:}
\begin{align}&
 \forall \delta >0: 
 \begin{bmatrix} \tilde{\Gamma}+\delta I & \breve{Z} \\ \breve{Z}^\top  & \Psi^\top \breve{Q} \Psi \end{bmatrix} \succeq 0 \Leftrightarrow  
  \begin{bmatrix} \tilde{\Gamma} & \breve{Z} \\ \breve{Z}^\top  & \Psi^\top \breve{Q} \Psi \end{bmatrix} \succeq 0,
\end{align}
where the forward direction (i.e., ``$\Rightarrow$") 
follows from the fact that the limit of a convergent sequence of positive semi-definite matrices is a positive semi-definite matrix, while the backward direction (i.e., ``$\Leftarrow$") holds since the summation of two positive semi-definite matrices is positive semi-definite. Finally, \syong{by} defining $Y \triangleq P \tilde{L}$ \syong{with} $\tilde{\Gamma} \succ 0$ and the LMIs in \eqref{eq:Q1}--\eqref{eq:Q3} and \eqref{eq:Q4}, we obtain the results for Class 0 functions.

\syong{Next, we consider system classes I--III by appealing to the fact that they are Class 0 systems with additional information about $M_{11},M_{12}$ and $M_{22}$:
\begin{itemize}
 \item Case I: 
   From Proposition \ref{prop:LiptodeltaQC}, 
   we can replace $M_{11},M_{12}$ and $M_{22}$ with $-I,0_{n \times n}$ and $L_f^2I$. 
    \item Case II: 
    By Definition \ref{defn:A}, we can substitute $M_{11},M_{12}$ and $M_{22}$ with $-I,\mathcal{A}$ and $-\mathcal{A}^\top \mathcal{A}$. 
\item Case III: 
Note that by Proposition \ref{prop:LPVtoLip}, an LPV function is Lipschitz continuous by Lipschitz constant $\tilde{\sigma}_m \triangleq \max_{i \in {1\dots N}} \|A^i\|$. Hence, 
we can apply Case I by replacing $L_f$ with $\tilde{\sigma}_m$. 
\end{itemize}}


Finally, suppose \syong{$\gamma <0$}. Then, the fact that $\kappa=0$ is not a \syong{valid} choice \syong{(as previously discussed)} 
implies that \syong{$\kappa \gamma <0$} 
and hence, the matrix in the left hand side of \eqref{eq:extension} always contains a strictly negative diagonal element \syong{$\kappa \gamma <0$}  and consequently, \eqref{eq:extension} can never hold, 
\syong{and thus, neither can} 
\eqref{eq:quad-matrix}.   
\QEDA

 \subsection{Proof of Proposition \ref{thm:stability-general}}
 \moham{To prove the necessity of \eqref{eq:sability-necessity-1}, we use contraposition. Suppose that the LMIs in \eqref{eq:sability-necessity-1} are feasible. Then, we will show that there exists a candidate Lyapunov function $\tilde{V}^{wn}_k=\tilde{x}^\top_{k|k}\tilde{P}\tilde{x}_{k|k}$, for some $\tilde{P} \succ 0$, such that $ \Delta \tilde{V}^{wn}_k \triangleq \tilde{V}^{wn}_{k+1}-\tilde{V}^{wn}_k \succ 0$ and hence, by the \emph{Lyapunov instability theorem} \cite[Theorem 3.3]{khalil2002nonlinear}, the error system is unstable. Therefore, the conditions in \eqref{eq:sability-necessity-1} are necessary for the stability of the observer. To do so, first note that 
 \begin{align} \label{eq:deltavtilda}
\hspace{-0.2cm}\Delta \tilde{V}^{wn}_k&\hspace{-0.05cm}=\Delta f^\top_k \Phi^\top (I-\tilde{L}C_2)^\top \tilde{P} (I-\tilde{L}C_2)\Phi \Delta f_k \\
\nonumber &\hspace{-0.35cm}+ \tilde{x}^\top_{k|k} (\Psi^\top \Phi^\top (I-\tilde{L}C_2)^\top \tilde{P} (I-\tilde{L}C_2)\Phi \Psi -\tilde{P})\tilde{x}_{k|k} \\
\nonumber &\hspace{-0.35cm}-2 \Delta f^\top_k  \Phi^\top (I-\tilde{L}C_2)^\top \tilde{P} (I-\tilde{L}C_2)\Phi \Psi \tilde{x}_{k|k}.
\end{align}
 Then, \eqref{eq:sability-necessity-1}, along with setting $\tilde{L}=\tilde{P}^{-1}\tilde{Y} \Rightarrow \tilde{Y} = \tilde{P}\tilde{L}$, defining $\tilde{S} \triangleq \tilde{P}-C^\top_2\tilde{Y}^\top -\tilde{Y}C_2$, $\hat{\Pi} \succeq 0$ and \syong{applying} Schur complement, 
 result in
 \begin{align} \label{eq:inequalities_tilde}
0 \prec \tilde{L}^\top \tilde{P} \tilde{L} \preceq \tilde{\Gamma} \preceq I,
\end{align}  
as well as 
\begin{align}
\label{eq:deltakey_tilde} &\Delta \tilde{V}^{wn}_k=\Delta f^\top_k \Phi^\top ( \tilde{S} + C^\top_2 \tilde{L}^\top \tilde{P} \tilde{L} C_2 )\Phi \Delta f_k \\
\nonumber &\ \ + \tilde{x}^\top_{k|k}(\Psi^\top \Phi^\top (\tilde{S} + C^\top_2 \tilde{L}^\top \tilde{P} \tilde{L} C_2)\Phi \Psi-\tilde{P})\tilde{x}_{k|k}\\
\nonumber &\ \ -2\Delta f^\top_k \Phi^\top \tilde{S}\Phi \Psi \tilde{x}_{k|k}\hspace{-0.1cm} -2\Delta f^\top_k \Phi^\top C^\top_2 \tilde{L}^\top \tilde{P} \tilde{L} C_2\Phi \Psi \tilde{x}_{k|k}.
\end{align}
 Then, \eqref{eq:inequalities_tilde}, \eqref{eq:deltakey_tilde} and Lemma \ref{lem:decouple} imply that there exists $\tilde{\eta} >0$
 such that 
 \begin{align}\label{eq:HELP}
 \begin{array}{rl}
 &\Delta \tilde{V}^{wn}_k\geq \Delta f^\top_k \Phi^\top ( \tilde{S} + \tilde{\eta} C^\top_2  C_2 )\Phi \Delta f_k \\
 &\ \ + \tilde{x}^\top_{k|k}(\Psi^\top \Phi^\top (\tilde{S} + \tilde{\eta}C^\top_2  C_2)\Phi \Psi-\tilde{P})\tilde{x}_{k|k}\\
&\ \ -2\Delta f^\top_k \Phi^\top \tilde{S}\Phi \Psi \tilde{x}_{k|k}\hspace{-0.1cm} -\Delta f^\top_k \Phi^\top C^\top_2C_2\Phi\Delta f_k  \\
&\ \ - \tilde{x}^\top_{k|k}\hspace{-.05cm}\Psi^\top\hspace{-.05cm}\Phi^\top \hspace{-.05cm} C^\top_2 \hspace{-.05cm}C_2\Phi \Psi \tilde{x}_{k|k}\triangleq \overline{\Delta} \tilde{V}^{wn}_k  \hspace{-.1cm}.
\end{array}
\end{align}
\syong{As with the proof of Theorem \ref{thm:quadratic_stability}, we first consider Class 0 systems, where by} 
plugging $ \tilde{\Pi}_{11}$, $ \tilde{\Pi}_{12}$ and $ \tilde{\Pi}_{22}$ given in \eqref{eq:gen-nec-lmi} into $\tilde{\Pi}$, as defined in \eqref{eq:sability-necessity-1}, \syong{we obtain:} 
\begin{align}\label{eq:decrement}
0 < [\Delta f^\top_k \ \tilde{x}^\top_{k|k}] \tilde{\Pi}[ \Delta f^\top_k \ \tilde{x}^\top_{k|k}]^\top = \overline{\Delta} \tilde{V}^{wn}_k-\delta_f
\end{align}
where 
\begin{align}\label{eq:decrement_positive}
\nonumber \delta_f &\triangleq \Delta f_k^\top \hspace{-0.1cm} M_{11} \Delta f_k+2\tilde{x}_{k|k}^\top M_{12}^\top \Delta f_k+\tilde{x}_{k|k}^\top M_{22}\tilde{x}_{k|k} \\
 & =\begin{bmatrix} \Delta f_k^\top &  \tilde{x}^\top_{k|k} \end{bmatrix} \mathcal{M} \begin{bmatrix} \Delta f_k^\top & \tilde{x}^\top_{k|k} \end{bmatrix}^\top \geq \gamma \geq 0.
 \end{align}
 Finally, 
 \syong{from \eqref{eq:HELP}, we have}  $\Delta \tilde{V}^{wn}_k \geq \overline{\Delta} \tilde{V}^{wn}_k > \delta_f \geq 0 \Rightarrow \Delta \tilde{V}^{wn}_k >0$.
 
 \syong{Furthermore, the proof for system classes I--III can be obtained from the above result for Class 0 with suitable values of $M_{11}$, $M_{12}$ and $M_{22}$ as discussed in the proof of Theorem \ref{thm:quadratic_stability}.}
 } 
 \QEDA
 }                              
 \subsection{Proof of Lemma \ref{lem:stability-existence}}
To show that uniform detectability is sufficient for existence of an observer, notice that for a Class \ref{class:convexcomb} function $f_k(\cdot)$, \eqref{eq:errors-dynamics} can be written as 
\begin{align}
 \nonumber \tilde{x}_{k|k}&= ( I - \tilde{L} C_2 ) \overline{A}_{k-1} \tilde{x}_{k-1|k-1} \\
 \label{eq:X_tilda} &+ ( I - \tilde{L} C_2 ) \tilde{w}_{k-1}-\tilde{L} \tilde{v}_{k-1}, 
\end{align}
where 
\begin{align*}
\tilde{w}_{k-1}  &\triangleq - ( I - G_2 M_2 C_2 )(G_1 M_1v_{1,k-1}-{w}_{k-1})\\
&\quad - G_2 M_2 {v}_{2,k}, \\
  \overline{A}_k &\triangleq \Phi ( \sum_{i=1}^N \lambda_{i,k} {A}^{i} - \Psi ), \quad \tilde{v}_{k-1} \triangleq {v}_{2,k}.
  \end{align*}
   Now, consider the following linear time-varying system without unknown inputs:
\begin{align} \label{eq:equivalentsys}
x_{k+1}=\overline{A}_k x_k + \tilde{w}_k, y_k=C_2 x_k + \tilde{v}_k.
\end{align}
Systems \eqref{eq:X_tilda} and \eqref{eq:equivalentsys} are equivalent from the viewpoint of estimation, since the estimation error equations for both problems are the same, hence they both have the same objective. Therefore, the pair $(\overline{A}_k,C_2)$ needs to be uniformly detectable such that the observer is stable \cite[Section 5]{Anderson.Moore.1981}.

Moreover, as for the necessity of the strong detectability of the constituent LTI systems, suppose for contradiction, that there exists a stable observer for system \eqref{eq:system} with any sequence $\{\lambda_{i,k}\}_{k=0}^\infty$ for all $i \in \{1,2,\dots,N\}$ that satisfies $ 0 \leq \lambda_{i,k} \leq1 , \sum_{i=1}^N \lambda_{i,k} =1, \forall k $, but one of the constituent linear time-invariant systems (e.g., $ ({A}^{j},G,{C},H) $) is not strongly detectable. Since the observer exists for any sequence of $\lambda_{i,k}$, that means that an observer also exists when $\lambda_{j,k}=1$ and $\lambda_{i,k}=0$, $\forall i \neq j$ for all $k$. However, we know from \cite{yong2018simultaneous} that strong detectability is necessary for the stability of the linear time-invariant system $ ({A}^{j},G,{C},H) $, which is a contradiction. Hence, the proof is complete.\QEDA
\subsection{Proof of Theorem \ref{thm:noise-attenuation-general}}
We use a similar approach as in the proof of Theorem \ref{thm:quadratic_stability}. 
First, 
consider the error dynamics with bounded noise signals \eqref{eq:errors-dynamics} and the candidate Lyapunov function $V^n_k \triangleq \tilde{x}^\top_{k|k} P \tilde{x}_{k|k}$. Observe that 
\begin{align} \label{eq:deltavn}
\Delta V^n_k \triangleq V^n_{k+1}-V^n_k = \Delta V^{wn}_k + \Delta r_k,
\end{align}
 where $V^{wn}_k$ is the Lyapunov function for the error dynamics without noise signals, defined in \eqref{eq:delta_V_wn}, and  
 \begin{align} \label{eq:deltrar}
\nonumber \Delta r_k &\triangleq2(\Delta f^\top_k-\tilde{x}^\top_{k|k} \Psi^\top )\Phi^\top (I-\tilde{L}C_2)^\top P \mathcal{W}(\tilde{L})\overline{w}_k\\
&+\overline{w}^\top_k \mathcal{W}(\tilde{L})^\top P \mathcal{W}(\tilde{L}) \overline{w}_k,
 \end{align}
 with $\Phi,\Psi,\overline{w}_k$ and $\mathcal{W}(\tilde{L})$ defined in Lemma \ref{lem:error-dynamics}. We will show for each system class we consider 
 that 
 \begin{align} \label{eq:delta-r-inequality}
 \overline{\Delta} r_k \triangleq \Delta r_k-\rho^2 \overline{w}^\top_k \overline{w}_k+\tilde{x}^\top_k(I-\alpha P) \tilde{x}_k \leq 0.
 \end{align}
  Then, by \eqref{eq:deltavn} and \eqref{eq:delta-r-inequality} in addition to the fact that $\Delta V^{wn}_k \leq -\alpha \tilde{x}^\top_k P \tilde{x}_k$ (follows from Theorem \ref{thm:quadratic_stability}), we have 
 \begin{align}\label{eq:delta-inequality}
 \Delta V^n_k \leq \rho^2 \overline{w}^\top_k \overline{w}_k-\tilde{x}^\top_{k|k} \tilde{x}_{k|k}.
 \end{align}
 Summing up both sides of \eqref{eq:delta-inequality} from zero to infinity, returns $V^n_{\infty}-V^n_0 \leq \rho^2 \textstyle{\sum}_{k=0}^{\infty} \overline{w}^\top_k \overline{w}_k - \textstyle{\sum}_{k=0}^{\infty} \tilde{x}^\top_{k|k} \tilde{x}_{k|k}=\rho^2 \textstyle{\sum}_{k=0}^{\infty} \vec{w}_{i}^\top \vec{w}_{i}- \textstyle{\sum}_{k=0}^{\infty} \tilde{x}^\top_{k|k} \tilde{x}_{k|k}$, where at each time step $k$, $\vec{w}^\top_k=\begin{bmatrix} w^\top_k & v^\top_k \end{bmatrix}^\top$. Then, it follows from setting the initial conditions to zero that $\textstyle{\sum}_{k=0}^{\infty} \tilde{x}^\top_{k|k} \tilde{x}_{k|k} \leq \rho^2 \textstyle{\sum}_{k=0}^{\infty} \vec{w}_{i}^\top \vec{w}_{i} $. 
 
Thus, it remains to show that \eqref{eq:delta-r-inequality} holds for each system class 0--III. 
Plugging the expression for $\mathcal{W}(\tilde{L})$ from Lemma \ref{lem:error-dynamics} into \eqref{eq:deltrar}, we obtain 
 \begin{align}\label{eq:Hinf_help}
 \nonumber  \overline{\Delta} r_k & = \tilde{x}^\top_{k|k}(I-\alpha P) \tilde{x}_{k|k}+ 2 (\Delta f_k-\Psi \tilde{x}_{k|k})^\top \Phi^\top (PR\\
 \nonumber   & \quad  -Y\Omega-C^\top_2 Y^\top R+ C^\top_2 \tilde{L}^\top P \tilde{L}\Omega)\overline{w}_k +\overline{w}^\top_k (R^\top P R \\
   & \quad -2\Omega^\top Y^\top R 
   +\Omega^\top \tilde{L}^\top P \tilde{L} \Omega - \rho^2I )\overline{w}_k,
   \end{align}
   On the other hand, note that $\Pi \succeq 0 \Leftrightarrow I-\Gamma \succeq 0$ and $\begin{bmatrix} \Gamma & Y^\top \\ Y & P \end{bmatrix} \succeq 0$, which by pre- and post-multiplication by $\begin{bmatrix} I & 0 \\ 0 & P^{-1} \end{bmatrix}$ and the fact that $Y=P\tilde{L}$, is equivalent to  $I-\Gamma \succeq 0$ and $\begin{bmatrix} \Gamma & \tilde{L}^\top \\ \tilde{L} & P^{-1} \end{bmatrix} \succeq 0$. Applying Schur complement to the latter, $\Pi \succeq 0$ is equivalent to 
\begin{align} \label{eq:inequalities}
0 \preceq \tilde{L}^\top P \tilde{L} \preceq \Gamma \preceq I.
\end{align} 
    Now, \eqref{eq:Hinf_help}, \eqref{eq:inequalities} and Lemma \ref{lem:decouple}, imply that:
 \begin{align}
 \nonumber   \overline{\Delta} r_k &\leq \overline{w}^\top_k (R^\top P R \hspace{-0.05cm}-\hspace{-0.05cm}2R^\top \hspace{-.1cm}Y \Omega \hspace{-0.05cm}+\hspace{-0.05cm}\Omega^\top \Gamma \Omega\hspace{-0.1cm}+\hspace{-.1cm}(\varepsilon^{-1}_{1}\hspace{-.1cm}+\hspace{-.1cm}\varepsilon^{-1}_{2})\Omega^\top \Omega \\
\nonumber    & -\hspace{-.1cm} \rho^2 I)\overline{w}_k\hspace{-.1cm}+\hspace{-.1cm}\tilde{x}^\top_{k|k}(I\hspace{-.1cm}-\hspace{-.1cm}\alpha P\hspace{-.1cm}+\hspace{-.1cm}\varepsilon_{1}\Psi^\top \Phi^\top C^\top_2 C_2 \Phi \Psi)\tilde{x}_{k|k}\\ 
  \nonumber  &+2 \Delta f^\top_k \Phi^\top (PR-Y\Omega-C^\top_2Y^\top R)\overline{w}_k \\ 
  \label{eq:Delta-r-bar}  &-2 \tilde{x}^\top_{k|k}\Psi^\top \Phi^\top (PR-Y\Omega -C^\top_2 Y^\top R)\overline{w}_k \\
  \nonumber &+\varepsilon_{2}\Delta f^\top_k (\Phi^\top C^\top_2 C_2 \Phi) \Delta f_k \triangleq \tilde{\Delta}r_k.
  \end{align}
  
  \syong{We first consider Class 0 systems, where the fact that $f_k(\cdot)$ is ($\mathcal{M},\gamma$)-QC with $\gamma \geq 0$ implies that 
  \begin{align*}
  &-\Delta f^\top_kM_{11} \Delta f_k-\tilde{x}^\top_{k|k}M_{22}\tilde{x}_{k|k}-2\Delta f_k^\top M_{12}^\top\tilde{x}_{k|k} \\
  &\triangleq \tilde{S} \leq -\gamma \leq 0,
  \end{align*}
  which, in addition to \eqref{eq:Delta-r-bar}, return $\overline{\Delta} r_k \leq \tilde{\Delta}r_k-\tilde{S}+\tilde{S}= -\zeta^\top \mathcal{N}\zeta+\tilde{S} \leq 0$, where $\zeta \triangleq \begin{bmatrix} \overline{w}^\top_k & \tilde{x}^\top_k & \Delta f^\top_k \end{bmatrix}^\top$ and $\mathcal{N}$ is the matrix in \eqref{eq:noise-attenuation-general} with its elements defined in \eqref{eq:elements-gen} and \eqref{eq:general_class_opt}. Finally, we can obtain the results for system classes I--III with suitable values of $M_{11}$, $M_{12}$ and $M_{22}$ as described in the proof of Theorem \ref{thm:quadratic_stability}.} 
\QEDA
\moham{\subsection{Proof of Theorem \ref{prop:radii-existence}}
Consider the noisy state error dynamics system in \eqref{eq:errors-dynamics}, where we treat the augmented noise signal $\overline{w}_k$ as an external input to the system. Recall from the proof of Theorem \ref{thm:noise-attenuation-general} that as a result of $\mathcal{H}_{\infty}$ observer design, the Lyapunov function $V^n_k \triangleq \tilde{x}^\top_{k|k} P \tilde{x}_{k|k}$ satisfies \eqref{eq:delta-inequality}, which is equivalent to $\Delta V^n_k \leq -\alpha_3(\|\tilde{x}_{k|k}\|)+\sigma(\|\overline{w}_k\|)$, with the $\mathcal{K}_{\infty}$-function\footnote{A function $\alpha : \mathbb{R}_+ \to \mathbb{R}_+$ is a $\mathcal{K}_{\infty}$-function if it is a $\mathcal{K}$-function and in addition $\alpha(s) \to \infty$ as $s \to \infty$.} $\alpha_3(r) \triangleq r^2$ and the $\mathcal{K}$-function $\sigma(s) \triangleq \eta^2 s^2$. This implies that $V^n_k$ is an ISS-Lyapunov function for system \eqref{eq:errors-dynamics} (cf. \cite[Definition 3.2]{jiang2001input}), i.e., \eqref{eq:errors-dynamics} admits an \syong{Input-to-State Stable (ISS)}-Lyapunov function. Equivalently, \eqref{eq:errors-dynamics} is  \syong{ISS and} 
UBIBS, and admits a $\mathcal{K}$-asymptotic gain by \cite[Theorem 1]{jiang2001input}. \QEDA 
}
\subsection{Proof of Theorem \ref{thm:error_bound}}
\moham{
\syong{In this theorem, we aim to find two upper bounds of $\|\tilde{x}_{k|k}\|$ such that} that $\|\tilde{x}_{k|k}\| \leq \overline{\delta}^x_{k,1}$ and $\|\tilde{x}_{k|k}\| \leq \overline{\delta}^x_{k,2}$. To 
\syong{derive the first upper bound,} notice that it follows from \eqref{eq:delta-inequality} that 
 \begin{align*}
\tilde{x}^\top_{k|k}P \tilde{x}_{k|k} \leq \rho^2 \overline{w}^\top_k \overline{w}_k+\tilde{x}^\top_{k-1|k-1}(P-I) \tilde{x}_{k-1|k-1},
 \end{align*}
\syong{and by applying Rayleigh's inequality, we obtain} 
  \begin{align*}
&\lambda_{\min}(P) \|\tilde{x}_{k|k}\|^2 \leq \rho^2 \overline{w}^\top_k \overline{w}_k+\lambda_{\max}(P-I) \|\tilde{x}_{k-1|k-1}\|^2 \\
& \Rightarrow \|\tilde{x}_{k|k}\| \leq \sqrt{\frac{\rho^2 \overline{w}^\top_k \overline{w}_k}{\lambda_{\min}(P)}+\frac{|\lambda_{\max}(P-I)|}{\lambda_{\min}(P)}\|\tilde{x}_{k-1|k-1}\|^2}.
 \end{align*}
 Repeating this \syong{procedure} 
 $k$ times and considering the fact that $\lambda_{\max}(P-I)=\lambda_{\max}(P)-1$ as a consequence of Weyl's Theorem \cite[\mohk{Theorem 4.3.1}]{horn2012matrix}, we obtain $\tilde{x}_{k|k} \leq \overline{\delta}^x_{k,1}$ \syong{with $\overline{\delta}^x_{k,1}$ given in \eqref{eq:delta1} and $\theta_1=\mohk{\frac{|\lambda_{\max}(P)-1|}{\lambda_{\min}(P)}}$}.
 }
 
\moh{
Next, \syong{we find the second upper bound for} 
$\|\tilde{x}_{k|k}\| \leq  \overline{\delta}^x_{k,2} \triangleq \delta^x_0 \theta_2^k +  \overline{\eta} \textstyle\sum_{i=1}^k \theta_2^{i-1}$ \syong{for system classes I--III}. }
\begin{enumerate}[(I)]
 \item If $f_k(\cdot)$ is a Class \ref{class:Lip} function, then, the result in \eqref{eq:stateradius} with $\theta$ defined in \eqref{eq:errors1}, directly follows from Lipschitz continuity of $f_k(\cdot)$, as well as applying triangle and sub-multiplicative inequalities for norms on  \eqref{eq:errors-dynamics}. Moreover, the result in \eqref{eq:inputradius} with $\beta$ defined in \eqref{eq:errors1}, is obtained by triangle and sub-multiplicative inequalities, \eqref{eq:augmenting-d}, \eqref{eq:inputerror111} and \eqref{eq:Inputerror222}.\\
 \item If $f_k(\cdot)$ is a Class \ref{class:A} function, then by Proposition \ref{prop:qcstar_Lip}, it is a Class \ref{class:Lip} function with $L_f=\sqrt{\lambda_{\max}(\mathcal{A}^\top \mathcal{A})}$. The rest of the proof is similar to the proof for {Class} \ref{class:Lip}.   
\item If $f_k(\cdot)$ is a Class \ref{class:convexcomb} function, we first 
find closed-form expressions for the state and input estimation errors through the following lemma.\\ 
\balance
\begin{lem}\label{lem:error_closedform}
	The state and input estimation errors are 
\vspace{-0.5cm}
\small	\begin{align*}
	 \tilde{x}_{k|k}\hspace{-0.05cm}&=\hspace{-0.05cm}\hspace{-0.05cm}\sum_{i=1}^k \prod_{j=0}^{i-2}A_{e,k-j}(\Psi \tilde{w}_{k-i} \hspace{-0.05cm}-\hspace{-0.05cm} \tilde{L} \tilde{v}_{k-i})+\prod_{j=0}^{k-1} A_{e,k-j}\tilde{x}_{0|0},\\ 
 \tilde{d}_{k-1}\hspace{-0.05cm}&=\hspace{-0.05cm}-\textstyle\sum_{i=1}^N \lambda_{i,k-1} (V_1 M_1 C_1 + V_2 M_2 C_2 A_{e,i}) \tilde{x}_{k-1|k-1}   \\ \nonumber & \quad +(V_2M_2C_2G_1M_1\hspace{-0.1cm}-\hspace{-0.1cm}V_1M_1)T_1 v_{k-1}
  \hspace{-0.1cm}-\hspace{-0.1cm}V_2 M_2 C_2 w_{k-1}\hspace{-0.1cm}\\
  & \quad -\hspace{-0.1cm}V_2M_2T_2 v_k. 
	\end{align*}
	\normalsize
\end{lem}

\noindent \begin{proof}
Starting from \eqref{eq:X_tilda} and applying simple induction return the results for the state errors. Then, the expression for the input errors follows from \eqref{eq:inputerror111}, \eqref{eq:Inputerror222} and \eqref{eq:augmenting-d}.
\end{proof}

Now, we are ready to show that $\|\tilde{x}_{k|k}\| \leq  \overline{\delta}^x_{k,2} \triangleq  \delta^x_0 \theta_2^k +  \overline{\eta} \textstyle\sum_{i=1}^k \theta_2^{i-1}$ 
for LPV (Class III) functions. First, we define
\begin{gather} \label{BekCektk}
\begin{array}{c}
 B_{e,k} \triangleq \textstyle\prod_{j=0}^{k-1} A_{e,k-j}, \\
   C^i_{e,k} \triangleq \textstyle\prod_{j=0}^{i-2} A_{e,k-j}, \
 \tilde{t}_k \triangleq  \Psi \tilde{w}_k - \tilde{L} \tilde{v}_k,
 \end{array}
    \end{gather} 
   for $ 1 \leq i \leq k $. Then,
  from Lemma \ref{lem:error_closedform}, we have
\begin{align}\label{eq:xtildanorm_inequal}
 \| \tilde{x}_{k|k}\| 
\leq \| B_{e,k} \| \| \tilde{x}_{0|0} \| +\| \textstyle\sum_{i=1}^k C^i_{e,k} \overline{t}_{k-i}  \|, 
\end{align}
by triangle inequality and submultiplicativity of norms. Moreover, by \syong{a} similar reasoning, we find 
\begin{align} 
\nonumber &\| B_{e,k} \| 
\hspace{-0.1cm}\leq \hspace{-0.1cm} \| \textstyle\prod_{j=0}^{k-1} \textstyle\sum_{i=1}^N \lambda^i_{k-j}  \Psi \Phi (A^i \hspace{-0.05cm}-\hspace{-0.05cm} G_1M_1C_1) \| 
\hspace{-0.1cm}\leq \hspace{-0.1cm}\syong{\theta_2^k} \hspace{-0.05cm}, \\
\label{eq:C_ek}&\|\sum_{i=1}^k C^i_{e,k} \overline{t}_{k-i}  \|
 \hspace{-0.1cm}\leq\hspace{-0.1cm} \sum_{i=1}^k \|C^i_{e,k} \| \| \overline{t}_{k-i} \|, \\
  \nonumber  &\| C^i_{e,k} \|  \leq 
 \textstyle\prod_{j=0}^{i-2} \| \textstyle\sum_{s=1}^N \lambda_{s,k-j} A_{e,s} \|  \leq \theta_2^{i-1}. 
\end{align}
Moreover, from \eqref{BekCektk}, {we have} 
\begin{align}
 \| \tilde{t}_{k-i}\| =\| \Re v_{k-i} +\Psi \Phi w_{k-i}\| \leq  \overline{\eta}, \label{eq:tbar_ineq}
\end{align}
with $\Re \triangleq -(\Psi \Phi G_1 M_1 T_1 +\Psi G_2 M_2 T_2 + \tilde{L} T_2)$. Then, from \eqref{eq:xtildanorm_inequal}--\eqref{eq:tbar_ineq}, we obtain  \eqref{eq:stateradius} with $\syong{\theta_2}$ and $\overline{\eta}$ defined in \eqref{eq:errors3}. 
  Furthermore, the result in \eqref{eq:inputradius} with $\beta$ and $\overline{\alpha}$ defined in \eqref{eq:errors3}, follows from applying Lemma \ref{lem:error_closedform}, as well as triangle inequality, the facts that $ 0 \leq \lambda_{i,k} \leq 1, \textstyle\sum_{i=1}^N \lambda_{i,k}=1 $ and sub-multiplicativity of matrix norms. 
 \end{enumerate}
 Further, the steady state values are obtained by taking the limit from both sides of \eqref{eq:delta1}, \eqref{eq:stateradius} and \eqref{eq:inputradius}, \syong{and assuming that $\theta \triangleq \min(\theta_1,\theta_2) <1$.  } 
 \QEDA
 \subsection{Proof of Corollary \ref{cor:convergence-lpv}}
 Clearly $\|A_{e,i}\|<1$ implies that $\theta <1$, which is a sufficient condition for the convergence of errors by Theorem \ref{thm:error_bound}. 
 \QEDA
 \moham{
 \subsection{Proof of 
 \syong{Proposition} \ref{lem:theta}}
 The stability of the observer and the noise attenuation level follow directly from Theorems \ref{lem:stability-existence} and \ref{thm:noise-attenuation-general}, respectively. Moreover, we will show that the additional LMIs
 \begin{align}
\label{eq:kappa1} &\kappa_1I \preceq P \preceq \kappa_2 I, \\
\label{eq:kappa2} &(\kappa_1 \geq 1, \kappa_2-\kappa_1 <1) \ \syong{\vee} 
\ (\kappa_2 \leq 1, \kappa_1 >0.5),
\end{align}
  imply that $\theta_1 \triangleq \syong{{\frac{|\lambda_{\max}(P)-1|}{\lambda_{\min}(P)}}} <1$, which guarantees that $\theta =\min (\theta_1,\theta_2) <1$ and \syong{therefore,} the radii upper sequences are convergent by Theorem  \ref{thm:error_bound}. To do so, we consider two cases:
 \begin{itemize}
 \item Case A) $\kappa_1 \geq 1, \kappa_2-\kappa_1 <1$. This, along with \eqref{eq:kappa1} result in $\lambda_{\max}(P) \geq \kappa_1 \geq 1$ and $|\lambda_{\max}(P)-1|=\lambda_{\max}(P)-1 \leq \kappa_2-1 < \kappa_1 \leq \lambda_{\min}(P)$. Hence, $\theta_1 \triangleq \syong{{\frac{|\lambda_{\max}(P)-1|}{\lambda_{\min}(P)}}} <1$.
  \item Case B) $\kappa_2 \leq 1, \kappa_1 >0.5$. This, in addition to \eqref{eq:kappa1} return $\lambda_{\max}(P) \leq \kappa_2 \leq 1$ and $|\lambda_{\max}(P)-1|=1-\lambda_{\max}(P) \leq 1-\kappa_1 < \kappa_1 \leq \lambda_{\min}(P)$. So, $\theta_1 \triangleq \syong{{\frac{|\lambda_{\max}(P)-1|}{\lambda_{\min}(P)}}} <1$. \QEDA
 \end{itemize}
 }
  \end{document}